\documentclass[aps,pra,twocolumn,groupedaddress,showpacs,notitlepage]{revtex4-1}
\usepackage[colorlinks,linkcolor={blue},citecolor={blue},urlcolor={blue}]{hyperref} 
\usepackage{graphicx}
\usepackage{subfigure}
\usepackage{amsmath}
\usepackage{braket}
\usepackage{soul} 
\usepackage[T1]{fontenc}
\usepackage[utf8]{inputenc}
\usepackage{float}
\usepackage{marvosym} 
\usepackage{wasysym} 
\usepackage{amssymb} 
\usepackage[normalem]{ulem}  

\makeatletter
\def\maketitle{
\@author@finish
\title@column\titleblock@produce
\suppressfloats[t]}
\makeatother

\newcommand{\bb}{\boldsymbol{b}}
\newcommand{\bk}{\boldsymbol{k}}

\newcommand{\bp}{\boldsymbol{p}}
\newcommand{\br}{\boldsymbol{r}}

\newcommand{\bv}{\boldsymbol{v}}

\newcommand{\bz}{\boldsymbol{z}}
\newcommand{\bA}{\boldsymbol{A}}
\newcommand{\bB}{\boldsymbol{B}}

\newcommand{\bR}{\boldsymbol{R}}

\newcommand{\bCalA}{\boldsymbol{\mathcal{A}}}
\newcommand{\bCalB}{\boldsymbol{\mathcal{B}}}
\newcommand{\CalA}{\mathcal{A}}

\newcommand{\CalE}{\mathcal{E}}
\newcommand{\CalH}{\mathcal{H}}

\newcommand{\bs}{\boldsymbol}

\newcommand{\hs}{\mathbb}
\newcommand{\bsigma}{\boldsymbol{\sigma}}
\newcommand{\bmu}{\boldsymbol{\mu}}

\newcommand{\bpi}{\boldsymbol{\pi}}

\newcommand{\hH}{\mathbb{H}}

\usepackage{xcolor}
\definecolor{mycolor}{RGB}{255,165,0}

\begin{document}



\title{Supersymmetry dictated topology in periodic gauge fields and realization in strained and twisted 2D materials}


\author{Dawei Zhai}
\email{dzhai@hku.hk}
\author{Zuzhang Lin}
\author{Wang Yao}
\email{wangyao@hku.hk}
\affiliation{New Cornerstone Science Laboratory, Department of Physics, The University of Hong Kong, Hong Kong, China}



\begin{abstract}

Supersymmetry (SUSY) of Hamiltonian dictates double degeneracy between a pair of superpartners (SPs) transformed by supercharge, except at zero energy where modes remain unpaired in many cases. 
Here we explore a SUSY of complete isospectrum between SPs -- with paired zero modes -- realized by 2D electrons in zero-flux periodic gauge fields, which can describe twisted or periodically strained 2D materials.
We find their low-energy sector containing zero (or threshold) modes must be topologically non-trivial, by proving that  Chern numbers of the two SPs have a finite difference dictated by the number of zero modes and energy dispersion in their
vicinity.
In $30^\circ$ twisted bilayer (double bilayer) transition metal dichalcogenides subject to periodic strain, we find one SP is topologically trivial in its lowest miniband, while the twin SP of identical dispersion 
has a Chern number of $1$ ($2$), in stark contrast to time-reversal partners that have to be simultaneously trivial or nontrivial.
For systems whose physical Hamiltonian corresponds to the square root of a SUSY Hamiltonian, such as twisted or strained bilayer graphene, we reveal that topological properties of the two SUSY SPs are transferred respectively to the conduction and valence bands, including the contrasted topology in the low-energy sector and identical topology in the high-energy sector. This offers a unified perspective for understanding topological properties in many flat-band systems described by such square-root models.
Both types of SUSY systems provide unique opportunities for exploring correlated and topological phases of matter.

\end{abstract}


\maketitle



\section{Introduction}

The discoveries of superconductivity and correlated insulating phases in twisted bilayer graphene (TBG)~\cite{TBGCaoYuan2018a,TBGCaoYuan2018b} have stimulated significant research interests in twisted and strained 2D materials that exhibit long-wavelength periodic spatial modulations. Many intriguing phenomena that have been observed in TBG~\cite{moireReviewEvaMacDonaldNatMater2020,moireReviewNatPhysBalents2020,moireReviewRubioNatPhys2021,moireReviewExptFolksNatRevMat2021,moireReviewJeanieLauNature2022} are made possible by its flat bands, which not only allow Coulomb interaction effects to stand out due to the narrow bandwidth but also host nontrivial band topology~\cite{TBGMagicAngleOriginPRL2019,DaiXiPseuodoFieldPRB2019}. 
Topological minibands with valley-opposite Chern numbers as dictated by time-reversal symmetry are also found in twisted homobilayer transition metal dichalcogenides (tTMDs)~\cite{WuMacDonaldPRL2019,HongyiNSR2020,ZhaiPRM2020,WuFengchengPRR2020,TwistedWSe2MagicFuLiang2021},
which underlie the observed fractional quantum anomalous Hall effects in tMoTe$_2$ when intrinsic ferromagnetism arises from Coulomb exchange~\cite{FCIMoTe2Eric2023,FCIMoTe2Jiaqi2023,FCIMoTe2ShanJie2023,MoTe2Park2023,FCIMoTe2PRX2023}. 
Interestingly, the electronic and topological properties of both TBG and tTMDs can be connected to periodic gauge fields. In TBG the relations between flat-bands and lowest Landau levels (LLs) are widely noted~\cite{TBGMagicAngleOriginPRL2019,DaiXiPseuodoFieldPRB2019,AshvinPRR2020,WangJiePRR2021,TBGLLPRB2021}, while in tTMDs the miniband topology can be attributed to an Abelian gauge field corresponding to the real-space Berry curvature acquired in the moir\'e~\cite{HongyiNSR2020,ZhaiPRM2020}.

Supersymmetry (SUSY) is an active topic of research in theoretical physics.
In quantum systems with SUSY~\cite{SUSYBookGeorg}, superpartners (SPs) with identical energy spectrum can emerge, reminiscent of the time-reversal partners represented by the two valleys in graphene or TMDs, whereas SPs are related by transformation known as supercharges. Initially proposed to relate fermions and bosons as SPs~\cite{SUSYPhysLett1973,SUSYNuclPhysB1974a,SUSYNuclPhysB1974b,SUSYNuclPhysB1975}, SUSY has been extended to encompass transformations between eigenstates of quantum Hamiltonians in general, providing a powerful tool across various frontiers of physics~\cite{SUSYBookGeorg}. 
Lately, there has been a surge of interest on the topological properties of lattice models that either exhibit SUSY or have indirect link to it~\cite{commentSUSYinSqrt}. Examples include topological mechanical lattices~\cite{SUSYTopoMechNatPhys2014,SUSYTopoMechPRR2019,SUSYLattice2022} and square-root topological materials~\cite{SqrtPRB2017,SqrtEzawaPRR2020,SqrtHatsugaiPRA2020,SqrtHatsugaiPRB2021}. A notable but not entirely unexpected finding is the two SPs can identically exhibit nontrivial topology in some high-energy sector~\cite{SUSYLattice2022,SqrtHatsugaiPRB2021}, which relies on the property that supercharges can transform SPs into each other at non-zero energy. This transformation, however, breaks down for the zero-energy modes, whose presence is required for defining an unbroken SUSY~\cite{SUSYBookGeorg,commentZeroModeSUSYbreaking}. Yet, the existing literature has mostly focused on models of either no zero mode or unpaired zero modes (e.g., a zero-energy flat band present only for one SP).

In this paper we present a systematic study on the topological properties of a SUSY of complete isospectrum between SPs, including paired zero modes, which can describe periodically strained TMDs on patterned substrate, as well as various flat-band superlattices, including e.g. TBG and strained bilayer graphene. We show that their low-energy sector containing zero (or threshold) mode must be topologically nontrivial, by proving a rigorous relation on the contrasted topology between SPs therein.
Specifically, in the lowest-energy bands of the two SPs that feature identical dispersion, we show a finite difference in their Chern numbers dictated by the number of zero modes and energy dispersion in their vicinity, which guarantees that at least one SP is topologically nontrivial in the low-energy sector. 
We propose realization of such SUSY in $30^\circ$ twisted bilayer (double bilayer) TMDs with substrate-patterned periodic strain, where one SP is topologically trivial while the other has finite Chern number $1$ ($2$), albeit their complete degeneracy in the lowest-energy miniband. This is in stark contrast to the band topology in the circumstance of double degeneracy dictated by time-reversal symmetry.

The distinct topology and quantum geometry of the isospectral SPs further suggests that spontaneous breaking of SUSY due to Coulomb interaction is accompanied by the emergence of correlation phenomena that are sensitive to topology or form factors of many-body interactions, e.g. fractional quantum anomalous Hall (FQAH) or composite Fermi liquid phases, versus charge density wave (CDW) or conventional Fermi liquid. 
This is also in sharp contrast to the  spontaneous symmetry breaking between time-reversal partners that comes in the form of magnetism to lower exchange energy.
At fractional filling of the lowest-energy miniband, we demonstrate the emergence of FQAH states and CDW in the twin SPs respectively based on exact diagonalization calculations.

For systems whose physical Hamiltonian corresponds to the square root of a SUSY Hamiltonian, such as TBG and strained bilayer graphene, we reveal that topological properties of the two SUSY SPs are transferred respectively to the conduction and valence bands of the square-root physical Hamiltonian, including the contrasted topology in the low-energy sector and identical topology in the high-energy sector. 
This provides unique opportunities for exploring topological excitons and optical phenomena in the context of topology-contrasted conduction and valence bands.
As many flat-band systems are described by such square-root models, our findings also offer a unified perspective for understanding their topological properties.

\section{Basics of SUSY}

\subsection{Pairing of superpartners between their positive energy states}
In this work we consider SUSY Hamiltonian of the form
\begin{equation}
\begin{aligned}
\CalH_{\text{susy}}=
\begin{pmatrix}
\hH_+&0\\0&\hH_-
\end{pmatrix}
=
\begin{pmatrix}
MM^{\dagger}&0\\0&M^{\dagger}M
\end{pmatrix}
\end{aligned},\label{Eq:Hsusy1}
\end{equation}
where $M$ denotes a generic matrix. 
Its connection to the SUSY quantum mechanics~\cite{SUSYBookGeorg} is revealed by rewriting
\begin{equation}
\CalH_{\text{susy}}=\{Q,\,Q^\dagger\},~\label{Eq:Hsusy=QQ}
\end{equation}
where
\begin{equation}
\begin{aligned}
Q=
\begin{pmatrix}
0&M\\0&0
\end{pmatrix}~~\text{and}~~
Q^\dagger=
\begin{pmatrix}
0&0\\M^{\dagger}&0
\end{pmatrix}
\end{aligned}\label{Eq:Q}
\end{equation}
are the complex supercharges (or SUSY generators).
One can readily verify that
\begin{equation}
	\{Q,\,Q\}=\{Q^\dagger,\,Q^\dagger\}=[\CalH_{\text{susy}},\,Q]=[\CalH_{\text{susy}},\,Q^\dagger]=0.
\end{equation}
These relations and Eq.~(\ref{Eq:Hsusy=QQ}) define the superalgebra and a SUSY quantum system~\cite{SUSYBookGeorg}. Such a SUSY system contains two parts ($\hH_{\pm}$), and particles characterized by $\hH_+\psi_{+}=\CalE_+\psi_{+}$ and $\hH_-\psi_{-}=\CalE_-\psi_{-}$ are considered as SPs of each other. The eigenvectors of $\CalH_{\text{susy}}$ can be expressed in terms of $\psi_\pm$ as $\Psi_+=(\psi_{+},\,0)^T$ and $\Psi_-=(0,\,\psi_{-})^T$.

A SUSY system has several characteristic properties.
First, as $\CalH_{\text{susy}}=\{Q,\,Q^\dagger\}$, its eigenenergies are non-negative. 
Second, the relation $[\CalH_{\text{susy}},\,Q]=[\CalH_{\text{susy}},\,Q^\dagger]=0$ dictates that any positive  eigenvalue, i.e. $\CalE>0$, must have a pair of eigenstates from the two SPs respectively, satisfying $\CalH_{\text{susy}}\Psi_{\pm}=\CalE\Psi_{\pm}$. They can transform into each other via the supercharges $\Psi_+=Q\Psi_-/\sqrt{\CalE}$ and $\Psi_-=Q^\dagger\Psi_+/\sqrt{\CalE}$,
or equivalently,
\begin{equation}
\psi_+=\frac{1}{\sqrt{\CalE}}M\psi_-,~~
\psi_-=\frac{1}{\sqrt{\CalE}}M^{\dagger}\psi_+. \label{Eq:SUSYWaveFunctionRelation}
\end{equation}

The zero (energy) modes, $\Psi_+^0=(\psi_{+}^{0},\,0)^T$ and $\Psi_-^0=(0,\,\psi_{-}^{0})^T$, of the SUSY system are the ones annihilated by the supercharges, $M^{\dagger}\psi_{+}^0=0$ and $M\psi_{-}^0=0$.
Thus the transformation connecting the SPs in Eq.~(\ref{Eq:SUSYWaveFunctionRelation}) fails for the zero modes. This indicates that the zero-energy modes may remain unpaired or even be absent~\cite{commentZeroModeSUSYbreaking}. 
The difference in the number of zero modes between the SPs is the Witten index, which can be employed to characterize the SUSY systems.

\subsection{Isospectrum in a zero-flux periodic gauge field}

A nontrivial scenario where SUSY emerges corresponds to 2D electrons in 
a gauge field. 
We consider $M$ and $M^\dagger$ given in terms of $(\pi_x,\,\pi_y)=\bp+e\bA$, where $\bp$ is the momentum operator and $\bA$ is the gauge potential:
\begin{equation}
\begin{aligned}
M=\alpha_{N} (\pi_x-i\pi_y)^N=\alpha_{N}\pi_-^N\\
M^{\dagger}=\alpha_{N} (\pi_x+i\pi_y)^N=\alpha_{N}\pi_+^N
\end{aligned}.\label{Eq:M_special}
\end{equation}
Here $N$ denotes a positive integer, $\alpha_{N}$ is a constant such that $MM^{\dagger}$ and $M^{\dagger}M$ have units of energy, and we have also defined $\pi_\pm=\pi_x\pm i\pi_y$. 
The SUSY Hamiltonian is then
\begin{equation}
\begin{aligned}
\CalH_{\text{susy}}=
\begin{pmatrix}
\hH_+&0\\0&\hH_-
\end{pmatrix}
=
\begin{pmatrix}
\alpha_{N}^{2} \pi_-^{N} \pi_+^{N}&0\\0&\alpha_{N}^{2} \pi_+^{N} \pi_-^{N}
\end{pmatrix}
\end{aligned}.~\label{Eq:SUSYHamN}
\end{equation}
The out-of-plane gauge field $\bB=B\hat{\bz}$ satisfies $[\pi_x,\,\pi_y]=-ie\hbar B$. 
We will adhere to the above conventions for assigning $\hH_+$ and $\hH_-$ throughout the manuscript.
In the following we will assume that the field is Abelian, the discussions will be extended to cover non-Abelian gauge fields later.

In the case of $N=1$, the model describes spin--$\frac{1}{2}$ 2D electrons in an out-of-plane magnetic field,
\begin{equation}
	M=\frac{1}{\sqrt{2m}}\pi_-,~~M^\dagger=\frac{1}{\sqrt{2m}}\pi_+~~(\text{when}~ N=1),\label{Eq:M_for_N=1}
\end{equation}
and
\begin{equation}
\begin{aligned}
\hH_\pm^{(N=1)}=\frac{1}{2m}\pi_\mp\pi_\pm=\frac{1}{2m}(\bp+e\bA)^2\pm\bmu_B\cdot\bB
\end{aligned},\label{Eq:SchrodingerWithB}
\end{equation}
where $\bmu_B=\frac{e\hbar}{2m}\hat{\bz}$ is the Bohr magneton.
Spin-up and spin-down electrons are SPs of each other in this case.

\begin{figure}[t]
	\centering
	\includegraphics[width=3.4in]{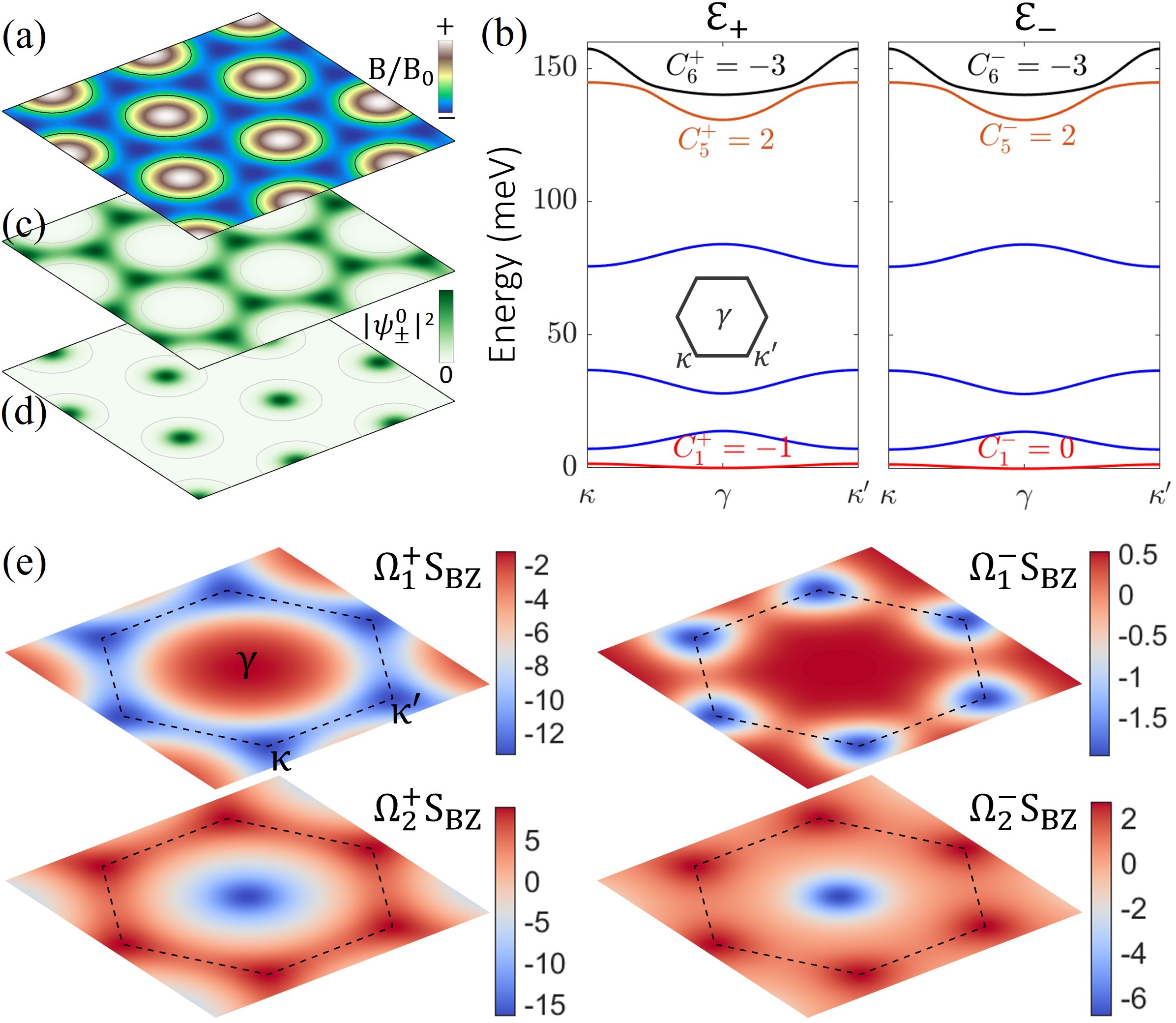}
	\caption{(a) A zero-flux periodic magnetic field, where black circles denote $B=0$ contours.	(b) Energy bands and Chern numbers $C_n^{\pm}$ in the case of $N=1$ for $\CalE_\pm$ from $\hH_{\pm}$ in Eq.~(\ref{Eq:SchrodingerWithB}). Blue bands have $C_n^{\pm}=0$. (c, d) Zero-mode distribution $|\psi_{\pm}^{0}|^2$ from $\CalE_+$ and $\CalE_-$, respectively. (e) Berry curvature of the first two bands of $\CalE_\pm$, the values are scaled by the area of the BZ $S_{BZ}$. The high-symmetry points $\gamma=(0,\,0)$, $\kappa=(-\frac{1}{2},\,-\frac{\sqrt{3}}{2})\frac{4\pi}{3L}$, and $\kappa'=(\frac{1}{2},\,-\frac{\sqrt{3}}{2})\frac{4\pi}{3L}$ are labeled in (b) and (e). Parameters: $B_0=120$ T, $L=14$ nm, $m=0.1$ free electron mass.}
	\label{Fig:QuadraticTriB}
\end{figure}

Information of the zero modes, i.e., their degeneracy and wave functions, can be obtained by solving $\pi_\pm\psi_{\pm}^0=0$, which depend on details of $\bB$. Here we distinguish two scenarios: 
\newline
(i) The situation of $\bB$ with nonzero flux is known: the zero modes are spin-polarized with its degeneracy determined by the flux [see e.g., Fig.~\ref{Fig:LLs_schematics}(a)], and notably, these features persist even when $\bB$ is spatially nonuniform~\cite{AharonovCasherPRA1979}.
\newline
(ii) The scenario of periodic $\bB$ with zero flux exhibits unique characteristics that have not been uncovered, and we will focus on them in this work. In this particular case, the two SPs each contains one unique zero mode, $\psi_{\pm}^{0}=c_{\pm}e^{\pm \phi}$,
where $c_{\pm}$ are normalization constants, and the Coulomb gauge has been used with $\phi$ satisfying $e\bA/\hbar=-\nabla\times(\phi\hat{\bz})$ and $\partial_{\br}^{2}\phi=eB/\hbar$.
Note that, although here we start with the case of $N=1$, the existence of such zero modes remains valid when $N>1$.
Similar zero modes have also been found in graphene with a periodic gauge field~\cite{SnymanPRB2009,PhongZeroMode} (also see Sec.~\ref{Sect:Sqrt}).
Therefore, the spectra of two SPs are exactly the same when $\bB$ is zero-flux periodic.
This isospectral behavior is confirmed by solving Eq.~(\ref{Eq:SchrodingerWithB}) in a zero-flux $\bB$ field of periodicity $L$ [Fig.~\ref{Fig:QuadraticTriB}(a)]: $\bB(\br)=\hat{z} B_0\sum_{i=1}^{3}\cos\left(\bb_{i}\cdot\br\right)$, $\bb_{1}=(\frac{\sqrt{3}}{2},\,-\frac{1}{2})\frac{4\pi}{\sqrt{3}L}$ 
and $\bb_{i}=C_3^{i-1}\bb_{1}$, where $C_3$ denotes the three-fold rotation operation.
As expected, $\CalE_+$ and $\CalE_-$ each has one zero mode located at the $\gamma$ point [Fig.~\ref{Fig:QuadraticTriB}(b)].


\section{Contrasted topology in the isospectrum}

Now we present the main findings of this work: the isospectral SPs in a zero-flux periodic $\bB$ field are contrasted by distinct topology in their lowest bands that contain the zero modes, while all other bands have identical Chern numbers.

To figure out the relation of band Chern numbers of the SPs, it is convenient to look at their Berry connection $\bR_n^{\pm}=i\braket{u_{n\bk}^{\pm}|\nabla_{\bk} u_{n\bk}^{\pm}}$, where $n=1,\,2,\cdots$ label the bands in ascending order of energy, and we denote the Bloch states of the SPs as $\ket{\psi_{\pm}}=e^{i\bk\cdot\br}\ket{u_{n\bk}^{\pm}}$. 
The Berry curvatures read $\Omega_{n}^{\pm}=[\nabla_{\bk}\times\bR_n^{\pm}]_z$, and their integrals over the Brillouin zone (BZ) yield the band Chern numbers $C_n^{\pm}=\int_{\text{BZ}}\Omega_{n}^{\pm}\,d^{2}\bk/(2\pi)$.
By using the relation in Eq.~(\ref{Eq:SUSYWaveFunctionRelation}), one finds that $\bR_n^{+}$ and $\bR_n^{-}$ differ by (see Sec.~S1 of the Supplementary Material~\cite{supp})
\begin{equation}
\begin{aligned}
\delta\bR_n&=\bR_n^{-}-\bR_n^{+}\\
&=\frac{i}{\CalE_{n\bk}}\braket{u_{n\bk}^+|M(\nabla_{\bk}M^{\dagger})|u_{n\bk}^+}-i\frac{\hbar\bv_{n\bk}}{2\CalE_{n\bk}}
\end{aligned}.~\label{Eq:BerryConnection}
\end{equation}
Here $\CalE_{n\bk}>0$ is a common eigenenergy shared by $\ket{u_{n\bk}^{\pm}}$, and $\bv_{n\bk}=(v_{n\bk}^{x},\,v_{n\bk}^{y})=\nabla_{\bk}\CalE_{n\bk}/\hbar$ is the band velocity. 
In the presence of zero modes, $\delta\bR_{n}$ becomes singular as $\CalE_{n\bk}\rightarrow0$, thus different topology is expected between the $n=1$ bands of the SPs.

Consider $M$ and $M^\dagger$ given in Eq.~(\ref{Eq:M_special}), one straightforwardly finds
\begin{equation}
\delta\bR_n=\frac{\hbar}{2\CalE_{n\bk}}(v_{n\bk}^{y},\,-v_{n\bk}^{x}).\label{Eq:BerryConnection_v}
\end{equation}
For any band except the first (i.e., $n>1$), since $\CalE_{n\bk}\ne0$, $\delta\bR_n$ is well-behaved and periodic in the entire BZ, so from Stokes' theorem, one readily identifies that the difference between the Chern numbers of the two SPs must vanish:
\begin{equation}
	C_n^{-} - C_n^{+} = 0,~~\text{for $n>1$}.\label{Eq:IdenticalC}
\end{equation}
Note that although the SPs have identical dispersion and Chern numbers, their Berry curvatures in general differ,
\begin{equation}
\Omega_{n}^{-}-\Omega_{n}^{+}=-\frac{\hbar}{2}\nabla\cdot\frac{\bv_{n\bk}}{\CalE_{n\bk}},~~\text{for $n>1$}
\end{equation}
as shown by the example in Fig.~\ref{Fig:QuadraticTriB}(e).
This is in contrast to the situation in some lattice models connected by SUSY, where identical Berry curvatures have been found~\cite{SUSYLattice2022}.
In the present case, Berry curvatures of the SPs can be identical throughout the BZ when the bands become flat with $\bv_{n\bk}\equiv0$. 

\begin{figure}[t]
	\centering
	\includegraphics[width=3.2in]{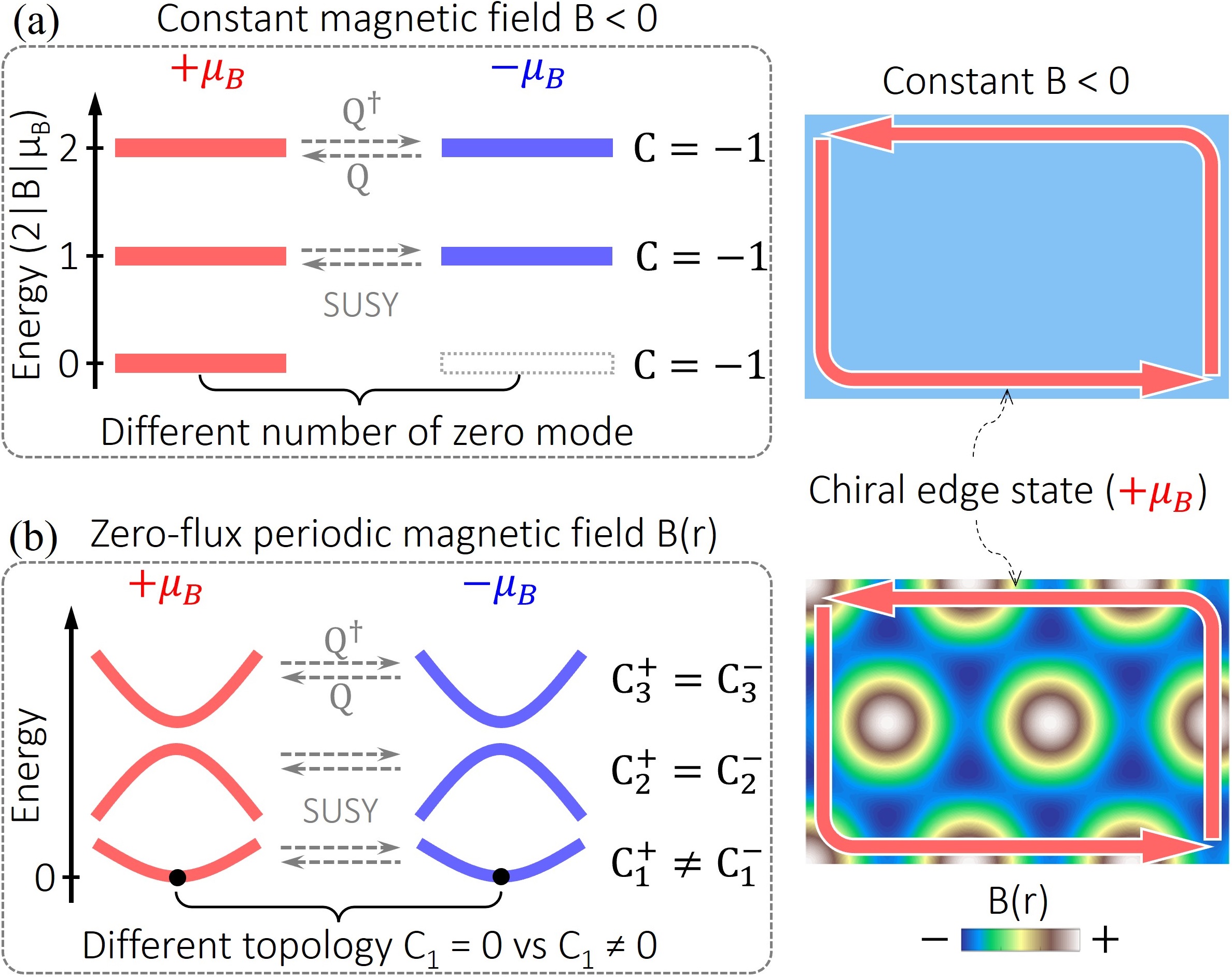}
	\caption{(a) Landau levels in a uniform field. The mismatched zero-energy level indicates unequal number of zero modes for the two spin species. (b) Fully spin-symmetric spectra in a zero-flux periodic field. The first bands of the two spin species each has one zero mode (black dot), but feature distinct band topology. Note that both (a) and (b) support fully spin-polarized chiral edge state inside the first energy gap.}
	\label{Fig:LLs_schematics}
\end{figure}

Paired bands of identical dispersion having the same topology as dictated by Eq.~(\ref{Eq:IdenticalC}) are intriguing but not that surprising, e.g., it is reminiscent of the pairing of Landau levels (LLs) in constant magnetic field [Fig.~\ref{Fig:LLs_schematics}(a)]. The results thus far, however, do not address the presence of the zero modes, which are crucial ingredients of SUSY~\cite{commentZeroModeSUSYbreaking}.

\begin{figure}[t]
	\centering
	\includegraphics[width=3.4in]{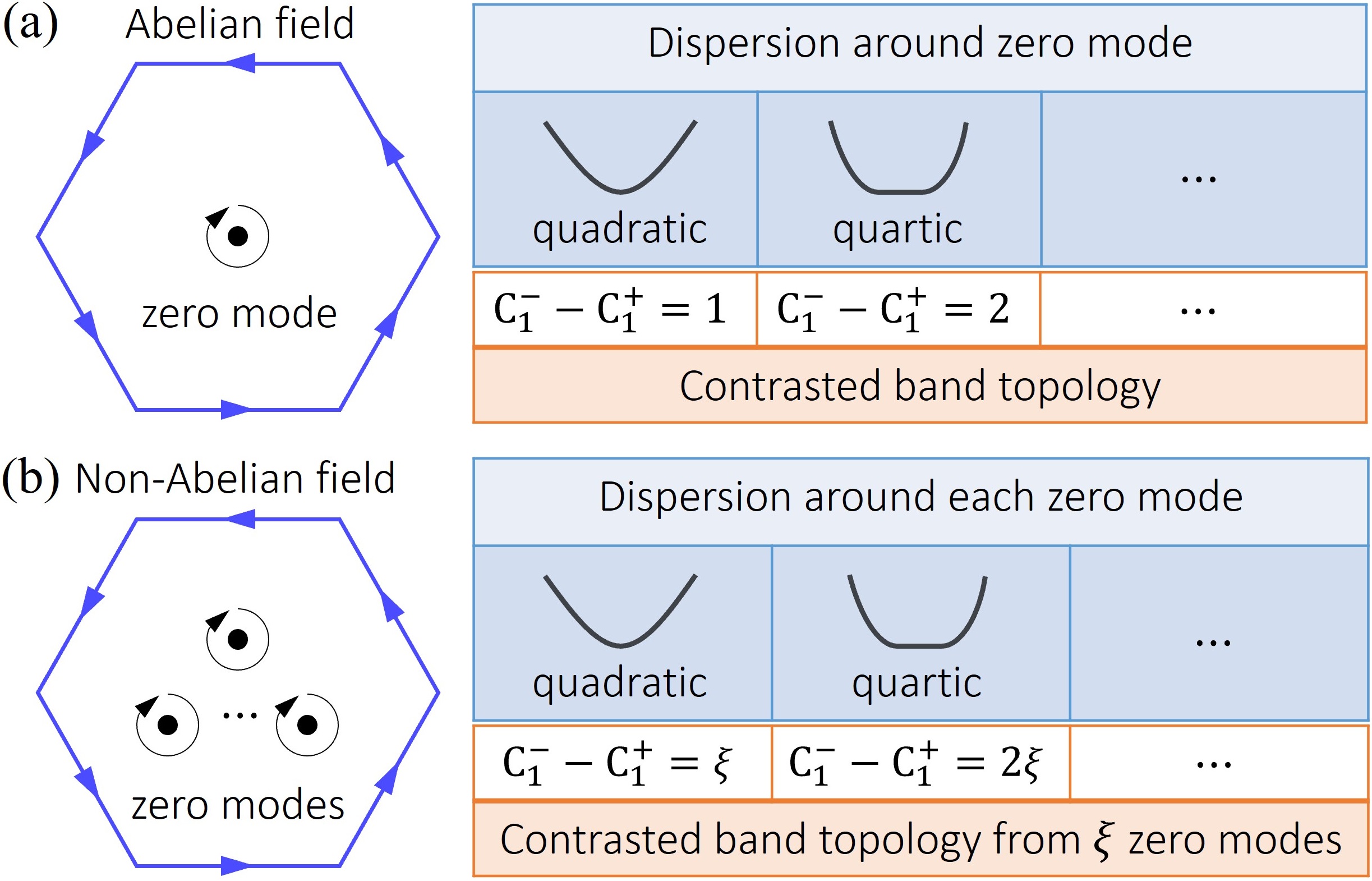}
	\caption{(a) Schematics of Brillouin zone (blue hexagon) with one zero mode (black dot) at the center for the case of an Abelian gauge field. The Chern number difference $C_1^- - C_1^+$ due to different low-energy dispersion around the zero mode is summarized. (b) Similar to case (a) but with multiple zero modes located at discrete positions, which might occur when the field is non-Abelian. The Chern number difference $C_1^- - C_1^+$ increases with the number of zero modes.}
	\label{Fig:TopologySummary}
\end{figure}

Remarkably, since Eq.~(\ref{Eq:BerryConnection}) is singular at the zero energy, topology of the $n=1$ bands of the SPs are distinct.
Like in the case of Fig.~\ref{Fig:QuadraticTriB}(b), let us assume for now that there is one zero mode on each of $\CalE_\pm$ and it locates at the $\gamma$ point.
As $\delta\bR_1$ is singular when $\bk\rightarrow\gamma$, while it is well-defined and periodic elsewhere, its line integral on a circle enclosing the zero mode will yield the difference of the Chern number [Fig.~\ref{Fig:TopologySummary}(a)]:
\begin{equation}
\begin{aligned}
C_{1}^{-}-C_{1}^{+}
=\frac{1}{2\pi}\oint_{\text{O}_\gamma}\delta\bR_{1}\cdot d\boldsymbol{l}
=\frac{1}{4\pi}\int^{2\pi}_{0}\frac{k}{\CalE_{1\bk}}\frac{\partial \CalE_{1\bk}}{\partial k}\,d\varphi
\end{aligned},\label{Eq:ChernNumberDiff}
\end{equation}
where $\text{O}_\gamma$ denotes a circle centered at $\gamma$. In the second equality, $\bk$ is understood to be on the circle and measured from $\gamma$, and $\varphi$ is its polar angle. 
Being the difference of two Chern numbers, the result must be an integer. 
\textit{In particular, when the low-energy dispersion $\CalE_{1\bk}\propto k^{2\hs{N}}$ around the zero mode, $\hs{N}$ a positive integer, we have},
\begin{equation}
	C_1^{-} - C_1^{+} = \hs{N}. \label{Eq:differN}
\end{equation}
This key finding of the work is schematically summarized in Fig.~\ref{Fig:TopologySummary}(a).
For typical 2D electron systems with quadratic dispersion at low energies, $C_{1}^{-}-C_{1}^{+}=1$. Although the individual value of $C_{1}^{\pm}$ cannot be determined here, a nonzero difference ($C_{1}^{-}-C_{1}^{+}\ne0$) guarantees that at least one of the bands is topologically nontrivial.

Figure~\ref{Fig:QuadraticTriB}(b) shows the band Chern numbers for
the example of Eq.~(\ref{Eq:SchrodingerWithB}) when the $\bB$ field is zero-flux periodic. 
As expected, $C_{n}^{+}= C_{n}^{-}$ when $n>1$. This part is reminiscent of LLs at nonzero energies in a uniform field [Fig.~\ref{Fig:LLs_schematics}(a)], but clear distinctions can be identified: here the Chern numbers can vary from band to band, and their values are not restricted to $\pm1$.
Most importantly, we find that in the lowest band $n=1$, one spin is topologically trivial ($C_{1}^{-}=0$) while its isospectral SP is nontrivial ($C_{1}^{+}=-1$), as dictated by Eq.~(\ref{Eq:differN}) for the quadratic dispersion. 
This contrasted topology [Fig.~\ref{Fig:QuadraticTriB}(b)] suggests that zero-flux periodic $\bB$ fields support spin-polarized chiral edge states in the gaps, akin to the situation in a uniform field. But importantly, this is made possible by the distinct topology in bands that have identical dispersion ($\CalE_{1\bk}^{+}=\CalE_{1\bk}^{-}$ but $C_1^{+}\ne C_1^{-}$), instead of the lack of zero-energy LL for one spin species in a uniform field. These significant distinctions between uniform and zero-flux periodic fields are schematically summarized in Fig.~\ref{Fig:LLs_schematics}(a) vs Fig.~\ref{Fig:LLs_schematics}(b).
For a general periodic $\bB$ field without net flux, one can speculate that the lowest bands of the SPs are always contrasted by Chern numbers $0$ vs $\pm1$, thus the presence of chiral edge states with a definite spin in the first gap is preserved when $\bB$ evolves from a uniform one [Fig.~\ref{Fig:LLs_schematics}(a)] to zero-flux periodic ones [Fig.~\ref{Fig:LLs_schematics}(b)].

Given one $\bB$ field, larger Chern number difference in the lowest band ($C_1^--C_1^+$) can be realized by considering the cases of $N>1$ in Eq.~(\ref{Eq:SUSYHamN}), i.e. with non-quadratic dispersions. An example of such scenarios will be discussed in Sec.~\ref{Sect:Expt4L}. This feature can be employed to achieve narrow energy bands with high Chern numbers.

The contrasted topology $C_1^{+}\ne C_1^{-}$ is also accompanied by distinct wave function distributions. The zero mode $\psi_{+}^{0}$ ($\psi_{-}^{0}$) distribution is strongly correlated with that of the $\bB$ field by primarily occupying regions where $B<0$ ($B>0$)~\cite{AharonovCasherPRA1979,SnymanPRB2009}. In fact, other states on the first band $\CalE_{1}^{+}$ ($\CalE_{1}^{-}$) also show similar distribution patterns.
This can be easily understood from the Zeeman term in Eq.~(\ref{Eq:SchrodingerWithB})~\cite{AlexNanoLett2017}: An electron with $+\bmu_{B}$ ($-\bmu_{B}$) has lower energy when it resides in the region with $B<0$ ($B>0$).
Therefore, in the case of Fig.~\ref{Fig:QuadraticTriB}(a), $|\psi_{+}^{0}|^2$ occupy the region of $B<0$, showing a honeycomb network [Fig.~\ref{Fig:QuadraticTriB}(c)], and the band Chern number $C_1^+=-1$ is consistent with the sign of the underlying field; while $|\psi_{-}^{0}|^2$ shows features of localized dots [Fig.~\ref{Fig:QuadraticTriB}(d)] with $C_1^-=0$. Importantly, profile of the zero-modes, i.e., connected or localized, which is readily known via the Zeeman term, serves as a indicator for quickly identifying the topology of the $n=1$ bands. For example, when the sign of the $\bB$ field in Fig.~\ref{Fig:QuadraticTriB}(a) is reversed, one can easily find $C_1^+=0$ vs $C_1^-=1$, thus the edge state in the first gap has opposite spin as expected and $C_{1}^{-}-C_{1}^{+}=1$ persists. It is interesting to notice that the Chern numbers change from $(C_1^+,\,C_1^-)=(-1,\,0)$ to $(C_1^+,\,C_1^-)=(0,\,1)$ on the lowest band after inverting the $\bB$ field, which is in contrast to the simple sign reversal of Chern numbers on the other bands, i.e., $C_{n}^{\pm}(\bB)=-C_{n}^{\pm}(-\bB)$ when $n>1$.

We now comment on the scenario of non-Abelian gauge field $\bB$. 
In non-Abelian cases, Eq.~(\ref{Eq:SUSYWaveFunctionRelation}) and the resultant Eq.~(\ref{Eq:BerryConnection_v}) remain valid at nonzero energies. 
However, it should be noted that each SP may have multiple zero modes when $\bB$ is non-Abelian (see example in Sec.~\ref{Sect:HsqrtExample}).
Nevertheless, the line integral around each zero mode --like that in Eq.~(\ref{Eq:ChernNumberDiff}) and Fig.~\ref{Fig:TopologySummary}(a)-- can be preformed. 
By adding up such individual contributions from all the zero modes, we obtain 
\begin{equation}
    C_{1}^{-}-C_{1}^{+}=\frac{1}{2\pi}\sum_{i}\oint_{\text{O}_i}\delta\bR_{1}\cdot d\boldsymbol{l},\label{Eq:DiffC_multizeros}
\end{equation}
where $\text{O}_i$ denotes a circle enclosing the $i$-th zero mode, and each summand is determined by Eq.~(\ref{Eq:differN}). Here we have assumed that the number of zero modes is \textit{limited} and they are \textit{discretely distributed} in the BZ. These results are schematically summarized in Fig.~\ref{Fig:TopologySummary}(b).
Therefore, the main findings discussed in the above remain valid, and they can be applied to many other systems, which we will discuss in Sec.~\ref{Sect:HsqrtExample}.

Section~S2 of the Supplementary Material~\cite{supp} presents more examples of band structures and Chern numbers by comparing to similar systems without SUSY. 
For models that do not have SUSY, the band structures change dramatically accompanied by topological transitions when the $\bB$ field is simply varied in intensity or period.
In contrast, such drastic changes are absent in systems with SUSY. These results indicate that SUSY systems could be advantageous for hosting stable electronic and topological properties, which are important for exploring exotic phases of matter and practical applications. With the intriguing feature of isospectra accompanied by contrasted band topology for the SPs, the SUSY systems also serve as important platforms for highlighting the quantum geometric effects in some physical processes, which sometimes coexist with `trivial' effects from the electronic properties of the system. For instance, it is proposed that the electron-phonon coupling has geometrical/topological contributions mixed with energetic effects that depend on energy dispersion of the system~\cite{YuJiabinPhonon}. As the energetic parts are expected to be identical for SPs, SUSY systems are promising candidates for testing the geometrical/topological effects in such scenarios.


\section{Realization in perodically strained 2D semiconductors}~\label{Sect:Expt}

Naively, one might expect that the discussed SUSY model [e.g., Eq.~(\ref{Eq:SchrodingerWithB})] and the intriguing feature of isospectra with contrasted band topology can be easily realized by applying a periodic magnetic field to 2D electron systems. However, there exist two obstacles in experimental implementations: (i) It is technically challenging to generate smooth and periodic magnetic field landscapes~\cite{PeriodicFieldDiracLiangFu2022}.
(ii) A real magnetic field can couple to magnetic moments of various origins (e.g., spin, orbital, and valley~\cite{VariousZeemanNatPhys2015}), which introduces extra Zeeman terms that break SUSY.
In the following, we will show that such difficulties can be circumvented by using periodically strained 2D semiconductors, e.g., TMDs. By employing strain-induced pseudo-magnetic field~\cite{StrainReviewPhysRep2010,StrainReviewPhysRep2016,StrainReviewRepProgPhys2017}, and taking the advantage that it only couples to the valley magnetic moment~\cite{DiXiaoValleyContrastingPRL2007}, we show that it is possible to realize the SUSY systems discussed earlier.


\subsection{$30^\circ$ twisted bilayer TMDs for $N=1$ SUSY with quardatic dispersion}

We will consider TMDs as an example below, but the discussions are also applicable for many other 2D semiconductors, e.g., graphene with a staggered potential or hBN. 
In monolayer TMDs, the low-energy carriers are governed by massive Dirac equations $h^{\text{1L}}_{\tau}=v\bsigma_{\tau}\cdot\bp+\Delta\sigma_{z}$ in two inequivalent valleys labeled as $\tau K$, where $\tau=\pm$ is the valley index, $v$ is the Fermi velocity, and $\bsigma_{\tau}=(\tau\sigma_{x},\,\sigma_{y})$~\cite{DiXiaoTMDsPRL2012}. 
By introducing nonuniform strain patterns, giant pseudo-magnetic fields opposite in the two valleys can be generated~\cite{StrainReviewPhysRep2010,StrainReviewPhysRep2016,StrainReviewRepProgPhys2017}. The Dirac Hamiltonian in the presence of strain pseudo-magnetic field becomes
\begin{equation}
H^{\text{D}}_{\tau}=v\bsigma_{\tau}\cdot(\bp+\tau e\bA)+\Delta\sigma_{z},~\label{Eq:HamDiracWithPseudoB}
\end{equation}
where $\tau\bA$ is the strain vector potential and its details will be discussed later. Note that there might also exist strain-induced scalar potentials. We discuss such terms in Sec.~S3 of Supplementary Material~\cite{supp}, and show that WSe$_2$ is an ideal candidate, in which such terms can be neglected. Importantly, profile of the pseudo-magnetic field, $\tau\bB=\nabla\times(\tau\bA)$, can be controlled by tuning the strain distribution. Strain engineering in 2D materials~\cite{QHEStrainGrapheneNatPhys2010,SubstrateApplPhysExpress2011,SubstrateNL2014,YuhangNL2017,Nadya2018,PhongPRL2022,ZhaiNanoLett2023} thus offers a solution to overcome the obstacle of generating patterned magnetic fields with zero flux.

The Dirac model $H^{\text{D}}_{\tau}$ describes charge carriers in both the conduction and valence bands. From experimental perspectives, for TMDs it is most natural to consider holes (h) near the valence band edge, which are described by
\begin{equation}
\begin{aligned}
H^{h}_{\tau}\psi=[\frac{1}{2m}(\bp+\tau e\bA)^2+\bmu_{\tau}^{h}\cdot(\tau\bB) ]\psi=\varepsilon\psi
\end{aligned}\label{Eq:HamiltonianHoleMonolayer}
\end{equation}
near the valence band edge in the limit of $\varepsilon\ll\Delta$, where $\bmu_{\tau}^{h}=-\tau\bmu_{B}$ is the valley magnetic moment~\cite{DiXiaoValleyContrastingPRL2007} for holes.
In this quadratic model, $\varepsilon$ is measured from the valence band edge and $m=\Delta/v^2$ is the effective mass.
Note that a magnetic moment in the form of Bohr magneton naturally appears here, which is a characteristic of the low-energy behavior of the massive Dirac model. More importantly, the strain pseudo-magnetic field can only couple to this valley magnetic moment, but not other sources (e.g., spin and orbital~\cite{VariousZeemanNatPhys2015}). Therefore, the second challenge mentioned above for realizing SUSY is resolved.


\subsubsection{Layer-valley pseudospin realization of superpartners}

We now propose a practical scheme to achieve SPs with contrasted band topology satisfying $C_1^{-}-C_1^{+}=1$. 
We first focus on the general ideas that are independent of details of the $\bB$ fields. 
The proposal employs a bilayer geometry with layer-opposite strain pseudo-magnetic fields, i.e., $\bB^{(1)}=-\bB^{(2)}$, where the superscript is layer index. Then holes from 
\textit{opposite valleys of opposite layers} form SPs [Fig.~\ref{Fig:SubstrateBilayer30deg}(b)]. To see this formally, we write the Hamiltonian in such configurations explicitly:
\begin{equation}
\begin{aligned}
H^{h}_{\tau}(\text{layer 1})&=\frac{1}{2m}\left(\bp+\tau e\bA^{(1)}\right)^2+\bmu_{\tau}^{h}\cdot\left(\tau\bB^{(1)}\right)\\
H^{h}_{-\tau}(\text{layer 2})&=\frac{1}{2m}\left(\bp+\tau e\bA^{(1)}\right)^2+\bmu_{-\tau}^{h}\cdot\left(\tau\bB^{(1)}\right)
\end{aligned},\label{Eq:SUSYMonolayer&MonolayerHoles}
\end{equation}
where  $-\bA^{(2)}$ and $-\bB^{(2)}$ have been replaced by $\bA^{(1)}$ and $\bB^{(1)}$ in layer 2, and interlayer coupling has been assumed to be negligible. Clearly, the only difference between the two lines lies in the magnetic moments $\bmu_{\tau}^{h}$ and $\bmu_{-\tau}^{h}$, which are opposite to each other but both have magnitude $\mu_B$. Such a bilayer geometry thus realizes the SUSY model in Eq.~(\ref{Eq:SchrodingerWithB}) with $M=[(p_x+\tau eA^{(1)}_x)-i(p_y+\tau eA^{(1)}_y)]/\sqrt{2m}$ and the assignment of the two lines of Eq.~(\ref{Eq:SUSYMonolayer&MonolayerHoles}) to $\hH_\pm$ is determined by the directions of $\bmu_{\tau}^{h}$ and $\bmu_{-\tau}^{h}$ [see Figs.~\ref{Fig:SubstrateBilayer30deg}(b, c)].

\subsubsection{Twist control of strain induced pseudo-magnetic field on patterned substrate}

\begin{figure}[t]
	\centering
	\includegraphics[width=3.4in]{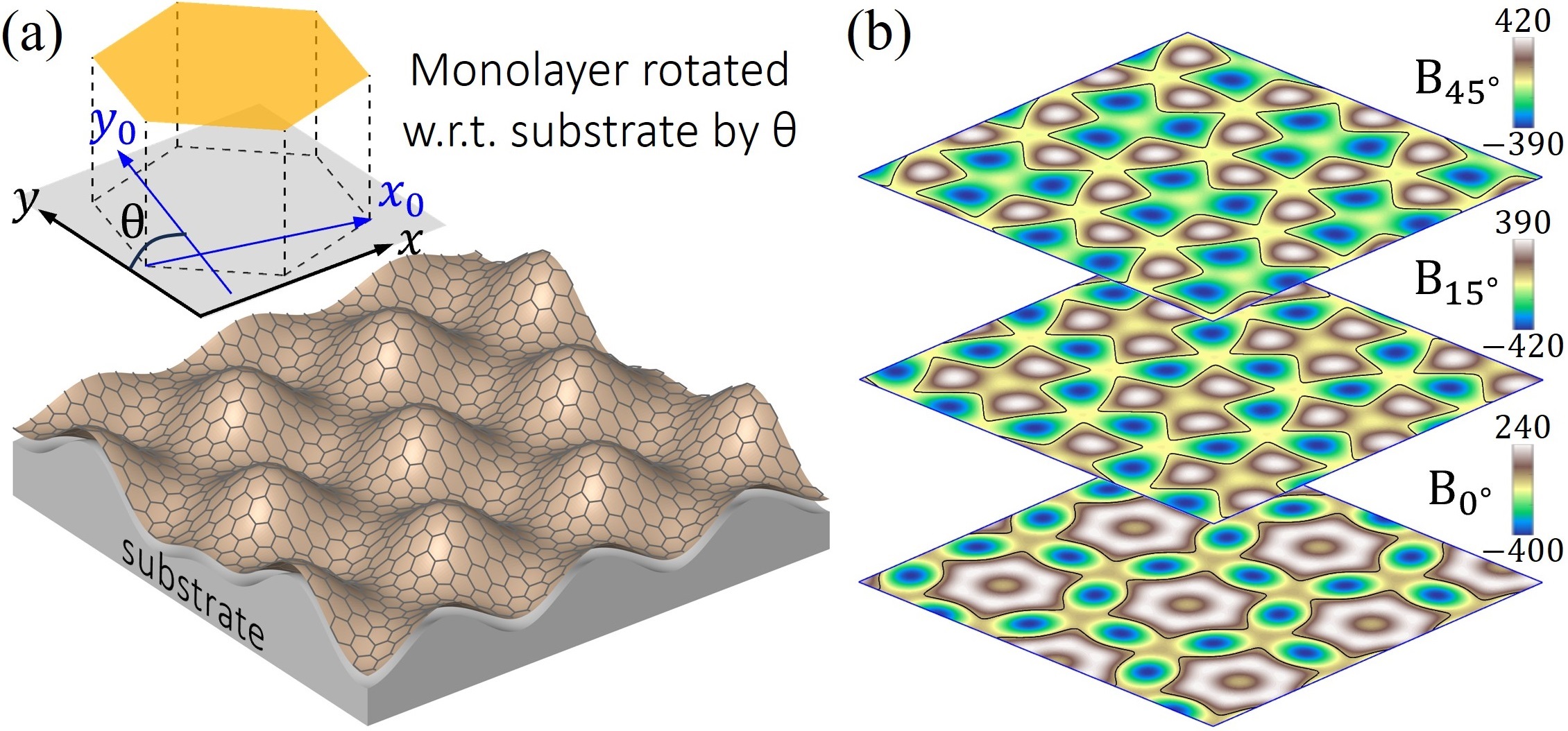}
	\caption{(a) Schematics of a honeycomb lattice on a patterned substrate, whose zigzag/armchair crystalline directions ($x_0/y_0$ axis in the inset) are rotated clockwise by $\theta$ degrees from the $x/y$ axis on the substrate. The inset shows the $\theta$ rotation of the $x_0y_0$ coordinate frame on the lattice (yellow hexagon) with respect to the $xy$ coordinate frame on the substrate (grey square). 
		(b) Pseudo-magnetic field (units: T) $B_\theta$ in the $+K$ valley for different $\theta$ with black curves denoting zero-field contours.
		Parameters: $z_0=0.6$ nm, $L=14$ nm, and $\Delta=0.825$ eV, $\hbar v=3.38$ eV$\cdot$\AA, $evA_0=2$ eV from WSe$_2$.
	}\label{Fig:SubstrateMonolayerTwist}
\end{figure}

Strained induced pseudo-magnetic field with the periodic landscape of Fig.~\ref{Fig:QuadraticTriB}(a) has been proposed in buckled graphene on NbSe$_2$~\cite{JinhaiNature2020}. Pseudo-magnetic fields with similar profiles can also appear due to lattice relaxation in the moir\'e of H-type bilayer TMDs~\cite{RelaxationTMDsFalkoPRL2020,KTLawQAHMoirePRL2022,commentScalarPotentialLatticeRelaxation}. However, strain originates from spontaneous atomic displacements in these scenarios,  thus it is challenging to tune the pseudo-magnetic field (e.g., magnitude, period, profile), if not impossible.
Alternatively, strain engineering can be implemented in a more controllable manner by using patterned substrates. This approach offers great flexibility in realizing different pseudo-magnetic fields, mostly by properly designing the substrates with various landscapes~\cite{SubstrateApplPhysExpress2011,SubstrateNL2014,YuhangNL2017,Nadya2018,PhongPRL2022,ZhaiNanoLett2023}. 

Here we highlight another important feature of substrate-assisted strain engineering: Given a pre-fabricated substrate with a fixed surface landscape, the periodic pseudo-magnetic field imprinted onto the deposited material can be controlled simply by rotating the material with respect to the substrate (Fig.~\ref{Fig:SubstrateMonolayerTwist}).
We will consider a TMD membrane following the surface of a patterned substrate modeled by $z(\br)=z_0\sum_{i=1}^{3}\cos(\bb_i\cdot\br+\frac{\pi}{12})$ as an example, where $\bb_{1}=(\frac{\sqrt{3}}{2},\,-\frac{1}{2})\frac{4\pi}{\sqrt{3}L}$
and $\bb_{i}=C_3^{i-1}\bb_{1}$. This height profile is shown schematically in Fig.~\ref{Fig:SubstrateMonolayerTwist}(a).
We note that in-plane atomic displacements on the lattice could also occur after the deposition. But their details strongly depend on the specific experimental settings, e.g., whether the membrane is freely relaxed, or clamped on the edges, or fixed on extrema of the patterned substrate, etc~\cite{RippledGrapheneEPL2008,QHEStrainGrapheneNatPhys2010,PhongPRL2022}. Here we neglect in-plane displacements for simplicity, which does not affect conclusions of SUSY and band topology in the following discussions.
The resultant strain tensor components on the lattice can then be obtained from its out-of-plane deformation, $\epsilon_{ij}=(\partial_{i}z)(\partial_{j}z)/2$, where $i,\,j\in\{x,\,y\}$.

\begin{figure}[t]
	\centering
	\includegraphics[width=3.4in]{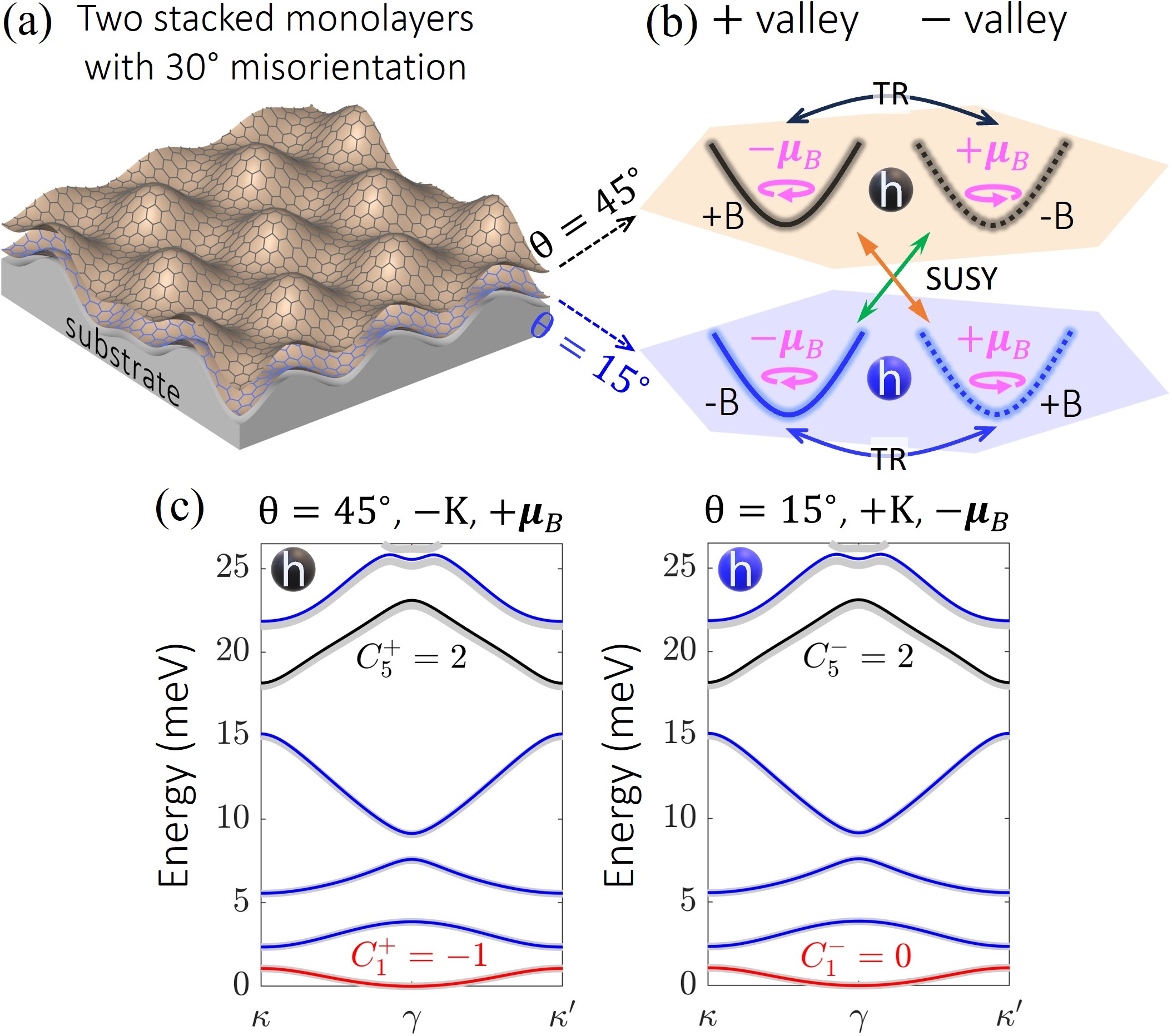}
	\caption{(a) Schematics of two $30^\circ$ misorientated honeycomb lattices on a patterned substrate, whose zigzag/armchair crystalline directions are rotated clockwise from the $x/y$ axis on the substrate by $\theta=15^\circ$ and $45^\circ$, respectively. The corresponding pseudo-magnetic field $B_\theta$ in the $+K$ valley are shown in Fig.~\ref{Fig:SubstrateMonolayerTwist}(b).
		(b) Schematics of Dirac cones (solid/dotted: $+K$/$-K$) and valley magnetic moments (curly arrows) for holes in the two $30^\circ$ misorientated layers (hexagon backgrounds). $\pm B$ in each column highlight the opposite pseudo-magnetic fields in the two layers. The letter `h' in spheres denotes holes.
		(c) Energy bands of holes and Chern numbers for the case of $\{\theta=45^\circ,\,-K\text{ valley}\}$ vs $\{\theta=15^\circ,\,+K\text{ valley}\}$. All the blue bands have zero Chern numbers. Grey curves in the background represent results from the Dirac model [Eq.~(\ref{Eq:HamDiracWithPseudoB})].
		Parameters: $z_0=0.6$ nm, $L=14$ nm, and $\Delta=0.825$ eV, $\hbar v=3.38$ eV$\cdot$\AA, $evA_0=2$ eV from WSe$_2$.}
	\label{Fig:SubstrateBilayer30deg}
\end{figure}

Crucially, the strain distribution on the lattice and the resultant pseudo-magnetic field depend on the alignment between the lattice and substrate. This is understandable as the local atomic bonds on the lattice are stressed differently when the alignment is varied, although the macroscopic topography of the membrane remains invariant [see e.g., Fig.~\ref{Fig:SubstrateBilayer30deg}(a)]. This feature underlies the twist-control of the pseudo-magnetic field by rotating the material with respect to the substrate.

To elaborate on this twist control of pseudo-magnetic field, we notice that there are two coordinate frames that are of interest [see Fig.~\ref{Fig:SubstrateMonolayerTwist}(a)]: (i) $xy$, fixed on the substrate, in which $z(\br)$ is defined; (ii) $x_0y_0$, fixed on lattice, with $x_0$ ($y_0$) along its zigzag (armchair) direction. The angle $\theta$ between the $x_0$ ($y_0$) and $x$ ($y$) axes characterizes the alignment of the lattice with respect to the substrate.
Without loss of generality, we assume that the lattice is rotated clockwise with respect to the substrate by $\theta$ degrees.
Here it is convenient to use the $xy$ coordinate frame fixed on the substrate, in which expressions of the out-of-plane deformation $z(\br)$ and the strain tensor components $\epsilon_{ij}$ remain independent of $\theta$. In this coordinate frame, the effect of twisting is manifested in the strain vector potential, which gains an explicit $\theta$-dependence~\cite{StrainDirectionPRL2013,ZhaiPRB2018}:
\begin{equation}
	\bA_{\theta}=R(3\theta)A_0(\epsilon_{xx}-\epsilon_{yy},\,-2\epsilon_{xy})^{T}.~\label{Eq:StrainVectorPotential}
\end{equation}
Here $R(3\theta)$ is the clockwise rotation matrix by $3\theta$ degrees and $A_0$ is a material-dependent constant (see Sec.~S3 of Supplementary Material~\cite{supp}).
When $\theta=0^\circ$, the strain vector potential reads $\bA_{0^\circ}=A_0(\epsilon_{xx}-\epsilon_{yy},-2\epsilon_{xy})^T$, which is commonly found in the literature as the zigzag direction is usually implicitly chosen as the $x$ axis~\cite{StrainReviewPhysRep2010,StrainReviewPhysRep2016,StrainReviewRepProgPhys2017}.
The twist control of the pseudo-magnetic field by rotating the lattice on the substrate is explicitly manifested in its $\theta$-dependence, $\tau\bB_{\theta}=\nabla\times(\tau\bA_{\theta})$.
Figure~\ref{Fig:SubstrateMonolayerTwist}(b) shows $\bB_{\theta}$ in the $xy$ coordinate frame in the $\tau=+$ valley for a few different $\theta$. One can see that both the magnitude and profile of the field change with $\theta$. In particular, one notices that $\bB_{15^\circ}$ and $\bB_{45^\circ}$ are effectively opposite (up to a rotation), which we will employ in the following.

\subsubsection{Isospectral bands with layer contrasted topology}

We note that stacked monolayers with $\sim30^\circ$ misorientation are decoupled around the charge neutrality point due to the large momentum separation of the band edges in the two layers~\cite{30degTBGNanoLett2020,30degTBGExptPRB2021,30degWSe2ChihKang}. In TMDs, interlayer coupling could occur via Umklapp scatterings at relatively small energies, but $\sim30$--$130$ meV deep below the valence band edge~\cite{30degWSe2ChihKang}. As the contrasted band topology in the lowest bands close to the zero energy is of particular interest here, we consider two decoupled layers of TMDs deposited  onto the substrate with $\theta=15^\circ$ and $\theta=45^\circ$ orientation respectively [Fig.~\ref{Fig:SubstrateBilayer30deg}(a)], which results in layer-opposite pseudo-magnetic fields [see Fig.~\ref{Fig:SubstrateMonolayerTwist}(b)]. In such a bilayer configuration, holes in \textit{opposite layers and opposite valleys} are governed by Eq.~(\ref{Eq:SUSYMonolayer&MonolayerHoles}), thus behave as SPs [Fig.~\ref{Fig:SubstrateBilayer30deg}(b)]. As there are two valleys in each layer, which are related by time-reversal (TR) symmetry, this configuration contains two sets of TR partners as well as two pairs of SPs with opposite Chern numbers.

Figure~\ref{Fig:SubstrateBilayer30deg}(c) presents the identical energy spectra for such SPs by considering two layers of WSe$_2$ stacked on the substrate as described above. As the $\pm$ valleys are connected by TR symmetry in the presence of pseudo-magnetic fields, only the bands from one valley in each layer are shown. The colored curves represent results from the hole model [Eq.~(\ref{Eq:HamiltonianHoleMonolayer})], which match perfectly with the grey curves from the Dirac model [Eq.~(\ref{Eq:HamDiracWithPseudoB})] shown in the background~\cite{commentDiracVsHole}. The first bands of the SPs exhibit distinct band Chern numbers $C_1^{+}=-1$ vs $C_1^{-}=0$. They satisfy $C_1^- - C_1^+ = 1$, which is expected for a quadratic band dispersion around $\gamma$ --location of the zero mode-- as dictated by Eq.~(\ref{Eq:differN}) [also see Fig.~\ref{Fig:TopologySummary}(a)]. Here the upper/lower layer with $\theta=45^\circ$/$15^\circ$ on the substrate has topologically nontrivial/trivial lowest bands, such topological properties obviously can be interchanged between the layers by changing the two layers' orientations.

\subsection{$30^\circ$ twisted double bilayer TMDs for $N=2$ SUSY with quartic dispersion}~\label{Sect:Expt4L}

\begin{figure}[t]
	\centering
	\includegraphics[width=3.4in]{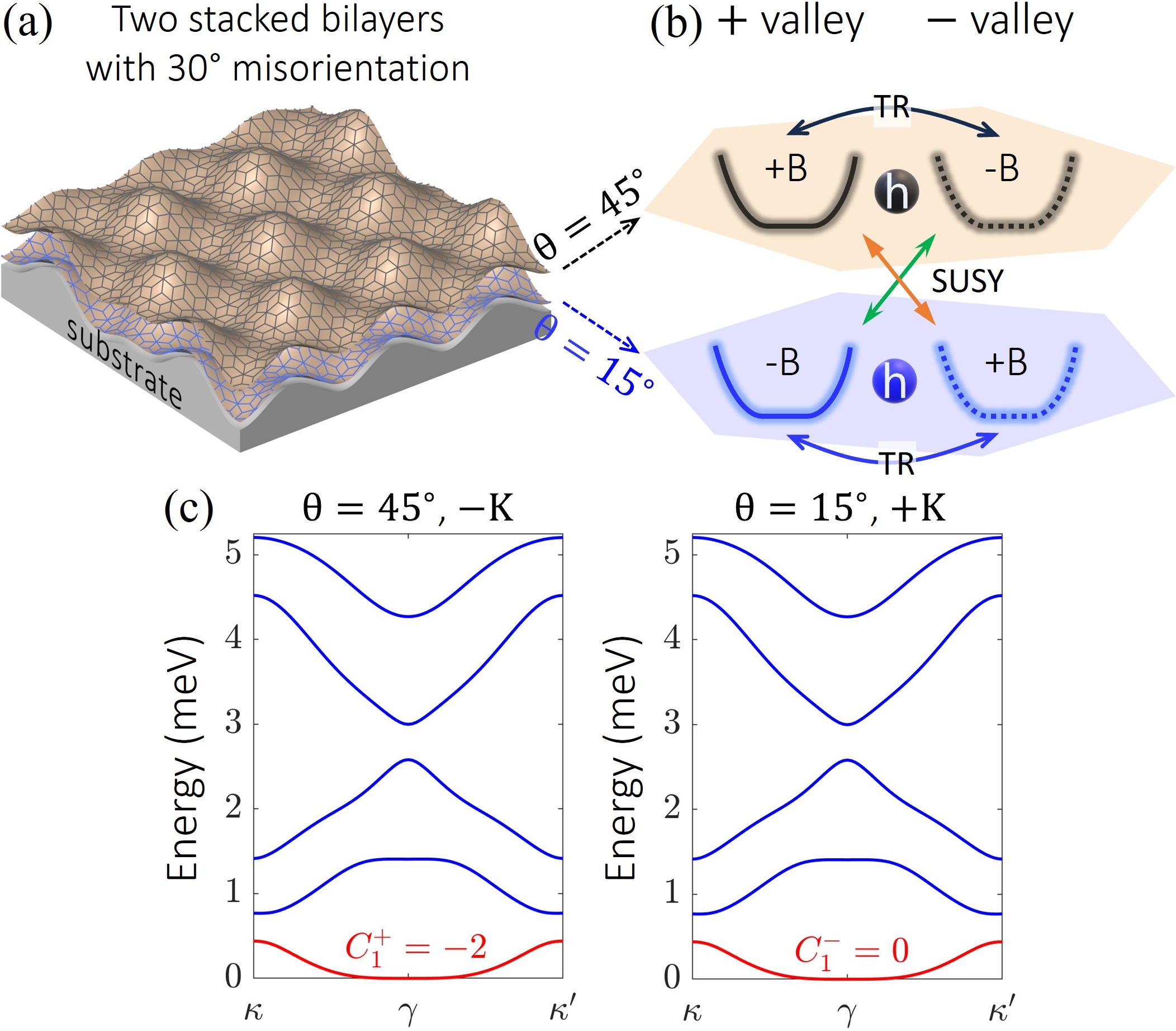}
	\caption{(a) Schematics of two $30^\circ$ misorientated rhombohedral bilayer TMDs on a patterned substrate, whose zigzag/armchair crystalline directions are rotated clockwise from the $x/y$ axis on the substrate by $\theta=15^\circ$ and $45^\circ$, respectively. The corresponding strain pseudo-magnetic field in the $+K$ valley are shown in Fig.~\ref{Fig:SubstrateMonolayerTwist}(b). (b) Schematics of quartic cones (solid/dotted: $+K$/$-K$) for holes in the two $30^\circ$ misorientated bilayers (hexagon backgrounds). $\pm B$ in each column highlight the opposite pseudo-magnetic fields in the two layers.  The letter `h' in spheres denotes holes.
	(c) Energy bands of holes for the case of $\{\theta=45^\circ,\,-K\text{ valley}\}$ vs $\{\theta=15^\circ,\,+K\text{ valley}\}$. All the blue bands have zero Chern numbers.
	Parameters: $z_0=0.6$ nm, $L=14$ nm, $t=0.1$ eV, and $\Delta=0.825$ eV, $\hbar v=3.38$ eV$\cdot$\AA, $evA_0=2$ eV from WSe$_2$.}
	\label{Fig:SubstrateBernalBilayerHole}
\end{figure}

In this section, we propose a physical system for the scenario of $N=2$ SUSY in Eq.~(\ref{Eq:SUSYHamN}). It allows us to obtain $C_1^- - C_1^+ = 2$ thanks to the quartic dispersion around the zero mode [Fig.~\ref{Fig:TopologySummary}(a)]. Such cases are useful for achieving topological bands with higher Chern numbers. The system is realized with the replacement of each monolayer in Fig.~\ref{Fig:SubstrateBilayer30deg}(a) by a bilayer, as shown schematically in Fig.~\ref{Fig:SubstrateBernalBilayerHole}(a). Notably, the isospectral SPs realized are still in stark contrast to the time-reversal partners, with one SP being topologically trivial ($C_1^- = 0$), and another nontrivial ($C_1^+ = -2$).

We first look at the basic ingredient of the setup-- a strained bilayer. For illustrative purposes, it suffices to consider a simple bilayer Hamiltonian
\begin{equation}
\mathcal{H}_{\tau}^{\text{2L}}=
\begin{pmatrix}
H_{\tau}^{\text{D}} &T\\T^\dagger&H_{\tau}^{\text{D}}
\end{pmatrix}
~\text{ with }~
T=
\begin{pmatrix}
0&0\\t&0
\end{pmatrix},\label{Eq:H2L}
\end{equation}
where $H_{\tau}^{\text{D}}=v\bsigma_{\tau}\cdot(\bp+\tau e\bA)+\Delta\sigma_{z}$ is the Dirac Hamiltonian for a strained monolayer, and $t$ is the interlayer coupling constant. This simple model can describe rhombohedral bilayer TMDs~\cite{BilayerTMDsDiXiao2018,TongQingjunNatPhys}, where $t\sim0.1$~eV from a recent experiment~\cite{BilayerTMDsInterlayerCouplingPRX2022}.
By focusing on holes, one can consider a low-energy model
\begin{equation}
\begin{aligned}
\CalH_{\tau}^{h}\psi=\frac{v^4}{2\Delta t^2}(\tau\pi_x^{\tau}+i\pi_y^{\tau})^2(\tau\pi_x^{\tau}-i\pi_y^{\tau})^2\psi=\varepsilon\psi,
\end{aligned}
\end{equation}
where $(\pi_{x}^{\tau},\,\pi_{y}^{\tau})=\bpi^{\tau}=\bp+\tau e\bA$. This equation is valid for $\varepsilon\ll t,\,\Delta$. One may notice that $\CalH_{\tau}^{h}$ can also describe holes in Bernal bilayer graphene by introducing an interlayer displacement field into the two-band model~\cite{BilayerGrapheneReviewMcCann}. In this case, $\Delta$ is replaced by the displacement field, and $t\sim0.4$~eV.
The Hamiltonian $\CalH_{\tau}^{h}$ resembles that in Eq.~(\ref{Eq:SUSYHamN}) with $N=2$, thus lays the basis for constructing the SUSY model.

By following a similar idea as in Fig.~\ref{Fig:SubstrateBilayer30deg}(a), the SUSY model in Eq.~(\ref{Eq:SUSYHamN}) with $N=2$ can be realized by using two copies of bilayer TMDs with layer-opposite strain pseudo-magnetic fields (i.e., $\bB^{(1)}=-\bB^{(2)}$). Then holes from \textit{opposite valleys of opposite bilayers} form SPs, for example,
\begin{equation}
\begin{aligned}
\CalH_{\tau=+}^{h}(\text{bilayer}~1)&=\frac{v^4}{2\Delta t^2}[\pi_+(\bA^{(1)})]^2[\pi_-(\bA^{(1)})]^2\\
\CalH_{\tau=-}^{h}(\text{bilayer}~2)&=\frac{v^4}{2\Delta t^2}[\pi_-(\bA^{(1)})]^2[\pi_+(\bA^{(1)})]^2
\end{aligned},~\label{Eq:HambiLbiLHole}
\end{equation}
where the notation $\pi_\pm(\bA^{(1)})=(p_x+eA_x^{(1)})\pm i (p_y+eA_y^{(1)})$ has been used, and $-\bA^{(2)}$ has been replaced by $\bA^{(1)}$ in layer 2. It now becomes clear that they correspond to $\hH_{\mp}$ in Eq.~(\ref{Eq:SUSYHamN}) with $N=2$.
In this form it is also evident that the low-energy dispersion will be quartic, thus one expects $C_1^- - C_1^+ = 2$ if there is one zero mode on the first band [Fig.~\ref{Fig:TopologySummary}(a)].

By depositing double bilayers with $\theta=15^\circ$ and $\theta=45^\circ$ respectively on the substrate [Fig.~\ref{Fig:SubstrateBernalBilayerHole}(a)], opposite pseudo-magnetic fields in the two bilayers can be achieved [see Fig.~\ref{Fig:SubstrateMonolayerTwist}(b)], thus realizing the model of Eq.~(\ref{Eq:HambiLbiLHole}).
Figure~\ref{Fig:SubstrateBernalBilayerHole}(c) shows representative energy bands for such SPs by considering double bilayers of WSe$_2$. The low-energy bands are obtained from Eq.~(\ref{Eq:H2L}) due to its simplicity. The band structure around the zero mode at $\gamma$ clearly shows a quartic dispersion [also compare with Fig.~\ref{Fig:SubstrateBilayer30deg}(c)]. The first bands of the SPs exhibit distinct band topology characterized by $C_1^{+}=-2$ vs $C_1^{-}=0$, which is consistent with $C_1^- - C_1^+ = 2$ due to the quartic dispersion around $\gamma$ [Eq.~(\ref{Eq:differN}) and Fig.~\ref{Fig:TopologySummary}(a)].

\subsection{Spontaneous breaking of SUSY by many-body interaction at fractional filling}

One notices that the lowest bands of strained TMDs, e.g., Fig.~\ref{Fig:SubstrateBilayer30deg}(c), are much narrower than their counterparts in graphene~\cite{PeetersPeriodicStrainPRB2020,Antonio2DMater2021a,Antonio2DMater2021b,PhongPRL2022,ZhaiNanoLett2023,EslamPeriodicStrainPRL2023}. Hence Coulomb interaction can play important roles when these bands are fractionally filled. Despite having identical energy dispersion, the SPs possess drastically different wave functions [see e.g., Fig.~\ref{Fig:QuadraticTriB}(c) vs Fig.~\ref{Fig:QuadraticTriB}(d)] that underlie their distinct topological and quantum geometrical properties, hence spontaneous SUSY breaking by Coulomb interaction can be anticipated.

\begin{figure}[t]
	\centering
	\includegraphics[width=3.4in]{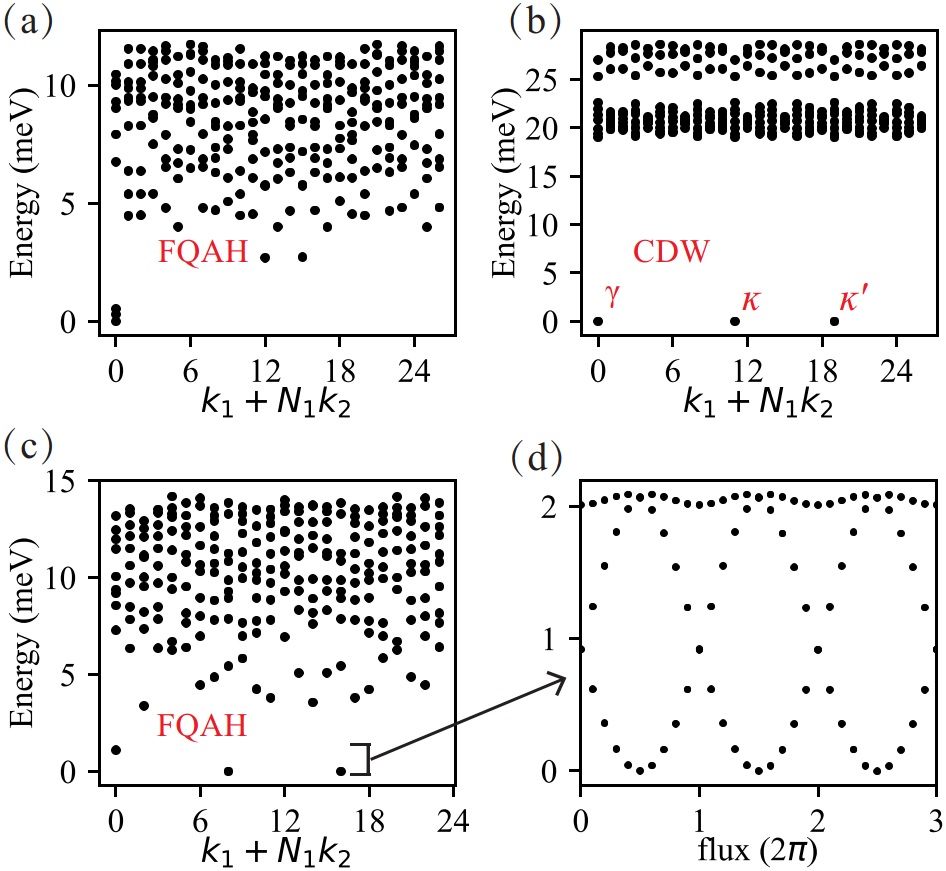}
	\caption{(a) Many-body spectrum at hole filling $\nu=2/3$ in the lowest band of a SP with $C_1^{+}=-1$ as shown in  Fig.~\ref{Fig:SubstrateBilayer30deg}(c). (b) The spectrum of its twin SP with $C_1^{-}=0$ at the same band filling. Exact diagonalizations in (a) and (b) are performed with a system size of $3\times9$ unit cells. (c) Many-body spectrum of the same scenario as in (a), obtained with a cluster size of $4\times6$ unit cells instead. (d) Ground-state spectral flow in (c) under flux insertion. Parameters: $\epsilon_r=2$, $d=30$~nm.}~\label{Fig:ED}
\end{figure}

Taking the system of Fig.~\ref{Fig:SubstrateBilayer30deg}(c) as an example, we project Coulomb interaction onto the lowest band of each SP and perform exact diagonalization calculations. The adopted Coulomb interaction before projection reads
\begin{equation}
H_{\text{int}}=\frac{1}{2S} \sum_{\bk,\, \bk',\, \bs{q}} V(\bs{q}) c_{\bk+\bs{q}}^{\dagger} c_{\bk'-\bs{q}}^{\dagger} c_{\bk'} c_{\bk},
\end{equation}
where $S$ is the system's area, $c_{\bk}^\dagger$ creates a plane wave with momentum $\bk$, $V(\bs{q})=e^2 \tanh{(qd)}/(2\epsilon_0 \epsilon_r q)$ is the dual-gate screened Coulomb potential,
$d$ is the distance from the material to the metallic gates, $\epsilon_0$ is the vacuum permittivity, and $\epsilon_r$ is the relative dielectric constant. With their distinct wave functions [see Fig.~\ref{Fig:QuadraticTriB}(c, d)], projection of the Coulomb potential to the lowest band leads to different form factors of the projected interactions for the twin SPs. 

Figure~\ref{Fig:ED}(a) and (b) present the many-body spectra as a function of crystal momentum obtained at filling factor $\nu=2/3$ for the two isospectral SPs with topologically nontrivial and trivial lowest band, respectively, with exact diagonalization calculations performed on a $3\times9$ unit-cell cluster put on a torus. The crystal momentum is assigned as $k_1/N_1 \bb_1+ k_2/N_2 \bb_2$ ($N_1=3$ and $N_2=9$ here) and labeled by an integer index $k_1+N_1 k_2$. Three nearly degenerate ground states can be identified in Fig.~\ref{Fig:ED}(a), which are separated from excited states by an energy gap of $\sim3$ meV. 
A similar feature of ground state degeneracy is also observed from the calculation with a different cluster size of $4\times6$ unit cells in Fig.~\ref{Fig:ED}(c).
This approximate ground state degeneracy is consistent with that of a fractional quantum Hall state on a torus~\cite{WenNiuPRB1990}. 
Additionally, upon magnetic flux insertion, these three ground states evolve into each other after $2\pi$ flux and return to themselves with a period of $6\pi$ [Fig.~\ref{Fig:ED}(d)].
These features provide strong evidence that the ground state at $\nu =2/3$ filling of this SP is FQAH state in nature. 
In stark contrast, the many-body spectrum at the same filling for its twin SP with topologically trivial lowest band is completely distinct [e.g., Fig.~\ref{Fig:ED}(b) vs Fig.~\ref{Fig:ED}(a)].
The spectrum in Fig.~\ref{Fig:ED}(b) shows the opening of a larger correlation gap, and has the character of the CDW phase, where the three degenerate ground states have momenta that can be folded back to the $\gamma$ point of a tripled unit cell~\cite{FuLiangFCIPRB2023,YuJiabinMoTe22023}.
The spontaneous SUSY breaking thus takes the form of developing distinct correlated phases for the two SPs with identical single-particle spectrum. 

The above results suggest that the SUSY systems are promising platforms for exploring correlation-driven effects that are sensitive to topology and quantum geometry, which are being intensively explored with the groundbreaking experimental discovery of FQAH effects in twisted bilayer MoTe$_2$~\cite{MoTe2Park2023,FCIMoTe2Jiaqi2023,FCIMoTe2ShanJie2023,FCIMoTe2PRX2023} and in pentalayer graphene/hBN moir\'e system~\cite{JuLongFCI}.


\section{SUSY dictated topology in square-root Hamiltonian and manifestions in flat-band superlattices}~\label{Sect:Sqrt}

We note that many 2D superlattice models that exhibit exact flat bands at magic angles, e.g., twisted bilayer graphene (TBG) in the chiral limit~\cite{TBGMagicAngleOriginPRL2019} and the strained $\Gamma$-valley model with quadratic band touching~\cite{StrainQuadraticBandTouchingSunKaiPRL2023}, as well as the square-root topological materials~\cite{SqrtPRB2017,SqrtEzawaPRR2020,SqrtHatsugaiPRA2020,SqrtHatsugaiPRB2021}, are described by Hamiltonian that can be considered as the `square root' of $\CalH_{\text{susy}}$ in Eq.~(\ref{Eq:Hsusy1}).
Here we show that such square-root Hamiltonian, $H_{\text{sqrt}}$, as defined below in Eq.~(\ref{Eq:Hsqrt}), exhibits intriguing electronic and topological properties that can be traced back to the SUSY systems. 
In particular, the isospectra of SPs underlies a  symmetric spectrum of $H_{\text{sqrt}}$, i.e., $E^{\text{sqrt}}_{\bk}\leftrightarrow-E^{\text{sqrt}}_{\bk}$, while the contrasted topologies of SPs, i.e., $C_1^+ \ne C_1^-$, are inherited by the first conduction and valence bands of $H_{\text{sqrt}}$.
The SUSY formalism discussed in earlier sections therefore connects many seemingly different systems and underlies the emergence of contrasted band topology.

\subsection{Linking electronic and topological properties of $H_{\text{sqrt}}$ to $\CalH_{\text{susy}}$}

\begin{figure*}[t]
	\centering
	\includegraphics[width=4.5in]{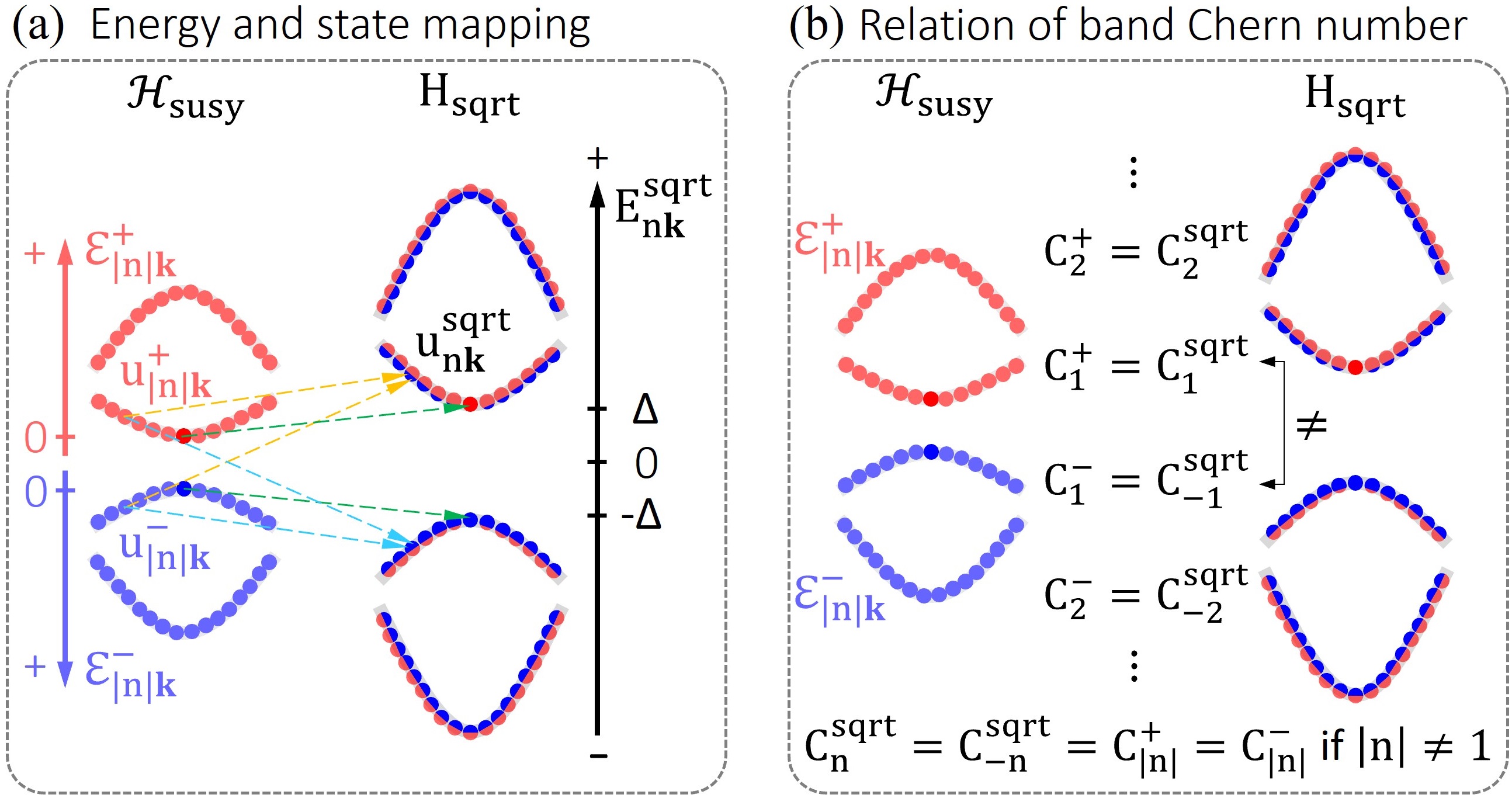}
	\caption{(a) Schematics of mapping from energy $\mathcal{E}_{|n|\bk}^{\pm}$ and states $u_{|n|\bk}^{\pm}$ of $\hH_{\pm}$ to those of $H_{\text{sqrt}}$. The energies are mapped into each other via $E_{n\bk}^{\text{sqrt}}=\pm(\mathcal{E}_{|n|\bk}^{\pm}+\Delta^2)^{1/2}$. The ${E}_{n\bk}^{\text{sqrt}}$ dispersions on the right column consist of dots whose red and blue portions represent the $u_{|n|\bk}^{+}$ and $u_{|n|\bk}^{-}$ components of $u_{n\bk}^{\text{sqrt}}$ respectively [see Eq.~(\ref{Eq:HsqrtUnk})]. The zero modes of $\hH_{+}$ ($\hH_{-}$) are exactly mapped to the threshold modes of $H_{\text{sqrt}}$ at energies $+\Delta$ ($-\Delta$). Other states of $H_{\text{sqrt}}$ involve a mixture of $u_{|n|\bk}^{+}$ and $u_{|n|\bk}^{-}$. (b) Relation of band Chern numbers of $H_{\text{sqrt}}$ (denoted as $C_n^{\text{sqrt}}$) to those of $\hH_{\pm}$ (denoted as $C_{|n|}^{\pm}$).}
	\label{Fig:SUSY2sqrt}
\end{figure*}

Here we define the square root of $\CalH_{\text{susy}}$ in Eq.~(\ref{Eq:Hsusy1}) as
\begin{equation}
H_{\text{sqrt}}=
\begin{pmatrix} 
\Delta \mathbb{I}&M\\M^{\dagger}&-\Delta \mathbb{I}
\end{pmatrix},\label{Eq:Hsqrt}
\end{equation}
where $\Delta$ is a constant and $\mathbb{I}$ is the identity matrix~\cite{commentMsquare}.
In the following we will assume $\Delta\ge0$, and the discussions are not limited to the forms of $M$ in Eq.~(\ref{Eq:M_special}).
$H_{\text{sqrt}}$ can be considered as the square root of $\CalH_{\text{susy}}$ in the sense that its square yields $\CalH_{\text{susy}}$ up to a constant:
\begin{equation}
H_{\text{sqrt}}^2=
\begin{pmatrix} 
MM^{\dagger}&0\\0&M^{\dagger}M 
\end{pmatrix}
+\Delta^2
=\CalH_{\text{susy}}+\Delta^2,\label{Eq:Hsq}
\end{equation}
where the identity matrix has been removed for simplicity.
As the SP Hamiltonian $\hH_{+}=MM^{\dagger}$ and $\hH_{-}=M^{\dagger}M$ have identical positive energies, which will be denoted as $\mathcal{E}>0$, Eq.~(\ref{Eq:Hsq}) dictates that the spectrum of $H_{\text{sqrt}}$ consists of $E^{\text{sqrt}}=\sqrt{\mathcal{E}+\Delta^2}>\Delta$, $E^{\text{sqrt}}=-\sqrt{\mathcal{E}+\Delta^2}<-\Delta$, and $E^{\text{sqrt}}=\pm\Delta$ if $\hH_{\pm}$ have zero modes. This points to a symmetric spectrum for $H_{\text{sqrt}}$ with the possible exception at energies $\pm\Delta$, where the eigenmodes are also dubbed as threshold modes.

When $\Delta=0$, the square-root Hamiltonian $H_{\text{sqrt}}$ has chiral symmetry, i.e., it satisfies $\{H_{\text{sqrt}},\,\Sigma_{z}\}=0$ with $\Sigma_z=\text{diag}(\mathbb{I},\,-\mathbb{I})$. Many lattice models consider such a case~\cite{SqrtEzawaPRR2020,SqrtHatsugaiPRA2020,SqrtHatsugaiPRB2021,SUSYLattice2022}. Chiral symmetry ensures a symmetric spectrum $E\leftrightarrow-E$, however, it should be noted that symmetric spectrum can also appear from SUSY in the absence of chiral symmetry when $\Delta\ne0$. 

When $\Delta\ne0$, the relation $\{H_{\text{sqrt}},\,[H_{\text{sqrt}},\,\Sigma_{z}]\}=0$ is satisfied, thus
$[H_{\text{sqrt}},\,\Sigma_{z}]\propto Q^{\dagger}-Q$ [see Eq.~(\ref{Eq:Q})] transforms eigenstates of $H_{\text{sqrt}}$ with opposite energies into each other~\cite{PedestrianEPL2009}. An exception occurs for the threshold modes: $Q$ and $Q^\dagger$ annihilate the threshold modes, $\Psi_{\Delta}^{\text{sqrt}}=(\psi_{+}^{0},\,0)^T$ and $\Psi_{-\Delta}^{\text{sqrt}}=(0,\,\psi_{-}^{0})^T$, which are also the zero modes of $\CalH_{\text{susy}}$ satisfying $M^{\dagger}\psi_{+}^0=0$ and $M\psi_{-}^0=0$.
Thus the energy levels at $E^{\text{sqrt}}=\Delta$ and $E^{\text{sqrt}}=-\Delta$ are not necessarily paired. 
However, intriguing features can emerge when $H_{\text{sqrt}}$ has symmetric threshold modes at both $\Delta$ and $-\Delta$, or equivalently, when $\hH_{+}$ and $\hH_{-}$ both have zero modes, which will be the focus now.

In the following we consider periodic systems. Bloch functions of the periodic $H_{\text{sqrt}}$ are also related to those of $\hH_{\pm}$. We denote the eigenvalue of the Bloch Hamiltonian $H_{\text{sqrt}}(\bk)$ as $E_{n\bk}^{\text{sqrt}}$, with $n=\pm1,\pm2,\cdots$ labeling conduction ($n>0$) and valence ($n<0$) bands in ascending order of energy magnitude. The eigenvalues of $\hH_{\pm}(\bk)$, which are non-negative, will be denoted as $\mathcal{E}_{|n|\bk}^{\pm}$ with $|n|=1,2,\cdots$, and they satisfy
\begin{equation}
	\mathcal{E}_{|n|\bk}^{\pm}=\left(E_{n\bk}^{\text{sqrt}}\right)^2-\Delta^2\ge0.
\end{equation}
Given the SP Bloch states $u_{|n|\bk}^{\pm}$ satisfying $\hH_{\pm}(\bk)u_{|n|\bk}^{\pm}=\mathcal{E}_{|n|\bk}^{\pm}u_{|n|\bk}^{\pm}$, one can obtain the normalized Bloch state of $H_{\text{sqrt}}(\bk)$ in terms of $u_{|n|\bk}^{\pm}$ as 
\begin{equation}
u_{n\bk}^{\text{sqrt}}
\equiv
\begin{pmatrix}
c_{n\bk}^{+}u_{|n|\bk}^{+}\\c_{n\bk}^{-}u_{|n|\bk}^{-}
\end{pmatrix}
=
\begin{pmatrix}
\sqrt{\frac{E_{n\bk}^{\text{sqrt}}+\Delta}{2E_{n\bk}^{\text{sqrt}}}}u_{|n|\bk}^{+}\\
\text{sgn}(E_{n\bk}^{\text{sqrt}})\sqrt{\frac{E_{n\bk}^{\text{sqrt}}-\Delta}{2E_{n\bk}^{\text{sqrt}}}}u_{|n|\bk}^{-}
\end{pmatrix}.\label{Eq:HsqrtUnk}
\end{equation}
In general, the two-component $u_{n\bk}^{\text{sqrt}}$ consists of both SP Bloch functions $u_{|n|\bk}^{\pm}$, except for the threshold modes at energy $E_{n\bk}^{\text{sqrt}}=\Delta$ ($-\Delta$), which are fully polarized on  $u_{|n|\bk}^{+}$ ($u_{|n|\bk}^{-}$) that correspond to the zero modes of $\hH_+$ ($\hH_-$).

When the bands are narrow, satisfying $|E_{n\bk}^{\text{sqrt}}|-\Delta\ll\Delta$, one finds $c_{n\bk}^{+}\approx1$ \& $c_{n\bk}^{-}\approx0$ ($c_{n\bk}^{+}\approx0$ \& $c_{n\bk}^{-}\approx-1$) in the conduction (valence) bands, thus $u_{n\bk}^{\text{sqrt}}$ is near fully polarized on the $u_{|n|\bk}^{+}$ ($u_{|n|\bk}^{-}$) component. 
In such cases, electrons in the conduction and valence bands of $H_{\text{sqrt}}$ can be well described by $H_{\text{sqrt}}^{c}=MM^\dagger/(2\Delta)$ and $H_{\text{sqrt}}^{v}=-M^\dagger M/(2\Delta)$, respectively. 

The mapping from energy $\mathcal{E}_{|n|\bk}^{\pm}$ and states $u_{|n|\bk}^{\pm}$ of $\hH_{\pm}$ to those of $H_{\text{sqrt}}$ are shown schematically in Fig.~\ref{Fig:SUSY2sqrt}(a). In the right column, the ${E}_{n\bk}^{\text{sqrt}}$ dispersions consist of dots whose red and blue portions represent the $u_{|n|\bk}^{+}$ and $u_{|n|\bk}^{-}$ components of $u_{n\bk}^{\text{sqrt}}$ respectively [see Eq.~(\ref{Eq:HsqrtUnk})]. Reciprocally, one can deduce the eigenenergies and eigenstates
of $\hH_{\pm}$ from $E_{n\bk}^{\text{sqrt}}$ and $u_{n\bk}^{\text{sqrt}}$. And usually diagonalizing $H_{\text{sqrt}}$ can be done more conveniently as $H_{\text{sqrt}}$ is linear in $M$ and $M^\dagger$.

Eq.~(\ref{Eq:HsqrtUnk}) further implies that the band topology of $H_{\text{sqrt}}$ are related to those of $\hH_{\pm}$. Recall that the states $u_{|n|\bk}^{\pm}$ are paired by Eq.~(\ref{Eq:SUSYWaveFunctionRelation}) when the energy is nonzero, 
this allows one to relate the Berry connection of $H_{\text{sqrt}}$ to those of $\hH_{\pm}$, whose details are given in Sec.~S5 of Supplementary Material~\cite{supp}. From such relations, we arrive at the following
connections between the band Chern numbers of $H_{\text{sqrt}}$ (denoted as $C_n^{\text{sqrt}}$) and those of $\hH_{\pm}$ (denoted as $C_{|n|}^{\pm}$):
\begin{equation}
C_n^{\text{sqrt}}
=
\begin{cases}
C_{|n|}^{+}=C_{|n|}^{-}&\text{if $|n|>1$}\\
C_{1}^{+}&\text{if $n=1$}\\
C_{1}^{-}&\text{if $n=-1$}
\end{cases}.\label{Eq:ChernNumberHsqrt}
\end{equation}
These relations, schematically summarized in Fig.~\ref{Fig:SUSY2sqrt}(b), are not limited to the forms of $M$ in Eq.~(\ref{Eq:M_special}) and independent of the value of $\Delta\ne0$ (with a fixed sign).
Therefore, the symmetric energy bands of $H_{\text{sqrt}}$ are accompanied by symmetric Chern numbers, i.e. $C_n^{\text{sqrt}}=C_{-n}^{\text{sqrt}}$, except for the first conduction ($n=1$) and valence ($n=-1$) bands that contain the threshold modes. 

Incidentally, in the case of $\Delta=0$, one readily notices that chiral symmetry also underlies the identical Chern numbers between the $|n| >1$ conduction and valence bands (the Chern numbers for individual $n=\pm1$ bands become ill-defined if there is degeneracy at the zero energy, they are equal if the zero modes are absent), but in the generic case this equality relation is traced back to Eq.~(\ref{Eq:IdenticalC})-- a consequence of SUSY.
And with the chiral symmetry at $\Delta\equiv0$, topology of the square-root Hamiltonian is also noted in some lattice models~\cite{SqrtPRB2017,SqrtEzawaPRR2020,SqrtHatsugaiPRA2020,SqrtHatsugaiPRB2021}.

When chiral symmetry is broken in the case of $\Delta\ne0$, our findings give insights on how {\it contrasted} topology of conduction and valence bands arises from the SUSY of the squared Hamiltonian, 
as dictated by the zero modes and the dispersion in their vicinity in the SUSY spectrum (e.g., Fig.~\ref{Fig:TopologySummary}).
With $\Delta>0$, the Chern numbers are well defined in the first conduction and valence bands of $H_{\text{sqrt}}$, and $C_{1}^{\text{sqrt}}=C_{1}^{+}$ and $C_{-1}^{\text{sqrt}}=C_{1}^{-}$ are satisfied. So $C_{1}^{\text{sqrt}}\ne C_{-1}^{\text{sqrt}}$ when the dispersion in the vicinity of threshold/zero mode determines a finite difference between $C_{1}^{+}$ and $C_{1}^{-}$ [see e.g., Eq.~(\ref{Eq:differN}) and Fig.~\ref{Fig:TopologySummary}].
The contrasted topology of the SPs of the squared Hamiltonian ($\CalH_{\text{susy}}$) in 
bands containing their zero modes
are therefore transferred to the first conduction and valence bands of $H_{\text{sqrt}}$ respectively [Fig.~\ref{Fig:SUSY2sqrt}(b)]. 


\subsection{Manifestions in flat-band superlattices}~\label{Sect:HsqrtExample}

Figure~S3 in the Supplementary Material~\cite{supp} confirms the above results with a simple example of $H_{\text{sqrt}}$ that describes massive Dirac fermions in a periodic magnetic field [see Fig.~\ref{Fig:QuadraticTriB}(a)]. Specifically, here $M=\pi_{-}$, and the associated SUSY system $\hH_{\pm}$ are given by Eq.~(\ref{Eq:SchrodingerWithB}). By comparing Fig.~S3(b) and Fig.~\ref{Fig:QuadraticTriB}(b), we find $C_{1}^{\text{sqrt}}=-1=C_{1}^{+}$ versus $C_{-1}^{\text{sqrt}}=0=C_{1}^{-}$, and $C_{n}^{\text{sqrt}}=C_{-n}^{\text{sqrt}}=C_{|n|}^{+}=C_{|n|}^{-}$ otherwise, which is consistent with Eq.~(\ref{Eq:ChernNumberHsqrt}) and Fig.~\ref{Fig:SUSY2sqrt}(b). 
This model of $H_{\text{sqrt}}$, and its TR counterpart, can be realized by using strained graphene with a staggered potential~\cite{ZhaiNanoLett2023,PhongPRL2022,PeriodicStrainSunKaiPRB2023,EslamPeriodicStrainPRL2023}.

TBG in the chiral limit~\cite{TBGMagicAngleOriginPRL2019} and strained bilayer graphene (SBG)~\cite{PeriodicStrainSunKaiPRB2023} are another two important examples where the contrasted topology of the conduction and valence minibands containing the threshold modes are dictated by Eq.~(\ref{Eq:differN}) and Eq.~(\ref{Eq:DiffC_multizeros}). On one hand, they serve as cases with non-Abelian gauge fields~\cite{TBGGaugeFieldPRL2012}, on the other hand, they both exhibit $C_{-1}^{\text{sqrt}}-C_{1}^{\text{sqrt}}=2$, but due to drastically different origins. They both can be formulated in the form of $H_{\text{sqrt}}=\sigma_{x}\otimes(p_x+e\CalA_x)+\sigma_{y}\otimes(p_y+e\CalA_y)$, as detailed in Supplementary Material~\cite{supp}. Unlike a usual magnetic field, here $\bCalA=(\CalA_x,\,\CalA_y)$ is a non-Abelian gauge potential due to interlayer coupling, and the associated non-Abelian gauge field reads $\bCalB=\nabla\times\bCalA+i\frac{e}{\hbar}[\CalA_{x},\,\CalA_{y}]\hat{\bz}$. They are matrices due to the layer degree of freedom in a hybridized bilayer. The associated SUSY Hamiltonian is given by the square of $H_{\text{sqrt}}$, i.e., $\hH_{\pm}=\frac{1}{2m}(\bp+e\bCalA)^2\pm\bmu_B\cdot\bCalB$. The results discussed earlier, e.g., contrasted band topology in $\hH_{\pm}$, thus can be applied directly. Let us focus on band topology and highlight the similarity/difference between TBG and SBG. In TBG, one expects $C_1^--C_1^+=2$ because each SP has two zero modes (at the Dirac points, and we assume the twist angle is not magic~\cite{BM,TBGMagicAngleOriginPRL2019}), around which the low-energy dispersion of $\hH_{\pm}$ is quadratic [see Fig.~\ref{Fig:TopologySummary}(b)]. While $C_1^--C_1^+=2$ is also expected in SBG, it instead results from one zero mode for each SP with quartic dispersion. If a sublattice staggered potential is added to introduce a finite $\Delta$ in $H_{\text{sqrt}}$, these results will manifest in the conduction and valence bands of $H_{\text{sqrt}}$ as $C_{-1}^{\text{sqrt}}-C_{+1}^{\text{sqrt}}=2$, which are consistent with numerical results (see Fig.~S4(c) and Fig.~S5(b) of Supplementary Material~\cite{supp}).

The strained $\Gamma$-valley model with quadratic band touching~\cite{StrainQuadraticBandTouchingSunKaiPRL2023} is an example that differs from all previous examples. Its Hamiltonian also has the form of $H_{\text{sqrt}}$, whereas $M=(p_x^2-p_y^2+f_1)-i(2p_xp_y+f_2)$ with $f_{1,2}(\br)$ being some space-dependent functions, thus the effects of strain cannot be understood in terms of a gauge field. Nevertheless, topological flat bands and magic angles --like those in TBG-- are also discovered. Our results [Fig.~\ref{Fig:SUSY2sqrt}(b)] show that its band topology are inherited from the associated $\CalH_{\text{susy}}$ (despite in absence of an effective gauge field), thus provide insights for understanding the emergence of nontrivial topology and the distinct Chern numbers in the bands that contain the zero/threshold modes.

In addition to being fundamentally intriguing, the contrasted band topology in the lowest conduction and valence bands could be important for hosting nontrivial physical phenomena. Firstly, similar correlation-driven states as those discussed in earlier sections can be expected with electron or hole doping when interaction is strong. Furthermore, the interplay of distinct topology in the conduction and valence bands could lead to other interesting states of matter. For example, it was predicted recently that topological excitons with a finite vorticity in momentum space can emerge in systems with topologically distinct conduction and valence bands, which can exhibit optical selection rules of geometrical/topological origin as well as lead to interesting nonlinear anomalous Hall effects~\cite{TopoExcitonXie}. The square-root systems discussed here serve as natural playgrounds for exploring topological exciton insulators and condensates.


\section{Summary and discussions}

In this work, we focus on SUSY systems that involve electrons in a zero-flux gauge field. We reveal that the SPs have isospectra, yet remarkably, the energy bands that contain the zero modes exhibit distinct topology [e.g., Fig.~\ref{Fig:QuadraticTriB}(b)]: the difference in their Chern numbers is determined by the number of zero modes and low-energy dispersion around them (Fig.~\ref{Fig:TopologySummary}). This puts forward a new scenario of band degeneracy, where the two degenerate partners can be topologically trivial and nontrivial respectively, in stark contrast to the scenario due to the time-reversal symmetry. Such SUSY models can be realized in multilayer 2D semiconductors by twist and strain engineering (e.g., Figs.~\ref{Fig:SubstrateBilayer30deg} and \ref{Fig:SubstrateBernalBilayerHole}). 
Spontaneous SUSY breaking due to Coulomb interaction can occur at fractional filling of the lowest bands of the SPs, which is accompanied by the emergence of distinct correlated phases for the two SPs.
We also generalize the discussions to systems whose physical Hamiltonian corresponds to the square root of a SUSY Hamiltonian, where the topological properties of the two SUSY SPs are transferred to the conduction and valence bands of the physical Hamiltonian respectively (Fig.~\ref{Fig:SUSY2sqrt}). As many of the flat-band systems are described by such square-root models, our findings offer a unifying picture to understand their topological properties.
It is conceivable that the knowledge of various SUSY systems and the zero modes~\cite{SUSYBookGeorg} will introduce novel insights and platforms for exploring the correlation-driven phenomena that are actively pursued in superlattice materials recently.
In this work $M$ and $M^\dagger$ in Eq.~(\ref{Eq:M_special}) are considered due to their relevance in many 2D materials. One could also explore similar contrasted properties in $\hH_{\pm}$ with other forms of $M$ and $M^\dagger$, where the SPs may exhibit other dispersion relation~\cite{NodalRingPRB2020} around the zero energy, or in SUSY systems with other structures of $\hH_{\pm}$~\cite{SUSYBookGeorg}.

Finally, we note that extra terms that break SUSY often exist in some of the discussed systems. For example, the model of TBG in the nonchiral limit contains an effective scalar potential~\cite{TBGGaugeFieldPRL2012}. However, provided such terms are not prominent, SUSY will only be weakly broken, thus the SUSY dictated topological properties are expected to remain unaffected in the nearly symmetric spectra. This is clearly manifested in TBG, in which the energy spectrum remain approximately symmetric at different twist angles~\cite{TBGKoshinoPRB2013} with SUSY dictated Chern numbers for the first conduction/valence bands.


\section{Data availability statement}

The numerical data generated by the custom codes that support the findings of this study are available upon reasonable request from the authors.

\section{Acknowledgment}

We thank Nancy Sandler, Bo Fu, Ci Li, Tixuan Tan, and Jie Wang for helpful discussions. The work is supported by Research Grant Council of Hong Kong SAR China (HKU SRFS2122-7S05 and AoE/P-701/20), and New Cornerstone Science Foundation.

\bibliography{StrainSuperlatticeRefs}

\begin{thebibliography}{88}%
\makeatletter
\providecommand \@ifxundefined [1]{%
 \@ifx{#1\undefined}
}%
\providecommand \@ifnum [1]{%
 \ifnum #1\expandafter \@firstoftwo
 \else \expandafter \@secondoftwo
 \fi
}%
\providecommand \@ifx [1]{%
 \ifx #1\expandafter \@firstoftwo
 \else \expandafter \@secondoftwo
 \fi
}%
\providecommand \natexlab [1]{#1}%
\providecommand \enquote  [1]{``#1''}%
\providecommand \bibnamefont  [1]{#1}%
\providecommand \bibfnamefont [1]{#1}%
\providecommand \citenamefont [1]{#1}%
\providecommand \href@noop [0]{\@secondoftwo}%
\providecommand \href [0]{\begingroup \@sanitize@url \@href}%
\providecommand \@href[1]{\@@startlink{#1}\@@href}%
\providecommand \@@href[1]{\endgroup#1\@@endlink}%
\providecommand \@sanitize@url [0]{\catcode `\\12\catcode `\$12\catcode
  `\&12\catcode `\#12\catcode `\^12\catcode `\_12\catcode `\%12\relax}%
\providecommand \@@startlink[1]{}%
\providecommand \@@endlink[0]{}%
\providecommand \url  [0]{\begingroup\@sanitize@url \@url }%
\providecommand \@url [1]{\endgroup\@href {#1}{\urlprefix }}%
\providecommand \urlprefix  [0]{URL }%
\providecommand \Eprint [0]{\href }%
\providecommand \doibase [0]{http://dx.doi.org/}%
\providecommand \selectlanguage [0]{\@gobble}%
\providecommand \bibinfo  [0]{\@secondoftwo}%
\providecommand \bibfield  [0]{\@secondoftwo}%
\providecommand \translation [1]{[#1]}%
\providecommand \BibitemOpen [0]{}%
\providecommand \bibitemStop [0]{}%
\providecommand \bibitemNoStop [0]{.\EOS\space}%
\providecommand \EOS [0]{\spacefactor3000\relax}%
\providecommand \BibitemShut  [1]{\csname bibitem#1\endcsname}%
\let\auto@bib@innerbib\@empty
\bibitem [{\citenamefont {Cao}\ \emph {et~al.}(2018{\natexlab{a}})\citenamefont
  {Cao}, \citenamefont {Fatemi}, \citenamefont {Fang}, \citenamefont
  {Watanabe}, \citenamefont {Taniguchi}, \citenamefont {Kaxiras},\ and\
  \citenamefont {Jarillo-Herrero}}]{TBGCaoYuan2018a}%
  \BibitemOpen
  \bibfield  {author} {\bibinfo {author} {\bibfnamefont {Y.}~\bibnamefont
  {Cao}}, \bibinfo {author} {\bibfnamefont {V.}~\bibnamefont {Fatemi}},
  \bibinfo {author} {\bibfnamefont {S.}~\bibnamefont {Fang}}, \bibinfo {author}
  {\bibfnamefont {K.}~\bibnamefont {Watanabe}}, \bibinfo {author}
  {\bibfnamefont {T.}~\bibnamefont {Taniguchi}}, \bibinfo {author}
  {\bibfnamefont {E.}~\bibnamefont {Kaxiras}}, \ and\ \bibinfo {author}
  {\bibfnamefont {P.}~\bibnamefont {Jarillo-Herrero}},\ }\href {\doibase
  10.1038/nature26160} {\bibfield  {journal} {\bibinfo  {journal} {Nature}\
  }\textbf {\bibinfo {volume} {556}},\ \bibinfo {pages} {43} (\bibinfo {year}
  {2018}{\natexlab{a}})}\BibitemShut {NoStop}%
\bibitem [{\citenamefont {Cao}\ \emph {et~al.}(2018{\natexlab{b}})\citenamefont
  {Cao}, \citenamefont {Fatemi}, \citenamefont {Demir}, \citenamefont {Fang},
  \citenamefont {Tomarken}, \citenamefont {Luo}, \citenamefont
  {Sanchez-Yamagishi}, \citenamefont {Watanabe}, \citenamefont {Taniguchi},
  \citenamefont {Kaxiras}, \citenamefont {Ashoori},\ and\ \citenamefont
  {Jarillo-Herrero}}]{TBGCaoYuan2018b}%
  \BibitemOpen
  \bibfield  {author} {\bibinfo {author} {\bibfnamefont {Y.}~\bibnamefont
  {Cao}}, \bibinfo {author} {\bibfnamefont {V.}~\bibnamefont {Fatemi}},
  \bibinfo {author} {\bibfnamefont {A.}~\bibnamefont {Demir}}, \bibinfo
  {author} {\bibfnamefont {S.}~\bibnamefont {Fang}}, \bibinfo {author}
  {\bibfnamefont {S.~L.}\ \bibnamefont {Tomarken}}, \bibinfo {author}
  {\bibfnamefont {J.~Y.}\ \bibnamefont {Luo}}, \bibinfo {author} {\bibfnamefont
  {J.~D.}\ \bibnamefont {Sanchez-Yamagishi}}, \bibinfo {author} {\bibfnamefont
  {K.}~\bibnamefont {Watanabe}}, \bibinfo {author} {\bibfnamefont
  {T.}~\bibnamefont {Taniguchi}}, \bibinfo {author} {\bibfnamefont
  {E.}~\bibnamefont {Kaxiras}}, \bibinfo {author} {\bibfnamefont {R.~C.}\
  \bibnamefont {Ashoori}}, \ and\ \bibinfo {author} {\bibfnamefont
  {P.}~\bibnamefont {Jarillo-Herrero}},\ }\href {\doibase 10.1038/nature26154}
  {\bibfield  {journal} {\bibinfo  {journal} {Nature}\ }\textbf {\bibinfo
  {volume} {556}},\ \bibinfo {pages} {80} (\bibinfo {year}
  {2018}{\natexlab{b}})}\BibitemShut {NoStop}%
\bibitem [{\citenamefont {Andrei}\ and\ \citenamefont
  {MacDonald}(2020)}]{moireReviewEvaMacDonaldNatMater2020}%
  \BibitemOpen
  \bibfield  {author} {\bibinfo {author} {\bibfnamefont {E.~Y.}\ \bibnamefont
  {Andrei}}\ and\ \bibinfo {author} {\bibfnamefont {A.~H.}\ \bibnamefont
  {MacDonald}},\ }\href {\doibase 10.1038/s41563-020-00840-0} {\bibfield
  {journal} {\bibinfo  {journal} {Nat. Mater.}\ }\textbf {\bibinfo {volume}
  {19}},\ \bibinfo {pages} {1265} (\bibinfo {year} {2020})}\BibitemShut
  {NoStop}%
\bibitem [{\citenamefont {Balents}\ \emph {et~al.}(2020)\citenamefont
  {Balents}, \citenamefont {Dean}, \citenamefont {Efetov},\ and\ \citenamefont
  {Young}}]{moireReviewNatPhysBalents2020}%
  \BibitemOpen
  \bibfield  {author} {\bibinfo {author} {\bibfnamefont {L.}~\bibnamefont
  {Balents}}, \bibinfo {author} {\bibfnamefont {C.~R.}\ \bibnamefont {Dean}},
  \bibinfo {author} {\bibfnamefont {D.~K.}\ \bibnamefont {Efetov}}, \ and\
  \bibinfo {author} {\bibfnamefont {A.~F.}\ \bibnamefont {Young}},\ }\href
  {\doibase 10.1038/s41567-020-0906-9} {\bibfield  {journal} {\bibinfo
  {journal} {Nat. Phys.}\ }\textbf {\bibinfo {volume} {16}},\ \bibinfo {pages}
  {725} (\bibinfo {year} {2020})}\BibitemShut {NoStop}%
\bibitem [{\citenamefont {Kennes}\ \emph {et~al.}(2021)\citenamefont {Kennes},
  \citenamefont {Claassen}, \citenamefont {Xian}, \citenamefont {Georges},
  \citenamefont {Millis}, \citenamefont {Hone}, \citenamefont {Dean},
  \citenamefont {Basov}, \citenamefont {Pasupathy},\ and\ \citenamefont
  {Rubio}}]{moireReviewRubioNatPhys2021}%
  \BibitemOpen
  \bibfield  {author} {\bibinfo {author} {\bibfnamefont {D.~M.}\ \bibnamefont
  {Kennes}}, \bibinfo {author} {\bibfnamefont {M.}~\bibnamefont {Claassen}},
  \bibinfo {author} {\bibfnamefont {L.}~\bibnamefont {Xian}}, \bibinfo {author}
  {\bibfnamefont {A.}~\bibnamefont {Georges}}, \bibinfo {author} {\bibfnamefont
  {A.~J.}\ \bibnamefont {Millis}}, \bibinfo {author} {\bibfnamefont
  {J.}~\bibnamefont {Hone}}, \bibinfo {author} {\bibfnamefont {C.~R.}\
  \bibnamefont {Dean}}, \bibinfo {author} {\bibfnamefont {D.~N.}\ \bibnamefont
  {Basov}}, \bibinfo {author} {\bibfnamefont {A.~N.}\ \bibnamefont
  {Pasupathy}}, \ and\ \bibinfo {author} {\bibfnamefont {A.}~\bibnamefont
  {Rubio}},\ }\href {\doibase 10.1038/s41567-020-01154-3} {\bibfield  {journal}
  {\bibinfo  {journal} {Nat. Phys.}\ }\textbf {\bibinfo {volume} {17}},\
  \bibinfo {pages} {155} (\bibinfo {year} {2021})}\BibitemShut {NoStop}%
\bibitem [{\citenamefont {Andrei}\ \emph {et~al.}(2021)\citenamefont {Andrei},
  \citenamefont {Efetov}, \citenamefont {Jarillo-Herrero}, \citenamefont
  {MacDonald}, \citenamefont {Mak}, \citenamefont {Senthil}, \citenamefont
  {Tutuc}, \citenamefont {Yazdani},\ and\ \citenamefont
  {Young}}]{moireReviewExptFolksNatRevMat2021}%
  \BibitemOpen
  \bibfield  {author} {\bibinfo {author} {\bibfnamefont {E.~Y.}\ \bibnamefont
  {Andrei}}, \bibinfo {author} {\bibfnamefont {D.~K.}\ \bibnamefont {Efetov}},
  \bibinfo {author} {\bibfnamefont {P.}~\bibnamefont {Jarillo-Herrero}},
  \bibinfo {author} {\bibfnamefont {A.~H.}\ \bibnamefont {MacDonald}}, \bibinfo
  {author} {\bibfnamefont {K.~F.}\ \bibnamefont {Mak}}, \bibinfo {author}
  {\bibfnamefont {T.}~\bibnamefont {Senthil}}, \bibinfo {author} {\bibfnamefont
  {E.}~\bibnamefont {Tutuc}}, \bibinfo {author} {\bibfnamefont
  {A.}~\bibnamefont {Yazdani}}, \ and\ \bibinfo {author} {\bibfnamefont
  {A.~F.}\ \bibnamefont {Young}},\ }\href {\doibase 10.1038/s41578-021-00284-1}
  {\bibfield  {journal} {\bibinfo  {journal} {Nat. Rev. Mater.}\ }\textbf
  {\bibinfo {volume} {6}},\ \bibinfo {pages} {201} (\bibinfo {year}
  {2021})}\BibitemShut {NoStop}%
\bibitem [{\citenamefont {Lau}\ \emph {et~al.}(2022)\citenamefont {Lau},
  \citenamefont {Bockrath}, \citenamefont {Mak},\ and\ \citenamefont
  {Zhang}}]{moireReviewJeanieLauNature2022}%
  \BibitemOpen
  \bibfield  {author} {\bibinfo {author} {\bibfnamefont {C.~N.}\ \bibnamefont
  {Lau}}, \bibinfo {author} {\bibfnamefont {M.~W.}\ \bibnamefont {Bockrath}},
  \bibinfo {author} {\bibfnamefont {K.~F.}\ \bibnamefont {Mak}}, \ and\
  \bibinfo {author} {\bibfnamefont {F.}~\bibnamefont {Zhang}},\ }\href
  {\doibase 10.1038/s41586-021-04173-z} {\bibfield  {journal} {\bibinfo
  {journal} {Nature}\ }\textbf {\bibinfo {volume} {602}},\ \bibinfo {pages}
  {41} (\bibinfo {year} {2022})}\BibitemShut {NoStop}%
\bibitem [{\citenamefont {Tarnopolsky}\ \emph {et~al.}(2019)\citenamefont
  {Tarnopolsky}, \citenamefont {Kruchkov},\ and\ \citenamefont
  {Vishwanath}}]{TBGMagicAngleOriginPRL2019}%
  \BibitemOpen
  \bibfield  {author} {\bibinfo {author} {\bibfnamefont {G.}~\bibnamefont
  {Tarnopolsky}}, \bibinfo {author} {\bibfnamefont {A.~J.}\ \bibnamefont
  {Kruchkov}}, \ and\ \bibinfo {author} {\bibfnamefont {A.}~\bibnamefont
  {Vishwanath}},\ }\href {\doibase 10.1103/PhysRevLett.122.106405} {\bibfield
  {journal} {\bibinfo  {journal} {Phys. Rev. Lett.}\ }\textbf {\bibinfo
  {volume} {122}},\ \bibinfo {pages} {106405} (\bibinfo {year}
  {2019})}\BibitemShut {NoStop}%
\bibitem [{\citenamefont {Liu}\ \emph {et~al.}(2019)\citenamefont {Liu},
  \citenamefont {Liu},\ and\ \citenamefont {Dai}}]{DaiXiPseuodoFieldPRB2019}%
  \BibitemOpen
  \bibfield  {author} {\bibinfo {author} {\bibfnamefont {J.}~\bibnamefont
  {Liu}}, \bibinfo {author} {\bibfnamefont {J.}~\bibnamefont {Liu}}, \ and\
  \bibinfo {author} {\bibfnamefont {X.}~\bibnamefont {Dai}},\ }\href {\doibase
  10.1103/PhysRevB.99.155415} {\bibfield  {journal} {\bibinfo  {journal} {Phys.
  Rev. B}\ }\textbf {\bibinfo {volume} {99}},\ \bibinfo {pages} {155415}
  (\bibinfo {year} {2019})}\BibitemShut {NoStop}%
\bibitem [{\citenamefont {Wu}\ \emph {et~al.}(2019)\citenamefont {Wu},
  \citenamefont {Lovorn}, \citenamefont {Tutuc}, \citenamefont {Martin},\ and\
  \citenamefont {MacDonald}}]{WuMacDonaldPRL2019}%
  \BibitemOpen
  \bibfield  {author} {\bibinfo {author} {\bibfnamefont {F.}~\bibnamefont
  {Wu}}, \bibinfo {author} {\bibfnamefont {T.}~\bibnamefont {Lovorn}}, \bibinfo
  {author} {\bibfnamefont {E.}~\bibnamefont {Tutuc}}, \bibinfo {author}
  {\bibfnamefont {I.}~\bibnamefont {Martin}}, \ and\ \bibinfo {author}
  {\bibfnamefont {A.~H.}\ \bibnamefont {MacDonald}},\ }\href {\doibase
  10.1103/PhysRevLett.122.086402} {\bibfield  {journal} {\bibinfo  {journal}
  {Phys. Rev. Lett.}\ }\textbf {\bibinfo {volume} {122}},\ \bibinfo {pages}
  {086402} (\bibinfo {year} {2019})}\BibitemShut {NoStop}%
\bibitem [{\citenamefont {Yu}\ \emph {et~al.}(2020)\citenamefont {Yu},
  \citenamefont {Chen},\ and\ \citenamefont {Yao}}]{HongyiNSR2020}%
  \BibitemOpen
  \bibfield  {author} {\bibinfo {author} {\bibfnamefont {H.}~\bibnamefont
  {Yu}}, \bibinfo {author} {\bibfnamefont {M.}~\bibnamefont {Chen}}, \ and\
  \bibinfo {author} {\bibfnamefont {W.}~\bibnamefont {Yao}},\ }\href {\doibase
  10.1093/nsr/nwz117} {\bibfield  {journal} {\bibinfo  {journal} {Natl. Sci.
  Rev.}\ }\textbf {\bibinfo {volume} {7}},\ \bibinfo {pages} {12} (\bibinfo
  {year} {2020})}\BibitemShut {NoStop}%
\bibitem [{\citenamefont {Zhai}\ and\ \citenamefont {Yao}(2020)}]{ZhaiPRM2020}%
  \BibitemOpen
  \bibfield  {author} {\bibinfo {author} {\bibfnamefont {D.}~\bibnamefont
  {Zhai}}\ and\ \bibinfo {author} {\bibfnamefont {W.}~\bibnamefont {Yao}},\
  }\href {\doibase 10.1103/PhysRevMaterials.4.094002} {\bibfield  {journal}
  {\bibinfo  {journal} {Phys. Rev. Materials}\ }\textbf {\bibinfo {volume}
  {4}},\ \bibinfo {pages} {094002} (\bibinfo {year} {2020})}\BibitemShut
  {NoStop}%
\bibitem [{\citenamefont {Pan}\ \emph {et~al.}(2020)\citenamefont {Pan},
  \citenamefont {Wu},\ and\ \citenamefont {Das~Sarma}}]{WuFengchengPRR2020}%
  \BibitemOpen
  \bibfield  {author} {\bibinfo {author} {\bibfnamefont {H.}~\bibnamefont
  {Pan}}, \bibinfo {author} {\bibfnamefont {F.}~\bibnamefont {Wu}}, \ and\
  \bibinfo {author} {\bibfnamefont {S.}~\bibnamefont {Das~Sarma}},\ }\href
  {\doibase 10.1103/PhysRevResearch.2.033087} {\bibfield  {journal} {\bibinfo
  {journal} {Phys. Rev. Res.}\ }\textbf {\bibinfo {volume} {2}},\ \bibinfo
  {pages} {033087} (\bibinfo {year} {2020})}\BibitemShut {NoStop}%
\bibitem [{\citenamefont {Devakul}\ \emph {et~al.}(2021)\citenamefont
  {Devakul}, \citenamefont {Cr{\'e}pel}, \citenamefont {Zhang},\ and\
  \citenamefont {Fu}}]{TwistedWSe2MagicFuLiang2021}%
  \BibitemOpen
  \bibfield  {author} {\bibinfo {author} {\bibfnamefont {T.}~\bibnamefont
  {Devakul}}, \bibinfo {author} {\bibfnamefont {V.}~\bibnamefont {Cr{\'e}pel}},
  \bibinfo {author} {\bibfnamefont {Y.}~\bibnamefont {Zhang}}, \ and\ \bibinfo
  {author} {\bibfnamefont {L.}~\bibnamefont {Fu}},\ }\href {\doibase
  10.1038/s41467-021-27042-9} {\bibfield  {journal} {\bibinfo  {journal} {Nat.
  Commun.}\ }\textbf {\bibinfo {volume} {12}},\ \bibinfo {pages} {6730}
  (\bibinfo {year} {2021})}\BibitemShut {NoStop}%
\bibitem [{\citenamefont {Anderson}\ \emph {et~al.}(2023)\citenamefont
  {Anderson}, \citenamefont {Fan}, \citenamefont {Cai}, \citenamefont
  {Holtzmann}, \citenamefont {Taniguchi}, \citenamefont {Watanabe},
  \citenamefont {Xiao}, \citenamefont {Yao},\ and\ \citenamefont
  {Xu}}]{FCIMoTe2Eric2023}%
  \BibitemOpen
  \bibfield  {author} {\bibinfo {author} {\bibfnamefont {E.}~\bibnamefont
  {Anderson}}, \bibinfo {author} {\bibfnamefont {F.-R.}\ \bibnamefont {Fan}},
  \bibinfo {author} {\bibfnamefont {J.}~\bibnamefont {Cai}}, \bibinfo {author}
  {\bibfnamefont {W.}~\bibnamefont {Holtzmann}}, \bibinfo {author}
  {\bibfnamefont {T.}~\bibnamefont {Taniguchi}}, \bibinfo {author}
  {\bibfnamefont {K.}~\bibnamefont {Watanabe}}, \bibinfo {author}
  {\bibfnamefont {D.}~\bibnamefont {Xiao}}, \bibinfo {author} {\bibfnamefont
  {W.}~\bibnamefont {Yao}}, \ and\ \bibinfo {author} {\bibfnamefont
  {X.}~\bibnamefont {Xu}},\ }\href {\doibase 10.1126/science.adg4268}
  {\bibfield  {journal} {\bibinfo  {journal} {Science}\ }\textbf {\bibinfo
  {volume} {381}},\ \bibinfo {pages} {325} (\bibinfo {year}
  {2023})}\BibitemShut {NoStop}%
\bibitem [{\citenamefont {Cai}\ \emph {et~al.}(2023)\citenamefont {Cai},
  \citenamefont {Anderson}, \citenamefont {Wang}, \citenamefont {Zhang},
  \citenamefont {Liu}, \citenamefont {Holtzmann}, \citenamefont {Zhang},
  \citenamefont {Fan}, \citenamefont {Taniguchi}, \citenamefont {Watanabe},
  \citenamefont {Ran}, \citenamefont {Cao}, \citenamefont {Fu}, \citenamefont
  {Xiao}, \citenamefont {Yao},\ and\ \citenamefont {Xu}}]{FCIMoTe2Jiaqi2023}%
  \BibitemOpen
  \bibfield  {author} {\bibinfo {author} {\bibfnamefont {J.}~\bibnamefont
  {Cai}}, \bibinfo {author} {\bibfnamefont {E.}~\bibnamefont {Anderson}},
  \bibinfo {author} {\bibfnamefont {C.}~\bibnamefont {Wang}}, \bibinfo {author}
  {\bibfnamefont {X.}~\bibnamefont {Zhang}}, \bibinfo {author} {\bibfnamefont
  {X.}~\bibnamefont {Liu}}, \bibinfo {author} {\bibfnamefont {W.}~\bibnamefont
  {Holtzmann}}, \bibinfo {author} {\bibfnamefont {Y.}~\bibnamefont {Zhang}},
  \bibinfo {author} {\bibfnamefont {F.}~\bibnamefont {Fan}}, \bibinfo {author}
  {\bibfnamefont {T.}~\bibnamefont {Taniguchi}}, \bibinfo {author}
  {\bibfnamefont {K.}~\bibnamefont {Watanabe}}, \bibinfo {author}
  {\bibfnamefont {Y.}~\bibnamefont {Ran}}, \bibinfo {author} {\bibfnamefont
  {T.}~\bibnamefont {Cao}}, \bibinfo {author} {\bibfnamefont {L.}~\bibnamefont
  {Fu}}, \bibinfo {author} {\bibfnamefont {D.}~\bibnamefont {Xiao}}, \bibinfo
  {author} {\bibfnamefont {W.}~\bibnamefont {Yao}}, \ and\ \bibinfo {author}
  {\bibfnamefont {X.}~\bibnamefont {Xu}},\ }\href {\doibase
  10.1038/s41586-023-06289-w} {\bibfield  {journal} {\bibinfo  {journal}
  {Nature}\ }\textbf {\bibinfo {volume} {622}},\ \bibinfo {pages} {63}
  (\bibinfo {year} {2023})}\BibitemShut {NoStop}%
\bibitem [{\citenamefont {Zeng}\ \emph {et~al.}(2023)\citenamefont {Zeng},
  \citenamefont {Xia}, \citenamefont {Kang}, \citenamefont {Zhu}, \citenamefont
  {Kn{\"u}ppel}, \citenamefont {Vaswani}, \citenamefont {Watanabe},
  \citenamefont {Taniguchi}, \citenamefont {Mak},\ and\ \citenamefont
  {Shan}}]{FCIMoTe2ShanJie2023}%
  \BibitemOpen
  \bibfield  {author} {\bibinfo {author} {\bibfnamefont {Y.}~\bibnamefont
  {Zeng}}, \bibinfo {author} {\bibfnamefont {Z.}~\bibnamefont {Xia}}, \bibinfo
  {author} {\bibfnamefont {K.}~\bibnamefont {Kang}}, \bibinfo {author}
  {\bibfnamefont {J.}~\bibnamefont {Zhu}}, \bibinfo {author} {\bibfnamefont
  {P.}~\bibnamefont {Kn{\"u}ppel}}, \bibinfo {author} {\bibfnamefont
  {C.}~\bibnamefont {Vaswani}}, \bibinfo {author} {\bibfnamefont
  {K.}~\bibnamefont {Watanabe}}, \bibinfo {author} {\bibfnamefont
  {T.}~\bibnamefont {Taniguchi}}, \bibinfo {author} {\bibfnamefont {K.~F.}\
  \bibnamefont {Mak}}, \ and\ \bibinfo {author} {\bibfnamefont
  {J.}~\bibnamefont {Shan}},\ }\href {\doibase 10.1038/s41586-023-06452-3}
  {\bibfield  {journal} {\bibinfo  {journal} {Nature}\ }\textbf {\bibinfo
  {volume} {622}},\ \bibinfo {pages} {69} (\bibinfo {year} {2023})}\BibitemShut
  {NoStop}%
\bibitem [{\citenamefont {Park}\ \emph {et~al.}(2023)\citenamefont {Park},
  \citenamefont {Cai}, \citenamefont {Anderson}, \citenamefont {Zhang},
  \citenamefont {Zhu}, \citenamefont {Liu}, \citenamefont {Wang}, \citenamefont
  {Holtzmann}, \citenamefont {Hu}, \citenamefont {Liu}, \citenamefont
  {Taniguchi}, \citenamefont {Watanabe}, \citenamefont {Chu}, \citenamefont
  {Cao}, \citenamefont {Fu}, \citenamefont {Yao}, \citenamefont {Chang},
  \citenamefont {Cobden}, \citenamefont {Xiao},\ and\ \citenamefont
  {Xu}}]{MoTe2Park2023}%
  \BibitemOpen
  \bibfield  {author} {\bibinfo {author} {\bibfnamefont {H.}~\bibnamefont
  {Park}}, \bibinfo {author} {\bibfnamefont {J.}~\bibnamefont {Cai}}, \bibinfo
  {author} {\bibfnamefont {E.}~\bibnamefont {Anderson}}, \bibinfo {author}
  {\bibfnamefont {Y.}~\bibnamefont {Zhang}}, \bibinfo {author} {\bibfnamefont
  {J.}~\bibnamefont {Zhu}}, \bibinfo {author} {\bibfnamefont {X.}~\bibnamefont
  {Liu}}, \bibinfo {author} {\bibfnamefont {C.}~\bibnamefont {Wang}}, \bibinfo
  {author} {\bibfnamefont {W.}~\bibnamefont {Holtzmann}}, \bibinfo {author}
  {\bibfnamefont {C.}~\bibnamefont {Hu}}, \bibinfo {author} {\bibfnamefont
  {Z.}~\bibnamefont {Liu}}, \bibinfo {author} {\bibfnamefont {T.}~\bibnamefont
  {Taniguchi}}, \bibinfo {author} {\bibfnamefont {K.}~\bibnamefont {Watanabe}},
  \bibinfo {author} {\bibfnamefont {J.-H.}\ \bibnamefont {Chu}}, \bibinfo
  {author} {\bibfnamefont {T.}~\bibnamefont {Cao}}, \bibinfo {author}
  {\bibfnamefont {L.}~\bibnamefont {Fu}}, \bibinfo {author} {\bibfnamefont
  {W.}~\bibnamefont {Yao}}, \bibinfo {author} {\bibfnamefont {C.-Z.}\
  \bibnamefont {Chang}}, \bibinfo {author} {\bibfnamefont {D.}~\bibnamefont
  {Cobden}}, \bibinfo {author} {\bibfnamefont {D.}~\bibnamefont {Xiao}}, \ and\
  \bibinfo {author} {\bibfnamefont {X.}~\bibnamefont {Xu}},\ }\href {\doibase
  10.1038/s41586-023-06536-0} {\bibfield  {journal} {\bibinfo  {journal}
  {Nature}\ }\textbf {\bibinfo {volume} {622}},\ \bibinfo {pages} {74}
  (\bibinfo {year} {2023})}\BibitemShut {NoStop}%
\bibitem [{\citenamefont {Xu}\ \emph {et~al.}(2023)\citenamefont {Xu},
  \citenamefont {Sun}, \citenamefont {Jia}, \citenamefont {Liu}, \citenamefont
  {Xu}, \citenamefont {Li}, \citenamefont {Gu}, \citenamefont {Watanabe},
  \citenamefont {Taniguchi}, \citenamefont {Tong}, \citenamefont {Jia},
  \citenamefont {Shi}, \citenamefont {Jiang}, \citenamefont {Zhang},
  \citenamefont {Liu},\ and\ \citenamefont {Li}}]{FCIMoTe2PRX2023}%
  \BibitemOpen
  \bibfield  {author} {\bibinfo {author} {\bibfnamefont {F.}~\bibnamefont
  {Xu}}, \bibinfo {author} {\bibfnamefont {Z.}~\bibnamefont {Sun}}, \bibinfo
  {author} {\bibfnamefont {T.}~\bibnamefont {Jia}}, \bibinfo {author}
  {\bibfnamefont {C.}~\bibnamefont {Liu}}, \bibinfo {author} {\bibfnamefont
  {C.}~\bibnamefont {Xu}}, \bibinfo {author} {\bibfnamefont {C.}~\bibnamefont
  {Li}}, \bibinfo {author} {\bibfnamefont {Y.}~\bibnamefont {Gu}}, \bibinfo
  {author} {\bibfnamefont {K.}~\bibnamefont {Watanabe}}, \bibinfo {author}
  {\bibfnamefont {T.}~\bibnamefont {Taniguchi}}, \bibinfo {author}
  {\bibfnamefont {B.}~\bibnamefont {Tong}}, \bibinfo {author} {\bibfnamefont
  {J.}~\bibnamefont {Jia}}, \bibinfo {author} {\bibfnamefont {Z.}~\bibnamefont
  {Shi}}, \bibinfo {author} {\bibfnamefont {S.}~\bibnamefont {Jiang}}, \bibinfo
  {author} {\bibfnamefont {Y.}~\bibnamefont {Zhang}}, \bibinfo {author}
  {\bibfnamefont {X.}~\bibnamefont {Liu}}, \ and\ \bibinfo {author}
  {\bibfnamefont {T.}~\bibnamefont {Li}},\ }\href {\doibase
  10.1103/PhysRevX.13.031037} {\bibfield  {journal} {\bibinfo  {journal} {Phys.
  Rev. X}\ }\textbf {\bibinfo {volume} {13}},\ \bibinfo {pages} {031037}
  (\bibinfo {year} {2023})}\BibitemShut {NoStop}%
\bibitem [{\citenamefont {Ledwith}\ \emph {et~al.}(2020)\citenamefont
  {Ledwith}, \citenamefont {Tarnopolsky}, \citenamefont {Khalaf},\ and\
  \citenamefont {Vishwanath}}]{AshvinPRR2020}%
  \BibitemOpen
  \bibfield  {author} {\bibinfo {author} {\bibfnamefont {P.~J.}\ \bibnamefont
  {Ledwith}}, \bibinfo {author} {\bibfnamefont {G.}~\bibnamefont
  {Tarnopolsky}}, \bibinfo {author} {\bibfnamefont {E.}~\bibnamefont {Khalaf}},
  \ and\ \bibinfo {author} {\bibfnamefont {A.}~\bibnamefont {Vishwanath}},\
  }\href {\doibase 10.1103/PhysRevResearch.2.023237} {\bibfield  {journal}
  {\bibinfo  {journal} {Phys. Rev. Res.}\ }\textbf {\bibinfo {volume} {2}},\
  \bibinfo {pages} {023237} (\bibinfo {year} {2020})}\BibitemShut {NoStop}%
\bibitem [{\citenamefont {Wang}\ \emph {et~al.}(2021)\citenamefont {Wang},
  \citenamefont {Zheng}, \citenamefont {Millis},\ and\ \citenamefont
  {Cano}}]{WangJiePRR2021}%
  \BibitemOpen
  \bibfield  {author} {\bibinfo {author} {\bibfnamefont {J.}~\bibnamefont
  {Wang}}, \bibinfo {author} {\bibfnamefont {Y.}~\bibnamefont {Zheng}},
  \bibinfo {author} {\bibfnamefont {A.~J.}\ \bibnamefont {Millis}}, \ and\
  \bibinfo {author} {\bibfnamefont {J.}~\bibnamefont {Cano}},\ }\href {\doibase
  10.1103/PhysRevResearch.3.023155} {\bibfield  {journal} {\bibinfo  {journal}
  {Phys. Rev. Res.}\ }\textbf {\bibinfo {volume} {3}},\ \bibinfo {pages}
  {023155} (\bibinfo {year} {2021})}\BibitemShut {NoStop}%
\bibitem [{\citenamefont {Popov}\ and\ \citenamefont
  {Milekhin}(2021)}]{TBGLLPRB2021}%
  \BibitemOpen
  \bibfield  {author} {\bibinfo {author} {\bibfnamefont {F.~K.}\ \bibnamefont
  {Popov}}\ and\ \bibinfo {author} {\bibfnamefont {A.}~\bibnamefont
  {Milekhin}},\ }\href {\doibase 10.1103/PhysRevB.103.155150} {\bibfield
  {journal} {\bibinfo  {journal} {Phys. Rev. B}\ }\textbf {\bibinfo {volume}
  {103}},\ \bibinfo {pages} {155150} (\bibinfo {year} {2021})}\BibitemShut
  {NoStop}%
\bibitem [{\citenamefont {Junker}(2019)}]{SUSYBookGeorg}%
  \BibitemOpen
  \bibfield  {author} {\bibinfo {author} {\bibfnamefont {G.}~\bibnamefont
  {Junker}},\ }\href {\doibase 10.1088/2053-2563/aae6d5} {\emph {\bibinfo
  {title} {Supersymmetric Methods in Quantum, Statistical and Solid State
  Physics}}},\ 2053-2563\ (\bibinfo  {publisher} {IOP Publishing},\ \bibinfo
  {year} {2019})\BibitemShut {NoStop}%
\bibitem [{\citenamefont {Volkov}\ and\ \citenamefont
  {Akulov}(1973)}]{SUSYPhysLett1973}%
  \BibitemOpen
  \bibfield  {author} {\bibinfo {author} {\bibfnamefont {D.}~\bibnamefont
  {Volkov}}\ and\ \bibinfo {author} {\bibfnamefont {V.}~\bibnamefont
  {Akulov}},\ }\href {\doibase https://doi.org/10.1016/0370-2693(73)90490-5}
  {\bibfield  {journal} {\bibinfo  {journal} {Phys. Lett. B}\ }\textbf
  {\bibinfo {volume} {46}},\ \bibinfo {pages} {109} (\bibinfo {year}
  {1973})}\BibitemShut {NoStop}%
\bibitem [{\citenamefont {Wess}\ and\ \citenamefont
  {Zumino}(1974{\natexlab{a}})}]{SUSYNuclPhysB1974a}%
  \BibitemOpen
  \bibfield  {author} {\bibinfo {author} {\bibfnamefont {J.}~\bibnamefont
  {Wess}}\ and\ \bibinfo {author} {\bibfnamefont {B.}~\bibnamefont {Zumino}},\
  }\href {\doibase https://doi.org/10.1016/0550-3213(74)90355-1} {\bibfield
  {journal} {\bibinfo  {journal} {Nucl. Phys. B}\ }\textbf {\bibinfo {volume}
  {70}},\ \bibinfo {pages} {39} (\bibinfo {year}
  {1974}{\natexlab{a}})}\BibitemShut {NoStop}%
\bibitem [{\citenamefont {Wess}\ and\ \citenamefont
  {Zumino}(1974{\natexlab{b}})}]{SUSYNuclPhysB1974b}%
  \BibitemOpen
  \bibfield  {author} {\bibinfo {author} {\bibfnamefont {J.}~\bibnamefont
  {Wess}}\ and\ \bibinfo {author} {\bibfnamefont {B.}~\bibnamefont {Zumino}},\
  }\href {\doibase https://doi.org/10.1016/0550-3213(74)90112-6} {\bibfield
  {journal} {\bibinfo  {journal} {Nucl. Phys. B}\ }\textbf {\bibinfo {volume}
  {78}},\ \bibinfo {pages} {1} (\bibinfo {year}
  {1974}{\natexlab{b}})}\BibitemShut {NoStop}%
\bibitem [{\citenamefont {Haag}\ \emph {et~al.}(1975)\citenamefont {Haag},
  \citenamefont {Łopuszański},\ and\ \citenamefont
  {Sohnius}}]{SUSYNuclPhysB1975}%
  \BibitemOpen
  \bibfield  {author} {\bibinfo {author} {\bibfnamefont {R.}~\bibnamefont
  {Haag}}, \bibinfo {author} {\bibfnamefont {J.~T.}\ \bibnamefont
  {Łopuszański}}, \ and\ \bibinfo {author} {\bibfnamefont {M.}~\bibnamefont
  {Sohnius}},\ }\href {\doibase https://doi.org/10.1016/0550-3213(75)90279-5}
  {\bibfield  {journal} {\bibinfo  {journal} {Nucl. Phys. B}\ }\textbf
  {\bibinfo {volume} {88}},\ \bibinfo {pages} {257} (\bibinfo {year}
  {1975})}\BibitemShut {NoStop}%
\bibitem [{com({\natexlab{a}})}]{commentSUSYinSqrt}%
  \BibitemOpen
  \href@noop {} {} ({\natexlab{a}}),\ \bibinfo {note} {for example, after
  squaring the Hamiltonian under consideration. In many works, this connection
  to SUSY in fact was not pointed out explicitly.}\BibitemShut {Stop}%
\bibitem [{\citenamefont {Kane}\ and\ \citenamefont
  {Lubensky}(2014)}]{SUSYTopoMechNatPhys2014}%
  \BibitemOpen
  \bibfield  {author} {\bibinfo {author} {\bibfnamefont {C.~L.}\ \bibnamefont
  {Kane}}\ and\ \bibinfo {author} {\bibfnamefont {T.~C.}\ \bibnamefont
  {Lubensky}},\ }\href {\doibase 10.1038/nphys2835} {\bibfield  {journal}
  {\bibinfo  {journal} {Nat. Phys.}\ }\textbf {\bibinfo {volume} {10}},\
  \bibinfo {pages} {39} (\bibinfo {year} {2014})}\BibitemShut {NoStop}%
\bibitem [{\citenamefont {Attig}\ \emph {et~al.}(2019)\citenamefont {Attig},
  \citenamefont {Roychowdhury}, \citenamefont {Lawler},\ and\ \citenamefont
  {Trebst}}]{SUSYTopoMechPRR2019}%
  \BibitemOpen
  \bibfield  {author} {\bibinfo {author} {\bibfnamefont {J.}~\bibnamefont
  {Attig}}, \bibinfo {author} {\bibfnamefont {K.}~\bibnamefont {Roychowdhury}},
  \bibinfo {author} {\bibfnamefont {M.~J.}\ \bibnamefont {Lawler}}, \ and\
  \bibinfo {author} {\bibfnamefont {S.}~\bibnamefont {Trebst}},\ }\href
  {\doibase 10.1103/PhysRevResearch.1.032047} {\bibfield  {journal} {\bibinfo
  {journal} {Phys. Rev. Res.}\ }\textbf {\bibinfo {volume} {1}},\ \bibinfo
  {pages} {032047} (\bibinfo {year} {2019})}\BibitemShut {NoStop}%
\bibitem [{\citenamefont {Roychowdhury}\ \emph {et~al.}(2022)\citenamefont
  {Roychowdhury}, \citenamefont {Attig}, \citenamefont {Trebst},\ and\
  \citenamefont {Lawler}}]{SUSYLattice2022}%
  \BibitemOpen
  \bibfield  {author} {\bibinfo {author} {\bibfnamefont {K.}~\bibnamefont
  {Roychowdhury}}, \bibinfo {author} {\bibfnamefont {J.}~\bibnamefont {Attig}},
  \bibinfo {author} {\bibfnamefont {S.}~\bibnamefont {Trebst}}, \ and\ \bibinfo
  {author} {\bibfnamefont {M.~J.}\ \bibnamefont {Lawler}},\ }\href
  {https://arxiv.org/abs/2207.09475} {\bibfield  {journal} {\bibinfo  {journal}
  {arXiv:2207.09475}\ } (\bibinfo {year} {2022})}\BibitemShut {NoStop}%
\bibitem [{\citenamefont {Arkinstall}\ \emph {et~al.}(2017)\citenamefont
  {Arkinstall}, \citenamefont {Teimourpour}, \citenamefont {Feng},
  \citenamefont {El-Ganainy},\ and\ \citenamefont {Schomerus}}]{SqrtPRB2017}%
  \BibitemOpen
  \bibfield  {author} {\bibinfo {author} {\bibfnamefont {J.}~\bibnamefont
  {Arkinstall}}, \bibinfo {author} {\bibfnamefont {M.~H.}\ \bibnamefont
  {Teimourpour}}, \bibinfo {author} {\bibfnamefont {L.}~\bibnamefont {Feng}},
  \bibinfo {author} {\bibfnamefont {R.}~\bibnamefont {El-Ganainy}}, \ and\
  \bibinfo {author} {\bibfnamefont {H.}~\bibnamefont {Schomerus}},\ }\href
  {\doibase 10.1103/PhysRevB.95.165109} {\bibfield  {journal} {\bibinfo
  {journal} {Phys. Rev. B}\ }\textbf {\bibinfo {volume} {95}},\ \bibinfo
  {pages} {165109} (\bibinfo {year} {2017})}\BibitemShut {NoStop}%
\bibitem [{\citenamefont {Ezawa}(2020)}]{SqrtEzawaPRR2020}%
  \BibitemOpen
  \bibfield  {author} {\bibinfo {author} {\bibfnamefont {M.}~\bibnamefont
  {Ezawa}},\ }\href {\doibase 10.1103/PhysRevResearch.2.033397} {\bibfield
  {journal} {\bibinfo  {journal} {Phys. Rev. Res.}\ }\textbf {\bibinfo {volume}
  {2}},\ \bibinfo {pages} {033397} (\bibinfo {year} {2020})}\BibitemShut
  {NoStop}%
\bibitem [{\citenamefont {Mizoguchi}\ \emph {et~al.}(2020)\citenamefont
  {Mizoguchi}, \citenamefont {Kuno},\ and\ \citenamefont
  {Hatsugai}}]{SqrtHatsugaiPRA2020}%
  \BibitemOpen
  \bibfield  {author} {\bibinfo {author} {\bibfnamefont {T.}~\bibnamefont
  {Mizoguchi}}, \bibinfo {author} {\bibfnamefont {Y.}~\bibnamefont {Kuno}}, \
  and\ \bibinfo {author} {\bibfnamefont {Y.}~\bibnamefont {Hatsugai}},\ }\href
  {\doibase 10.1103/PhysRevA.102.033527} {\bibfield  {journal} {\bibinfo
  {journal} {Phys. Rev. A}\ }\textbf {\bibinfo {volume} {102}},\ \bibinfo
  {pages} {033527} (\bibinfo {year} {2020})}\BibitemShut {NoStop}%
\bibitem [{\citenamefont {Yoshida}\ \emph {et~al.}(2021)\citenamefont
  {Yoshida}, \citenamefont {Mizoguchi}, \citenamefont {Kuno},\ and\
  \citenamefont {Hatsugai}}]{SqrtHatsugaiPRB2021}%
  \BibitemOpen
  \bibfield  {author} {\bibinfo {author} {\bibfnamefont {T.}~\bibnamefont
  {Yoshida}}, \bibinfo {author} {\bibfnamefont {T.}~\bibnamefont {Mizoguchi}},
  \bibinfo {author} {\bibfnamefont {Y.}~\bibnamefont {Kuno}}, \ and\ \bibinfo
  {author} {\bibfnamefont {Y.}~\bibnamefont {Hatsugai}},\ }\href {\doibase
  10.1103/PhysRevB.103.235130} {\bibfield  {journal} {\bibinfo  {journal}
  {Phys. Rev. B}\ }\textbf {\bibinfo {volume} {103}},\ \bibinfo {pages}
  {235130} (\bibinfo {year} {2021})}\BibitemShut {NoStop}%
\bibitem [{com({\natexlab{b}})}]{commentZeroModeSUSYbreaking}%
  \BibitemOpen
  \href@noop {} {} ({\natexlab{b}}),\ \bibinfo {note} {the zero modes play the
  important role of characterizing whether SUSY is broken or unbroken: SUSY is
  unbroken (broken) if zero modes are present (absent). Such terminology is
  based on the viewpoint that symmetries are ``spontaneously broken'' if they
  do not leave the vacuum invariant. Here broken SUSY means the ground states
  are not annihilated by the supercharges.}\BibitemShut {Stop}%
\bibitem [{\citenamefont {Aharonov}\ and\ \citenamefont
  {Casher}(1979)}]{AharonovCasherPRA1979}%
  \BibitemOpen
  \bibfield  {author} {\bibinfo {author} {\bibfnamefont {Y.}~\bibnamefont
  {Aharonov}}\ and\ \bibinfo {author} {\bibfnamefont {A.}~\bibnamefont
  {Casher}},\ }\href {\doibase 10.1103/PhysRevA.19.2461} {\bibfield  {journal}
  {\bibinfo  {journal} {Phys. Rev. A}\ }\textbf {\bibinfo {volume} {19}},\
  \bibinfo {pages} {2461} (\bibinfo {year} {1979})}\BibitemShut {NoStop}%
\bibitem [{\citenamefont {Snyman}(2009)}]{SnymanPRB2009}%
  \BibitemOpen
  \bibfield  {author} {\bibinfo {author} {\bibfnamefont {I.}~\bibnamefont
  {Snyman}},\ }\href {\doibase 10.1103/PhysRevB.80.054303} {\bibfield
  {journal} {\bibinfo  {journal} {Phys. Rev. B}\ }\textbf {\bibinfo {volume}
  {80}},\ \bibinfo {pages} {054303} (\bibinfo {year} {2009})}\BibitemShut
  {NoStop}%
\bibitem [{\citenamefont {Phong}\ and\ \citenamefont
  {Mele}(2023)}]{PhongZeroMode}%
  \BibitemOpen
  \bibfield  {author} {\bibinfo {author} {\bibfnamefont {V.~T.}\ \bibnamefont
  {Phong}}\ and\ \bibinfo {author} {\bibfnamefont {E.~J.}\ \bibnamefont
  {Mele}},\ }\href {https://arxiv.org/abs/2310.05913} {\bibfield  {journal}
  {\bibinfo  {journal} {arXiv:2310.05913}\ } (\bibinfo {year}
  {2023})}\BibitemShut {NoStop}%
\bibitem [{sup()}]{supp}%
  \BibitemOpen
  \href@noop {} {}\bibinfo {note} {The Supplemental Material contains
  derivations of Eqs. (10), (11), and (13), energy bands for cases without
  SUSY, strain effects in 2D materials, relation of Berry connection of the
  SUSY Hamiltonian to its square root, results of Dirac fermions in a zero-flux
  gauge field, non-Abelian gauge fields in twisted bilayer graphene and
  strained bilayer graphene, and details of exact diagonalization.}\BibitemShut
  {Stop}%
\bibitem [{\citenamefont {Georgi}\ \emph {et~al.}(2017)\citenamefont {Georgi},
  \citenamefont {Nemes-Incze}, \citenamefont {Carrillo-Bastos}, \citenamefont
  {Faria}, \citenamefont {Viola~Kusminskiy}, \citenamefont {Zhai},
  \citenamefont {Schneider}, \citenamefont {Subramaniam}, \citenamefont
  {Mashoff}, \citenamefont {Freitag}, \citenamefont {Liebmann}, \citenamefont
  {Pratzer}, \citenamefont {Wirtz}, \citenamefont {Woods}, \citenamefont
  {Gorbachev}, \citenamefont {Cao}, \citenamefont {Novoselov}, \citenamefont
  {Sandler},\ and\ \citenamefont {Morgenstern}}]{AlexNanoLett2017}%
  \BibitemOpen
  \bibfield  {author} {\bibinfo {author} {\bibfnamefont {A.}~\bibnamefont
  {Georgi}}, \bibinfo {author} {\bibfnamefont {P.}~\bibnamefont {Nemes-Incze}},
  \bibinfo {author} {\bibfnamefont {R.}~\bibnamefont {Carrillo-Bastos}},
  \bibinfo {author} {\bibfnamefont {D.}~\bibnamefont {Faria}}, \bibinfo
  {author} {\bibfnamefont {S.}~\bibnamefont {Viola~Kusminskiy}}, \bibinfo
  {author} {\bibfnamefont {D.}~\bibnamefont {Zhai}}, \bibinfo {author}
  {\bibfnamefont {M.}~\bibnamefont {Schneider}}, \bibinfo {author}
  {\bibfnamefont {D.}~\bibnamefont {Subramaniam}}, \bibinfo {author}
  {\bibfnamefont {T.}~\bibnamefont {Mashoff}}, \bibinfo {author} {\bibfnamefont
  {N.~M.}\ \bibnamefont {Freitag}}, \bibinfo {author} {\bibfnamefont
  {M.}~\bibnamefont {Liebmann}}, \bibinfo {author} {\bibfnamefont
  {M.}~\bibnamefont {Pratzer}}, \bibinfo {author} {\bibfnamefont
  {L.}~\bibnamefont {Wirtz}}, \bibinfo {author} {\bibfnamefont {C.~R.}\
  \bibnamefont {Woods}}, \bibinfo {author} {\bibfnamefont {R.~V.}\ \bibnamefont
  {Gorbachev}}, \bibinfo {author} {\bibfnamefont {Y.}~\bibnamefont {Cao}},
  \bibinfo {author} {\bibfnamefont {K.~S.}\ \bibnamefont {Novoselov}}, \bibinfo
  {author} {\bibfnamefont {N.}~\bibnamefont {Sandler}}, \ and\ \bibinfo
  {author} {\bibfnamefont {M.}~\bibnamefont {Morgenstern}},\ }\href {\doibase
  10.1021/acs.nanolett.6b04870} {\bibfield  {journal} {\bibinfo  {journal}
  {Nano Lett.}\ }\textbf {\bibinfo {volume} {17}},\ \bibinfo {pages} {2240}
  (\bibinfo {year} {2017})}\BibitemShut {NoStop}%
\bibitem [{\citenamefont {Yu}\ \emph {et~al.}(2024{\natexlab{a}})\citenamefont
  {Yu}, \citenamefont {Ciccarino}, \citenamefont {Bianco}, \citenamefont
  {Errea}, \citenamefont {Narang},\ and\ \citenamefont
  {Bernevig}}]{YuJiabinPhonon}%
  \BibitemOpen
  \bibfield  {author} {\bibinfo {author} {\bibfnamefont {J.}~\bibnamefont
  {Yu}}, \bibinfo {author} {\bibfnamefont {C.~J.}\ \bibnamefont {Ciccarino}},
  \bibinfo {author} {\bibfnamefont {R.}~\bibnamefont {Bianco}}, \bibinfo
  {author} {\bibfnamefont {I.}~\bibnamefont {Errea}}, \bibinfo {author}
  {\bibfnamefont {P.}~\bibnamefont {Narang}}, \ and\ \bibinfo {author}
  {\bibfnamefont {B.~A.}\ \bibnamefont {Bernevig}},\ }\href {\doibase
  10.1038/s41567-024-02486-0} {\bibfield  {journal} {\bibinfo  {journal} {Nat.
  Phys.}\ }\textbf {\bibinfo {volume} {20}},\ \bibinfo {pages} {1262} (\bibinfo
  {year} {2024}{\natexlab{a}})}\BibitemShut {NoStop}%
\bibitem [{\citenamefont {Dong}\ \emph {et~al.}(2022)\citenamefont {Dong},
  \citenamefont {Wang},\ and\ \citenamefont
  {Fu}}]{PeriodicFieldDiracLiangFu2022}%
  \BibitemOpen
  \bibfield  {author} {\bibinfo {author} {\bibfnamefont {J.}~\bibnamefont
  {Dong}}, \bibinfo {author} {\bibfnamefont {J.}~\bibnamefont {Wang}}, \ and\
  \bibinfo {author} {\bibfnamefont {L.}~\bibnamefont {Fu}},\ }\href
  {https://arxiv.org/abs/2208.10516} {\bibfield  {journal} {\bibinfo  {journal}
  {arXiv:2208.10516}\ } (\bibinfo {year} {2022})}\BibitemShut {NoStop}%
\bibitem [{\citenamefont {Aivazian}\ \emph {et~al.}(2015)\citenamefont
  {Aivazian}, \citenamefont {Gong}, \citenamefont {Jones}, \citenamefont {Chu},
  \citenamefont {Yan}, \citenamefont {Mandrus}, \citenamefont {Zhang},
  \citenamefont {Cobden}, \citenamefont {Yao},\ and\ \citenamefont
  {Xu}}]{VariousZeemanNatPhys2015}%
  \BibitemOpen
  \bibfield  {author} {\bibinfo {author} {\bibfnamefont {G.}~\bibnamefont
  {Aivazian}}, \bibinfo {author} {\bibfnamefont {Z.}~\bibnamefont {Gong}},
  \bibinfo {author} {\bibfnamefont {A.~M.}\ \bibnamefont {Jones}}, \bibinfo
  {author} {\bibfnamefont {R.-L.}\ \bibnamefont {Chu}}, \bibinfo {author}
  {\bibfnamefont {J.}~\bibnamefont {Yan}}, \bibinfo {author} {\bibfnamefont
  {D.~G.}\ \bibnamefont {Mandrus}}, \bibinfo {author} {\bibfnamefont
  {C.}~\bibnamefont {Zhang}}, \bibinfo {author} {\bibfnamefont
  {D.}~\bibnamefont {Cobden}}, \bibinfo {author} {\bibfnamefont
  {W.}~\bibnamefont {Yao}}, \ and\ \bibinfo {author} {\bibfnamefont
  {X.}~\bibnamefont {Xu}},\ }\href {\doibase 10.1038/nphys3201} {\bibfield
  {journal} {\bibinfo  {journal} {Nat. Phys.}\ }\textbf {\bibinfo {volume}
  {11}},\ \bibinfo {pages} {148} (\bibinfo {year} {2015})}\BibitemShut
  {NoStop}%
\bibitem [{\citenamefont {Vozmediano}\ \emph {et~al.}(2010)\citenamefont
  {Vozmediano}, \citenamefont {Katsnelson},\ and\ \citenamefont
  {Guinea}}]{StrainReviewPhysRep2010}%
  \BibitemOpen
  \bibfield  {author} {\bibinfo {author} {\bibfnamefont {M.}~\bibnamefont
  {Vozmediano}}, \bibinfo {author} {\bibfnamefont {M.}~\bibnamefont
  {Katsnelson}}, \ and\ \bibinfo {author} {\bibfnamefont {F.}~\bibnamefont
  {Guinea}},\ }\href {\doibase https://doi.org/10.1016/j.physrep.2010.07.003}
  {\bibfield  {journal} {\bibinfo  {journal} {Phys. Rep.}\ }\textbf {\bibinfo
  {volume} {496}},\ \bibinfo {pages} {109} (\bibinfo {year}
  {2010})}\BibitemShut {NoStop}%
\bibitem [{\citenamefont {Amorim}\ \emph {et~al.}(2016)\citenamefont {Amorim},
  \citenamefont {Cortijo}, \citenamefont {{de Juan}}, \citenamefont {Grushin},
  \citenamefont {Guinea}, \citenamefont {Gutiérrez-Rubio}, \citenamefont
  {Ochoa}, \citenamefont {Parente}, \citenamefont {Roldán}, \citenamefont
  {San-Jose}, \citenamefont {Schiefele}, \citenamefont {Sturla},\ and\
  \citenamefont {Vozmediano}}]{StrainReviewPhysRep2016}%
  \BibitemOpen
  \bibfield  {author} {\bibinfo {author} {\bibfnamefont {B.}~\bibnamefont
  {Amorim}}, \bibinfo {author} {\bibfnamefont {A.}~\bibnamefont {Cortijo}},
  \bibinfo {author} {\bibfnamefont {F.}~\bibnamefont {{de Juan}}}, \bibinfo
  {author} {\bibfnamefont {A.}~\bibnamefont {Grushin}}, \bibinfo {author}
  {\bibfnamefont {F.}~\bibnamefont {Guinea}}, \bibinfo {author} {\bibfnamefont
  {A.}~\bibnamefont {Gutiérrez-Rubio}}, \bibinfo {author} {\bibfnamefont
  {H.}~\bibnamefont {Ochoa}}, \bibinfo {author} {\bibfnamefont
  {V.}~\bibnamefont {Parente}}, \bibinfo {author} {\bibfnamefont
  {R.}~\bibnamefont {Roldán}}, \bibinfo {author} {\bibfnamefont
  {P.}~\bibnamefont {San-Jose}}, \bibinfo {author} {\bibfnamefont
  {J.}~\bibnamefont {Schiefele}}, \bibinfo {author} {\bibfnamefont
  {M.}~\bibnamefont {Sturla}}, \ and\ \bibinfo {author} {\bibfnamefont
  {M.}~\bibnamefont {Vozmediano}},\ }\href {\doibase
  https://doi.org/10.1016/j.physrep.2015.12.006} {\bibfield  {journal}
  {\bibinfo  {journal} {Phys. Rep.}\ }\textbf {\bibinfo {volume} {617}},\
  \bibinfo {pages} {1} (\bibinfo {year} {2016})}\BibitemShut {NoStop}%
\bibitem [{\citenamefont {Naumis}\ \emph {et~al.}(2017)\citenamefont {Naumis},
  \citenamefont {Barraza-Lopez}, \citenamefont {Oliva-Leyva},\ and\
  \citenamefont {Terrones}}]{StrainReviewRepProgPhys2017}%
  \BibitemOpen
  \bibfield  {author} {\bibinfo {author} {\bibfnamefont {G.~G.}\ \bibnamefont
  {Naumis}}, \bibinfo {author} {\bibfnamefont {S.}~\bibnamefont
  {Barraza-Lopez}}, \bibinfo {author} {\bibfnamefont {M.}~\bibnamefont
  {Oliva-Leyva}}, \ and\ \bibinfo {author} {\bibfnamefont {H.}~\bibnamefont
  {Terrones}},\ }\href {\doibase 10.1088/1361-6633/aa74ef} {\bibfield
  {journal} {\bibinfo  {journal} {Rep. Prog. Phys.}\ }\textbf {\bibinfo
  {volume} {80}},\ \bibinfo {pages} {096501} (\bibinfo {year}
  {2017})}\BibitemShut {NoStop}%
\bibitem [{\citenamefont {Xiao}\ \emph {et~al.}(2007)\citenamefont {Xiao},
  \citenamefont {Yao},\ and\ \citenamefont
  {Niu}}]{DiXiaoValleyContrastingPRL2007}%
  \BibitemOpen
  \bibfield  {author} {\bibinfo {author} {\bibfnamefont {D.}~\bibnamefont
  {Xiao}}, \bibinfo {author} {\bibfnamefont {W.}~\bibnamefont {Yao}}, \ and\
  \bibinfo {author} {\bibfnamefont {Q.}~\bibnamefont {Niu}},\ }\href {\doibase
  10.1103/PhysRevLett.99.236809} {\bibfield  {journal} {\bibinfo  {journal}
  {Phys. Rev. Lett.}\ }\textbf {\bibinfo {volume} {99}},\ \bibinfo {pages}
  {236809} (\bibinfo {year} {2007})}\BibitemShut {NoStop}%
\bibitem [{\citenamefont {Xiao}\ \emph {et~al.}(2012)\citenamefont {Xiao},
  \citenamefont {Liu}, \citenamefont {Feng}, \citenamefont {Xu},\ and\
  \citenamefont {Yao}}]{DiXiaoTMDsPRL2012}%
  \BibitemOpen
  \bibfield  {author} {\bibinfo {author} {\bibfnamefont {D.}~\bibnamefont
  {Xiao}}, \bibinfo {author} {\bibfnamefont {G.-B.}\ \bibnamefont {Liu}},
  \bibinfo {author} {\bibfnamefont {W.}~\bibnamefont {Feng}}, \bibinfo {author}
  {\bibfnamefont {X.}~\bibnamefont {Xu}}, \ and\ \bibinfo {author}
  {\bibfnamefont {W.}~\bibnamefont {Yao}},\ }\href {\doibase
  10.1103/PhysRevLett.108.196802} {\bibfield  {journal} {\bibinfo  {journal}
  {Phys. Rev. Lett.}\ }\textbf {\bibinfo {volume} {108}},\ \bibinfo {pages}
  {196802} (\bibinfo {year} {2012})}\BibitemShut {NoStop}%
\bibitem [{\citenamefont {Guinea}\ \emph {et~al.}(2010)\citenamefont {Guinea},
  \citenamefont {Katsnelson},\ and\ \citenamefont
  {Geim}}]{QHEStrainGrapheneNatPhys2010}%
  \BibitemOpen
  \bibfield  {author} {\bibinfo {author} {\bibfnamefont {F.}~\bibnamefont
  {Guinea}}, \bibinfo {author} {\bibfnamefont {M.~I.}\ \bibnamefont
  {Katsnelson}}, \ and\ \bibinfo {author} {\bibfnamefont {A.~K.}\ \bibnamefont
  {Geim}},\ }\href {\doibase 10.1038/nphys1420} {\bibfield  {journal} {\bibinfo
   {journal} {Nat. Phys.}\ }\textbf {\bibinfo {volume} {6}},\ \bibinfo {pages}
  {30} (\bibinfo {year} {2010})}\BibitemShut {NoStop}%
\bibitem [{\citenamefont {Tomori}\ \emph {et~al.}(2011)\citenamefont {Tomori},
  \citenamefont {Kanda}, \citenamefont {Goto}, \citenamefont {Ootuka},
  \citenamefont {Tsukagoshi}, \citenamefont {Moriyama}, \citenamefont
  {Watanabe},\ and\ \citenamefont {Tsuya}}]{SubstrateApplPhysExpress2011}%
  \BibitemOpen
  \bibfield  {author} {\bibinfo {author} {\bibfnamefont {H.}~\bibnamefont
  {Tomori}}, \bibinfo {author} {\bibfnamefont {A.}~\bibnamefont {Kanda}},
  \bibinfo {author} {\bibfnamefont {H.}~\bibnamefont {Goto}}, \bibinfo {author}
  {\bibfnamefont {Y.}~\bibnamefont {Ootuka}}, \bibinfo {author} {\bibfnamefont
  {K.}~\bibnamefont {Tsukagoshi}}, \bibinfo {author} {\bibfnamefont
  {S.}~\bibnamefont {Moriyama}}, \bibinfo {author} {\bibfnamefont
  {E.}~\bibnamefont {Watanabe}}, \ and\ \bibinfo {author} {\bibfnamefont
  {D.}~\bibnamefont {Tsuya}},\ }\href {\doibase 10.1143/apex.4.075102}
  {\bibfield  {journal} {\bibinfo  {journal} {Appl. Phys. Express}\ }\textbf
  {\bibinfo {volume} {4}},\ \bibinfo {pages} {075102} (\bibinfo {year}
  {2011})}\BibitemShut {NoStop}%
\bibitem [{\citenamefont {Reserbat-Plantey}\ \emph {et~al.}(2014)\citenamefont
  {Reserbat-Plantey}, \citenamefont {Kalita}, \citenamefont {Han},
  \citenamefont {Ferlazzo}, \citenamefont {Autier-Laurent}, \citenamefont
  {Komatsu}, \citenamefont {Li}, \citenamefont {Weil}, \citenamefont {Ralko},
  \citenamefont {Marty}, \citenamefont {Guaron}, \citenamefont {Bendiab},
  \citenamefont {Bouchiat},\ and\ \citenamefont {Bouchiat}}]{SubstrateNL2014}%
  \BibitemOpen
  \bibfield  {author} {\bibinfo {author} {\bibfnamefont {A.}~\bibnamefont
  {Reserbat-Plantey}}, \bibinfo {author} {\bibfnamefont {D.}~\bibnamefont
  {Kalita}}, \bibinfo {author} {\bibfnamefont {Z.}~\bibnamefont {Han}},
  \bibinfo {author} {\bibfnamefont {L.}~\bibnamefont {Ferlazzo}}, \bibinfo
  {author} {\bibfnamefont {S.}~\bibnamefont {Autier-Laurent}}, \bibinfo
  {author} {\bibfnamefont {K.}~\bibnamefont {Komatsu}}, \bibinfo {author}
  {\bibfnamefont {C.}~\bibnamefont {Li}}, \bibinfo {author} {\bibfnamefont
  {R.}~\bibnamefont {Weil}}, \bibinfo {author} {\bibfnamefont {A.}~\bibnamefont
  {Ralko}}, \bibinfo {author} {\bibfnamefont {L.}~\bibnamefont {Marty}},
  \bibinfo {author} {\bibfnamefont {S.}~\bibnamefont {Guaron}}, \bibinfo
  {author} {\bibfnamefont {N.}~\bibnamefont {Bendiab}}, \bibinfo {author}
  {\bibfnamefont {H.}~\bibnamefont {Bouchiat}}, \ and\ \bibinfo {author}
  {\bibfnamefont {V.}~\bibnamefont {Bouchiat}},\ }\href {\doibase
  10.1021/nl5016552} {\bibfield  {journal} {\bibinfo  {journal} {Nano Lett.}\
  }\textbf {\bibinfo {volume} {14}},\ \bibinfo {pages} {5044} (\bibinfo {year}
  {2014})}\BibitemShut {NoStop}%
\bibitem [{\citenamefont {Jiang}\ \emph {et~al.}(2017)\citenamefont {Jiang},
  \citenamefont {Mao}, \citenamefont {Duan}, \citenamefont {Lai}, \citenamefont
  {Watanabe}, \citenamefont {Taniguchi},\ and\ \citenamefont
  {Andrei}}]{YuhangNL2017}%
  \BibitemOpen
  \bibfield  {author} {\bibinfo {author} {\bibfnamefont {Y.}~\bibnamefont
  {Jiang}}, \bibinfo {author} {\bibfnamefont {J.}~\bibnamefont {Mao}}, \bibinfo
  {author} {\bibfnamefont {J.}~\bibnamefont {Duan}}, \bibinfo {author}
  {\bibfnamefont {X.}~\bibnamefont {Lai}}, \bibinfo {author} {\bibfnamefont
  {K.}~\bibnamefont {Watanabe}}, \bibinfo {author} {\bibfnamefont
  {T.}~\bibnamefont {Taniguchi}}, \ and\ \bibinfo {author} {\bibfnamefont
  {E.~Y.}\ \bibnamefont {Andrei}},\ }\href {\doibase
  10.1021/acs.nanolett.6b05228} {\bibfield  {journal} {\bibinfo  {journal}
  {Nano Lett.}\ }\textbf {\bibinfo {volume} {17}},\ \bibinfo {pages} {2839}
  (\bibinfo {year} {2017})}\BibitemShut {NoStop}%
\bibitem [{\citenamefont {Zhang}\ \emph
  {et~al.}(2018{\natexlab{a}})\citenamefont {Zhang}, \citenamefont {Kim},
  \citenamefont {Gilbert},\ and\ \citenamefont {Mason}}]{Nadya2018}%
  \BibitemOpen
  \bibfield  {author} {\bibinfo {author} {\bibfnamefont {Y.}~\bibnamefont
  {Zhang}}, \bibinfo {author} {\bibfnamefont {Y.}~\bibnamefont {Kim}}, \bibinfo
  {author} {\bibfnamefont {M.~J.}\ \bibnamefont {Gilbert}}, \ and\ \bibinfo
  {author} {\bibfnamefont {N.}~\bibnamefont {Mason}},\ }\href {\doibase
  10.1038/s41699-018-0076-0} {\bibfield  {journal} {\bibinfo  {journal} {npj 2D
  Mater. Appl.}\ }\textbf {\bibinfo {volume} {2}},\ \bibinfo {pages} {31}
  (\bibinfo {year} {2018}{\natexlab{a}})}\BibitemShut {NoStop}%
\bibitem [{\citenamefont {Phong}\ and\ \citenamefont
  {Mele}(2022)}]{PhongPRL2022}%
  \BibitemOpen
  \bibfield  {author} {\bibinfo {author} {\bibfnamefont {V.~T.}\ \bibnamefont
  {Phong}}\ and\ \bibinfo {author} {\bibfnamefont {E.~J.}\ \bibnamefont
  {Mele}},\ }\href {\doibase 10.1103/PhysRevLett.128.176406} {\bibfield
  {journal} {\bibinfo  {journal} {Phys. Rev. Lett.}\ }\textbf {\bibinfo
  {volume} {128}},\ \bibinfo {pages} {176406} (\bibinfo {year}
  {2022})}\BibitemShut {NoStop}%
\bibitem [{\citenamefont {Mahmud}\ \emph {et~al.}(2023)\citenamefont {Mahmud},
  \citenamefont {Zhai},\ and\ \citenamefont {Sandler}}]{ZhaiNanoLett2023}%
  \BibitemOpen
  \bibfield  {author} {\bibinfo {author} {\bibfnamefont {M.~T.}\ \bibnamefont
  {Mahmud}}, \bibinfo {author} {\bibfnamefont {D.}~\bibnamefont {Zhai}}, \ and\
  \bibinfo {author} {\bibfnamefont {N.}~\bibnamefont {Sandler}},\ }\href
  {\doibase 10.1021/acs.nanolett.3c02513} {\bibfield  {journal} {\bibinfo
  {journal} {Nano Lett.}\ }\textbf {\bibinfo {volume} {23}},\ \bibinfo {pages}
  {7725} (\bibinfo {year} {2023})}\BibitemShut {NoStop}%
\bibitem [{\citenamefont {Mao}\ \emph {et~al.}(2020)\citenamefont {Mao},
  \citenamefont {Milovanovi{\'{c}}}, \citenamefont {Andelkovi{\'{c}}},
  \citenamefont {Lai}, \citenamefont {Cao}, \citenamefont {Watanabe},
  \citenamefont {Taniguchi}, \citenamefont {Covaci}, \citenamefont {Peeters},
  \citenamefont {Geim}, \citenamefont {Jiang},\ and\ \citenamefont
  {Andrei}}]{JinhaiNature2020}%
  \BibitemOpen
  \bibfield  {author} {\bibinfo {author} {\bibfnamefont {J.}~\bibnamefont
  {Mao}}, \bibinfo {author} {\bibfnamefont {S.~P.}\ \bibnamefont
  {Milovanovi{\'{c}}}}, \bibinfo {author} {\bibfnamefont {M.}~\bibnamefont
  {Andelkovi{\'{c}}}}, \bibinfo {author} {\bibfnamefont {X.}~\bibnamefont
  {Lai}}, \bibinfo {author} {\bibfnamefont {Y.}~\bibnamefont {Cao}}, \bibinfo
  {author} {\bibfnamefont {K.}~\bibnamefont {Watanabe}}, \bibinfo {author}
  {\bibfnamefont {T.}~\bibnamefont {Taniguchi}}, \bibinfo {author}
  {\bibfnamefont {L.}~\bibnamefont {Covaci}}, \bibinfo {author} {\bibfnamefont
  {F.~M.}\ \bibnamefont {Peeters}}, \bibinfo {author} {\bibfnamefont {A.~K.}\
  \bibnamefont {Geim}}, \bibinfo {author} {\bibfnamefont {Y.}~\bibnamefont
  {Jiang}}, \ and\ \bibinfo {author} {\bibfnamefont {E.~Y.}\ \bibnamefont
  {Andrei}},\ }\href {\doibase 10.1038/s41586-020-2567-3} {\bibfield  {journal}
  {\bibinfo  {journal} {Nature}\ }\textbf {\bibinfo {volume} {584}},\ \bibinfo
  {pages} {215} (\bibinfo {year} {2020})}\BibitemShut {NoStop}%
\bibitem [{\citenamefont {Enaldiev}\ \emph {et~al.}(2020)\citenamefont
  {Enaldiev}, \citenamefont {Z\'olyomi}, \citenamefont {Yelgel}, \citenamefont
  {Magorrian},\ and\ \citenamefont {Fal'ko}}]{RelaxationTMDsFalkoPRL2020}%
  \BibitemOpen
  \bibfield  {author} {\bibinfo {author} {\bibfnamefont {V.~V.}\ \bibnamefont
  {Enaldiev}}, \bibinfo {author} {\bibfnamefont {V.}~\bibnamefont {Z\'olyomi}},
  \bibinfo {author} {\bibfnamefont {C.}~\bibnamefont {Yelgel}}, \bibinfo
  {author} {\bibfnamefont {S.~J.}\ \bibnamefont {Magorrian}}, \ and\ \bibinfo
  {author} {\bibfnamefont {V.~I.}\ \bibnamefont {Fal'ko}},\ }\href {\doibase
  10.1103/PhysRevLett.124.206101} {\bibfield  {journal} {\bibinfo  {journal}
  {Phys. Rev. Lett.}\ }\textbf {\bibinfo {volume} {124}},\ \bibinfo {pages}
  {206101} (\bibinfo {year} {2020})}\BibitemShut {NoStop}%
\bibitem [{\citenamefont {Xie}\ \emph {et~al.}(2022)\citenamefont {Xie},
  \citenamefont {Zhang}, \citenamefont {Hu}, \citenamefont {Mak},\ and\
  \citenamefont {Law}}]{KTLawQAHMoirePRL2022}%
  \BibitemOpen
  \bibfield  {author} {\bibinfo {author} {\bibfnamefont {Y.-M.}\ \bibnamefont
  {Xie}}, \bibinfo {author} {\bibfnamefont {C.-P.}\ \bibnamefont {Zhang}},
  \bibinfo {author} {\bibfnamefont {J.-X.}\ \bibnamefont {Hu}}, \bibinfo
  {author} {\bibfnamefont {K.~F.}\ \bibnamefont {Mak}}, \ and\ \bibinfo
  {author} {\bibfnamefont {K.~T.}\ \bibnamefont {Law}},\ }\href {\doibase
  10.1103/PhysRevLett.128.026402} {\bibfield  {journal} {\bibinfo  {journal}
  {Phys. Rev. Lett.}\ }\textbf {\bibinfo {volume} {128}},\ \bibinfo {pages}
  {026402} (\bibinfo {year} {2022})}\BibitemShut {NoStop}%
\bibitem [{com({\natexlab{c}})}]{commentScalarPotentialLatticeRelaxation}%
  \BibitemOpen
  \href@noop {} {} ({\natexlab{c}}),\ \bibinfo {note} {in addition to the
  pseudo magnetic field, there is also a scalar potential in the moir\'e
  lattice. It is usually different from the Zeeman term discussed in Eq.~(9) in
  both magnitude and profile. Therefore, although pseudo magnetic field can
  emerge in the moir\'e of H-type bilayer TMDs, SUSY is usually
  absent}\BibitemShut {NoStop}%
\bibitem [{\citenamefont {Wehling}\ \emph {et~al.}(2008)\citenamefont
  {Wehling}, \citenamefont {Balatsky}, \citenamefont {Tsvelik}, \citenamefont
  {Katsnelson},\ and\ \citenamefont {Lichtenstein}}]{RippledGrapheneEPL2008}%
  \BibitemOpen
  \bibfield  {author} {\bibinfo {author} {\bibfnamefont {T.~O.}\ \bibnamefont
  {Wehling}}, \bibinfo {author} {\bibfnamefont {A.~V.}\ \bibnamefont
  {Balatsky}}, \bibinfo {author} {\bibfnamefont {A.~M.}\ \bibnamefont
  {Tsvelik}}, \bibinfo {author} {\bibfnamefont {M.~I.}\ \bibnamefont
  {Katsnelson}}, \ and\ \bibinfo {author} {\bibfnamefont {A.~I.}\ \bibnamefont
  {Lichtenstein}},\ }\href {\doibase 10.1209/0295-5075/84/17003} {\bibfield
  {journal} {\bibinfo  {journal} {EPL}\ }\textbf {\bibinfo {volume} {84}},\
  \bibinfo {pages} {17003} (\bibinfo {year} {2008})}\BibitemShut {NoStop}%
\bibitem [{\citenamefont {Jiang}\ \emph {et~al.}(2013)\citenamefont {Jiang},
  \citenamefont {Low}, \citenamefont {Chang}, \citenamefont {Katsnelson},\ and\
  \citenamefont {Guinea}}]{StrainDirectionPRL2013}%
  \BibitemOpen
  \bibfield  {author} {\bibinfo {author} {\bibfnamefont {Y.}~\bibnamefont
  {Jiang}}, \bibinfo {author} {\bibfnamefont {T.}~\bibnamefont {Low}}, \bibinfo
  {author} {\bibfnamefont {K.}~\bibnamefont {Chang}}, \bibinfo {author}
  {\bibfnamefont {M.~I.}\ \bibnamefont {Katsnelson}}, \ and\ \bibinfo {author}
  {\bibfnamefont {F.}~\bibnamefont {Guinea}},\ }\href {\doibase
  10.1103/PhysRevLett.110.046601} {\bibfield  {journal} {\bibinfo  {journal}
  {Phys. Rev. Lett.}\ }\textbf {\bibinfo {volume} {110}},\ \bibinfo {pages}
  {046601} (\bibinfo {year} {2013})}\BibitemShut {NoStop}%
\bibitem [{\citenamefont {Zhai}\ and\ \citenamefont
  {Sandler}(2018)}]{ZhaiPRB2018}%
  \BibitemOpen
  \bibfield  {author} {\bibinfo {author} {\bibfnamefont {D.}~\bibnamefont
  {Zhai}}\ and\ \bibinfo {author} {\bibfnamefont {N.}~\bibnamefont {Sandler}},\
  }\href {\doibase 10.1103/PhysRevB.98.165437} {\bibfield  {journal} {\bibinfo
  {journal} {Phys. Rev. B}\ }\textbf {\bibinfo {volume} {98}},\ \bibinfo
  {pages} {165437} (\bibinfo {year} {2018})}\BibitemShut {NoStop}%
\bibitem [{\citenamefont {Pezzini}\ \emph {et~al.}(2020)\citenamefont
  {Pezzini}, \citenamefont {Mišeikis}, \citenamefont {Piccinini},
  \citenamefont {Forti}, \citenamefont {Pace}, \citenamefont {Engelke},
  \citenamefont {Rossella}, \citenamefont {Watanabe}, \citenamefont
  {Taniguchi}, \citenamefont {Kim},\ and\ \citenamefont
  {Coletti}}]{30degTBGNanoLett2020}%
  \BibitemOpen
  \bibfield  {author} {\bibinfo {author} {\bibfnamefont {S.}~\bibnamefont
  {Pezzini}}, \bibinfo {author} {\bibfnamefont {V.}~\bibnamefont {Mišeikis}},
  \bibinfo {author} {\bibfnamefont {G.}~\bibnamefont {Piccinini}}, \bibinfo
  {author} {\bibfnamefont {S.}~\bibnamefont {Forti}}, \bibinfo {author}
  {\bibfnamefont {S.}~\bibnamefont {Pace}}, \bibinfo {author} {\bibfnamefont
  {R.}~\bibnamefont {Engelke}}, \bibinfo {author} {\bibfnamefont
  {F.}~\bibnamefont {Rossella}}, \bibinfo {author} {\bibfnamefont
  {K.}~\bibnamefont {Watanabe}}, \bibinfo {author} {\bibfnamefont
  {T.}~\bibnamefont {Taniguchi}}, \bibinfo {author} {\bibfnamefont
  {P.}~\bibnamefont {Kim}}, \ and\ \bibinfo {author} {\bibfnamefont
  {C.}~\bibnamefont {Coletti}},\ }\href {\doibase 10.1021/acs.nanolett.0c00172}
  {\bibfield  {journal} {\bibinfo  {journal} {Nano Lett.}\ }\textbf {\bibinfo
  {volume} {20}},\ \bibinfo {pages} {3313} (\bibinfo {year}
  {2020})}\BibitemShut {NoStop}%
\bibitem [{\citenamefont {Piccinini}\ \emph {et~al.}(2021)\citenamefont
  {Piccinini}, \citenamefont {Mi\ifmmode~\check{s}\else \v{s}\fi{}eikis},
  \citenamefont {Watanabe}, \citenamefont {Taniguchi}, \citenamefont
  {Coletti},\ and\ \citenamefont {Pezzini}}]{30degTBGExptPRB2021}%
  \BibitemOpen
  \bibfield  {author} {\bibinfo {author} {\bibfnamefont {G.}~\bibnamefont
  {Piccinini}}, \bibinfo {author} {\bibfnamefont {V.}~\bibnamefont
  {Mi\ifmmode~\check{s}\else \v{s}\fi{}eikis}}, \bibinfo {author}
  {\bibfnamefont {K.}~\bibnamefont {Watanabe}}, \bibinfo {author}
  {\bibfnamefont {T.}~\bibnamefont {Taniguchi}}, \bibinfo {author}
  {\bibfnamefont {C.}~\bibnamefont {Coletti}}, \ and\ \bibinfo {author}
  {\bibfnamefont {S.}~\bibnamefont {Pezzini}},\ }\href {\doibase
  10.1103/PhysRevB.104.L241410} {\bibfield  {journal} {\bibinfo  {journal}
  {Phys. Rev. B}\ }\textbf {\bibinfo {volume} {104}},\ \bibinfo {pages}
  {L241410} (\bibinfo {year} {2021})}\BibitemShut {NoStop}%
\bibitem [{\citenamefont {Li}\ \emph {et~al.}(2024)\citenamefont {Li},
  \citenamefont {Zhang}, \citenamefont {Ha}, \citenamefont {Lin}, \citenamefont
  {Dong}, \citenamefont {Gao}, \citenamefont {Liu}, \citenamefont {Liu},
  \citenamefont {Ryu}, \citenamefont {Kim}, \citenamefont {Jozwiak},
  \citenamefont {Bostwick}, \citenamefont {Watanabe}, \citenamefont
  {Taniguchi}, \citenamefont {Kousa}, \citenamefont {Li}, \citenamefont
  {Rotenberg}, \citenamefont {Khalaf}, \citenamefont {Robinson}, \citenamefont
  {Giustino},\ and\ \citenamefont {Shih}}]{30degWSe2ChihKang}%
  \BibitemOpen
  \bibfield  {author} {\bibinfo {author} {\bibfnamefont {Y.}~\bibnamefont
  {Li}}, \bibinfo {author} {\bibfnamefont {F.}~\bibnamefont {Zhang}}, \bibinfo
  {author} {\bibfnamefont {V.-A.}\ \bibnamefont {Ha}}, \bibinfo {author}
  {\bibfnamefont {Y.-C.}\ \bibnamefont {Lin}}, \bibinfo {author} {\bibfnamefont
  {C.}~\bibnamefont {Dong}}, \bibinfo {author} {\bibfnamefont {Q.}~\bibnamefont
  {Gao}}, \bibinfo {author} {\bibfnamefont {Z.}~\bibnamefont {Liu}}, \bibinfo
  {author} {\bibfnamefont {X.}~\bibnamefont {Liu}}, \bibinfo {author}
  {\bibfnamefont {S.~H.}\ \bibnamefont {Ryu}}, \bibinfo {author} {\bibfnamefont
  {H.}~\bibnamefont {Kim}}, \bibinfo {author} {\bibfnamefont {C.}~\bibnamefont
  {Jozwiak}}, \bibinfo {author} {\bibfnamefont {A.}~\bibnamefont {Bostwick}},
  \bibinfo {author} {\bibfnamefont {K.}~\bibnamefont {Watanabe}}, \bibinfo
  {author} {\bibfnamefont {T.}~\bibnamefont {Taniguchi}}, \bibinfo {author}
  {\bibfnamefont {B.}~\bibnamefont {Kousa}}, \bibinfo {author} {\bibfnamefont
  {X.}~\bibnamefont {Li}}, \bibinfo {author} {\bibfnamefont {E.}~\bibnamefont
  {Rotenberg}}, \bibinfo {author} {\bibfnamefont {E.}~\bibnamefont {Khalaf}},
  \bibinfo {author} {\bibfnamefont {J.~A.}\ \bibnamefont {Robinson}}, \bibinfo
  {author} {\bibfnamefont {F.}~\bibnamefont {Giustino}}, \ and\ \bibinfo
  {author} {\bibfnamefont {C.-K.}\ \bibnamefont {Shih}},\ }\href {\doibase
  10.1038/s41586-023-06904-w} {\bibfield  {journal} {\bibinfo  {journal}
  {Nature}\ }\textbf {\bibinfo {volume} {625}},\ \bibinfo {pages} {494}
  (\bibinfo {year} {2024})}\BibitemShut {NoStop}%
\bibitem [{com({\natexlab{d}})}]{commentDiracVsHole}%
  \BibitemOpen
  \href@noop {} {} ({\natexlab{d}}),\ \bibinfo {note} {to compare the results,
  we have flipped the sign of the valence band eneriges from the Dirac model
  and shift the lowest value to 0.}\BibitemShut {Stop}%
\bibitem [{\citenamefont {Zhang}\ \emph
  {et~al.}(2018{\natexlab{b}})\citenamefont {Zhang}, \citenamefont {Shan},\
  and\ \citenamefont {Xiao}}]{BilayerTMDsDiXiao2018}%
  \BibitemOpen
  \bibfield  {author} {\bibinfo {author} {\bibfnamefont {X.}~\bibnamefont
  {Zhang}}, \bibinfo {author} {\bibfnamefont {W.-Y.}\ \bibnamefont {Shan}}, \
  and\ \bibinfo {author} {\bibfnamefont {D.}~\bibnamefont {Xiao}},\ }\href
  {\doibase 10.1103/PhysRevLett.120.077401} {\bibfield  {journal} {\bibinfo
  {journal} {Phys. Rev. Lett.}\ }\textbf {\bibinfo {volume} {120}},\ \bibinfo
  {pages} {077401} (\bibinfo {year} {2018}{\natexlab{b}})}\BibitemShut
  {NoStop}%
\bibitem [{\citenamefont {Tong}\ \emph {et~al.}(2017)\citenamefont {Tong},
  \citenamefont {Yu}, \citenamefont {Zhu}, \citenamefont {Wang}, \citenamefont
  {Xu},\ and\ \citenamefont {Yao}}]{TongQingjunNatPhys}%
  \BibitemOpen
  \bibfield  {author} {\bibinfo {author} {\bibfnamefont {Q.}~\bibnamefont
  {Tong}}, \bibinfo {author} {\bibfnamefont {H.}~\bibnamefont {Yu}}, \bibinfo
  {author} {\bibfnamefont {Q.}~\bibnamefont {Zhu}}, \bibinfo {author}
  {\bibfnamefont {Y.}~\bibnamefont {Wang}}, \bibinfo {author} {\bibfnamefont
  {X.}~\bibnamefont {Xu}}, \ and\ \bibinfo {author} {\bibfnamefont
  {W.}~\bibnamefont {Yao}},\ }\href {\doibase 10.1038/nphys3968} {\bibfield
  {journal} {\bibinfo  {journal} {Nat. Phys.}\ }\textbf {\bibinfo {volume}
  {13}},\ \bibinfo {pages} {356} (\bibinfo {year} {2017})}\BibitemShut
  {NoStop}%
\bibitem [{\citenamefont {Liang}\ \emph {et~al.}(2022)\citenamefont {Liang},
  \citenamefont {Yang}, \citenamefont {Wu}, \citenamefont {Dadap},
  \citenamefont {Watanabe}, \citenamefont {Taniguchi},\ and\ \citenamefont
  {Ye}}]{BilayerTMDsInterlayerCouplingPRX2022}%
  \BibitemOpen
  \bibfield  {author} {\bibinfo {author} {\bibfnamefont {J.}~\bibnamefont
  {Liang}}, \bibinfo {author} {\bibfnamefont {D.}~\bibnamefont {Yang}},
  \bibinfo {author} {\bibfnamefont {J.}~\bibnamefont {Wu}}, \bibinfo {author}
  {\bibfnamefont {J.~I.}\ \bibnamefont {Dadap}}, \bibinfo {author}
  {\bibfnamefont {K.}~\bibnamefont {Watanabe}}, \bibinfo {author}
  {\bibfnamefont {T.}~\bibnamefont {Taniguchi}}, \ and\ \bibinfo {author}
  {\bibfnamefont {Z.}~\bibnamefont {Ye}},\ }\href {\doibase
  10.1103/PhysRevX.12.041005} {\bibfield  {journal} {\bibinfo  {journal} {Phys.
  Rev. X}\ }\textbf {\bibinfo {volume} {12}},\ \bibinfo {pages} {041005}
  (\bibinfo {year} {2022})}\BibitemShut {NoStop}%
\bibitem [{\citenamefont {McCann}\ and\ \citenamefont
  {Koshino}(2013)}]{BilayerGrapheneReviewMcCann}%
  \BibitemOpen
  \bibfield  {author} {\bibinfo {author} {\bibfnamefont {E.}~\bibnamefont
  {McCann}}\ and\ \bibinfo {author} {\bibfnamefont {M.}~\bibnamefont
  {Koshino}},\ }\href {\doibase 10.1088/0034-4885/76/5/056503} {\bibfield
  {journal} {\bibinfo  {journal} {Rep. Prog. Phys.}\ }\textbf {\bibinfo
  {volume} {76}},\ \bibinfo {pages} {056503} (\bibinfo {year}
  {2013})}\BibitemShut {NoStop}%
\bibitem [{\citenamefont {Milovanovi\ifmmode~\acute{c}\else \'{c}\fi{}}\ \emph
  {et~al.}(2020)\citenamefont {Milovanovi\ifmmode~\acute{c}\else \'{c}\fi{}},
  \citenamefont {An\dj{}elkovi\ifmmode~\acute{c}\else \'{c}\fi{}},
  \citenamefont {Covaci},\ and\ \citenamefont
  {Peeters}}]{PeetersPeriodicStrainPRB2020}%
  \BibitemOpen
  \bibfield  {author} {\bibinfo {author} {\bibfnamefont {S.~P.}\ \bibnamefont
  {Milovanovi\ifmmode~\acute{c}\else \'{c}\fi{}}}, \bibinfo {author}
  {\bibfnamefont {M.}~\bibnamefont {An\dj{}elkovi\ifmmode~\acute{c}\else
  \'{c}\fi{}}}, \bibinfo {author} {\bibfnamefont {L.}~\bibnamefont {Covaci}}, \
  and\ \bibinfo {author} {\bibfnamefont {F.~M.}\ \bibnamefont {Peeters}},\
  }\href {\doibase 10.1103/PhysRevB.102.245427} {\bibfield  {journal} {\bibinfo
   {journal} {Phys. Rev. B}\ }\textbf {\bibinfo {volume} {102}},\ \bibinfo
  {pages} {245427} (\bibinfo {year} {2020})}\BibitemShut {NoStop}%
\bibitem [{\citenamefont {Manesco}\ \emph {et~al.}(2021)\citenamefont
  {Manesco}, \citenamefont {Lado}, \citenamefont {Ribeiro}, \citenamefont
  {Weber},\ and\ \citenamefont {Jr}}]{Antonio2DMater2021a}%
  \BibitemOpen
  \bibfield  {author} {\bibinfo {author} {\bibfnamefont {A.~L.~R.}\
  \bibnamefont {Manesco}}, \bibinfo {author} {\bibfnamefont {J.~L.}\
  \bibnamefont {Lado}}, \bibinfo {author} {\bibfnamefont {E.~V.~S.}\
  \bibnamefont {Ribeiro}}, \bibinfo {author} {\bibfnamefont {G.}~\bibnamefont
  {Weber}}, \ and\ \bibinfo {author} {\bibfnamefont {D.~R.}\ \bibnamefont
  {Jr}},\ }\href {\doibase 10.1088/2053-1583/abbc5f} {\bibfield  {journal}
  {\bibinfo  {journal} {2D Mater.}\ }\textbf {\bibinfo {volume} {8}},\ \bibinfo
  {pages} {015011} (\bibinfo {year} {2021})}\BibitemShut {NoStop}%
\bibitem [{\citenamefont {Manesco}\ and\ \citenamefont
  {Lado}(2021)}]{Antonio2DMater2021b}%
  \BibitemOpen
  \bibfield  {author} {\bibinfo {author} {\bibfnamefont {A.~L.~R.}\
  \bibnamefont {Manesco}}\ and\ \bibinfo {author} {\bibfnamefont {J.~L.}\
  \bibnamefont {Lado}},\ }\href {\doibase 10.1088/2053-1583/ac0b48} {\bibfield
  {journal} {\bibinfo  {journal} {2D Mater.}\ }\textbf {\bibinfo {volume}
  {8}},\ \bibinfo {pages} {035057} (\bibinfo {year} {2021})}\BibitemShut
  {NoStop}%
\bibitem [{\citenamefont {Gao}\ \emph {et~al.}(2023)\citenamefont {Gao},
  \citenamefont {Dong}, \citenamefont {Ledwith}, \citenamefont {Parker},\ and\
  \citenamefont {Khalaf}}]{EslamPeriodicStrainPRL2023}%
  \BibitemOpen
  \bibfield  {author} {\bibinfo {author} {\bibfnamefont {Q.}~\bibnamefont
  {Gao}}, \bibinfo {author} {\bibfnamefont {J.}~\bibnamefont {Dong}}, \bibinfo
  {author} {\bibfnamefont {P.}~\bibnamefont {Ledwith}}, \bibinfo {author}
  {\bibfnamefont {D.}~\bibnamefont {Parker}}, \ and\ \bibinfo {author}
  {\bibfnamefont {E.}~\bibnamefont {Khalaf}},\ }\href {\doibase
  10.1103/PhysRevLett.131.096401} {\bibfield  {journal} {\bibinfo  {journal}
  {Phys. Rev. Lett.}\ }\textbf {\bibinfo {volume} {131}},\ \bibinfo {pages}
  {096401} (\bibinfo {year} {2023})}\BibitemShut {NoStop}%
\bibitem [{\citenamefont {Wen}\ and\ \citenamefont
  {Niu}(1990)}]{WenNiuPRB1990}%
  \BibitemOpen
  \bibfield  {author} {\bibinfo {author} {\bibfnamefont {X.~G.}\ \bibnamefont
  {Wen}}\ and\ \bibinfo {author} {\bibfnamefont {Q.}~\bibnamefont {Niu}},\
  }\href {\doibase 10.1103/PhysRevB.41.9377} {\bibfield  {journal} {\bibinfo
  {journal} {Phys. Rev. B}\ }\textbf {\bibinfo {volume} {41}},\ \bibinfo
  {pages} {9377} (\bibinfo {year} {1990})}\BibitemShut {NoStop}%
\bibitem [{\citenamefont {Reddy}\ \emph {et~al.}(2023)\citenamefont {Reddy},
  \citenamefont {Alsallom}, \citenamefont {Zhang}, \citenamefont {Devakul},\
  and\ \citenamefont {Fu}}]{FuLiangFCIPRB2023}%
  \BibitemOpen
  \bibfield  {author} {\bibinfo {author} {\bibfnamefont {A.~P.}\ \bibnamefont
  {Reddy}}, \bibinfo {author} {\bibfnamefont {F.}~\bibnamefont {Alsallom}},
  \bibinfo {author} {\bibfnamefont {Y.}~\bibnamefont {Zhang}}, \bibinfo
  {author} {\bibfnamefont {T.}~\bibnamefont {Devakul}}, \ and\ \bibinfo
  {author} {\bibfnamefont {L.}~\bibnamefont {Fu}},\ }\href {\doibase
  10.1103/PhysRevB.108.085117} {\bibfield  {journal} {\bibinfo  {journal}
  {Phys. Rev. B}\ }\textbf {\bibinfo {volume} {108}},\ \bibinfo {pages}
  {085117} (\bibinfo {year} {2023})}\BibitemShut {NoStop}%
\bibitem [{\citenamefont {Yu}\ \emph {et~al.}(2024{\natexlab{b}})\citenamefont
  {Yu}, \citenamefont {Herzog-Arbeitman}, \citenamefont {Wang}, \citenamefont
  {Vafek}, \citenamefont {Bernevig},\ and\ \citenamefont
  {Regnault}}]{YuJiabinMoTe22023}%
  \BibitemOpen
  \bibfield  {author} {\bibinfo {author} {\bibfnamefont {J.}~\bibnamefont
  {Yu}}, \bibinfo {author} {\bibfnamefont {J.}~\bibnamefont
  {Herzog-Arbeitman}}, \bibinfo {author} {\bibfnamefont {M.}~\bibnamefont
  {Wang}}, \bibinfo {author} {\bibfnamefont {O.}~\bibnamefont {Vafek}},
  \bibinfo {author} {\bibfnamefont {B.~A.}\ \bibnamefont {Bernevig}}, \ and\
  \bibinfo {author} {\bibfnamefont {N.}~\bibnamefont {Regnault}},\ }\href
  {\doibase 10.1103/PhysRevB.109.045147} {\bibfield  {journal} {\bibinfo
  {journal} {Phys. Rev. B}\ }\textbf {\bibinfo {volume} {109}},\ \bibinfo
  {pages} {045147} (\bibinfo {year} {2024}{\natexlab{b}})}\BibitemShut
  {NoStop}%
\bibitem [{\citenamefont {Lu}\ \emph {et~al.}(2024)\citenamefont {Lu},
  \citenamefont {Han}, \citenamefont {Yao}, \citenamefont {Reddy},
  \citenamefont {Yang}, \citenamefont {Seo}, \citenamefont {Watanabe},
  \citenamefont {Taniguchi}, \citenamefont {Fu},\ and\ \citenamefont
  {Ju}}]{JuLongFCI}%
  \BibitemOpen
  \bibfield  {author} {\bibinfo {author} {\bibfnamefont {Z.}~\bibnamefont
  {Lu}}, \bibinfo {author} {\bibfnamefont {T.}~\bibnamefont {Han}}, \bibinfo
  {author} {\bibfnamefont {Y.}~\bibnamefont {Yao}}, \bibinfo {author}
  {\bibfnamefont {A.~P.}\ \bibnamefont {Reddy}}, \bibinfo {author}
  {\bibfnamefont {J.}~\bibnamefont {Yang}}, \bibinfo {author} {\bibfnamefont
  {J.}~\bibnamefont {Seo}}, \bibinfo {author} {\bibfnamefont {K.}~\bibnamefont
  {Watanabe}}, \bibinfo {author} {\bibfnamefont {T.}~\bibnamefont {Taniguchi}},
  \bibinfo {author} {\bibfnamefont {L.}~\bibnamefont {Fu}}, \ and\ \bibinfo
  {author} {\bibfnamefont {L.}~\bibnamefont {Ju}},\ }\href
  {https://www.nature.com/articles/s41586-023-07010-7} {\bibfield  {journal}
  {\bibinfo  {journal} {Nature}\ }\textbf {\bibinfo {volume} {626}},\ \bibinfo
  {pages} {759} (\bibinfo {year} {2024})}\BibitemShut {NoStop}%
\bibitem [{\citenamefont {Wan}\ \emph {et~al.}(2023{\natexlab{a}})\citenamefont
  {Wan}, \citenamefont {Sarkar}, \citenamefont {Lin},\ and\ \citenamefont
  {Sun}}]{StrainQuadraticBandTouchingSunKaiPRL2023}%
  \BibitemOpen
  \bibfield  {author} {\bibinfo {author} {\bibfnamefont {X.}~\bibnamefont
  {Wan}}, \bibinfo {author} {\bibfnamefont {S.}~\bibnamefont {Sarkar}},
  \bibinfo {author} {\bibfnamefont {S.-Z.}\ \bibnamefont {Lin}}, \ and\
  \bibinfo {author} {\bibfnamefont {K.}~\bibnamefont {Sun}},\ }\href {\doibase
  10.1103/PhysRevLett.130.216401} {\bibfield  {journal} {\bibinfo  {journal}
  {Phys. Rev. Lett.}\ }\textbf {\bibinfo {volume} {130}},\ \bibinfo {pages}
  {216401} (\bibinfo {year} {2023}{\natexlab{a}})}\BibitemShut {NoStop}%
\bibitem [{com({\natexlab{e}})}]{commentMsquare}%
  \BibitemOpen
  \href@noop {} {} ({\natexlab{e}}),\ \bibinfo {note} {in general $M$ does not
  need to be a square matrix, here we assume it for simplicity.}\BibitemShut
  {Stop}%
\bibitem [{\citenamefont {Kailasvuori}(2009)}]{PedestrianEPL2009}%
  \BibitemOpen
  \bibfield  {author} {\bibinfo {author} {\bibfnamefont {J.}~\bibnamefont
  {Kailasvuori}},\ }\href {\doibase 10.1209/0295-5075/87/47008} {\bibfield
  {journal} {\bibinfo  {journal} {EPL}\ }\textbf {\bibinfo {volume} {87}},\
  \bibinfo {pages} {47008} (\bibinfo {year} {2009})}\BibitemShut {NoStop}%
\bibitem [{\citenamefont {Wan}\ \emph {et~al.}(2023{\natexlab{b}})\citenamefont
  {Wan}, \citenamefont {Sarkar}, \citenamefont {Sun},\ and\ \citenamefont
  {Lin}}]{PeriodicStrainSunKaiPRB2023}%
  \BibitemOpen
  \bibfield  {author} {\bibinfo {author} {\bibfnamefont {X.}~\bibnamefont
  {Wan}}, \bibinfo {author} {\bibfnamefont {S.}~\bibnamefont {Sarkar}},
  \bibinfo {author} {\bibfnamefont {K.}~\bibnamefont {Sun}}, \ and\ \bibinfo
  {author} {\bibfnamefont {S.-Z.}\ \bibnamefont {Lin}},\ }\href {\doibase
  10.1103/PhysRevB.108.125129} {\bibfield  {journal} {\bibinfo  {journal}
  {Phys. Rev. B}\ }\textbf {\bibinfo {volume} {108}},\ \bibinfo {pages}
  {125129} (\bibinfo {year} {2023}{\natexlab{b}})}\BibitemShut {NoStop}%
\bibitem [{\citenamefont {San-Jose}\ \emph {et~al.}(2012)\citenamefont
  {San-Jose}, \citenamefont {Gonz\'alez},\ and\ \citenamefont
  {Guinea}}]{TBGGaugeFieldPRL2012}%
  \BibitemOpen
  \bibfield  {author} {\bibinfo {author} {\bibfnamefont {P.}~\bibnamefont
  {San-Jose}}, \bibinfo {author} {\bibfnamefont {J.}~\bibnamefont
  {Gonz\'alez}}, \ and\ \bibinfo {author} {\bibfnamefont {F.}~\bibnamefont
  {Guinea}},\ }\href {\doibase 10.1103/PhysRevLett.108.216802} {\bibfield
  {journal} {\bibinfo  {journal} {Phys. Rev. Lett.}\ }\textbf {\bibinfo
  {volume} {108}},\ \bibinfo {pages} {216802} (\bibinfo {year}
  {2012})}\BibitemShut {NoStop}%
\bibitem [{\citenamefont {Bistritzer}\ and\ \citenamefont
  {MacDonald}(2011)}]{BM}%
  \BibitemOpen
  \bibfield  {author} {\bibinfo {author} {\bibfnamefont {R.}~\bibnamefont
  {Bistritzer}}\ and\ \bibinfo {author} {\bibfnamefont {A.~H.}\ \bibnamefont
  {MacDonald}},\ }\href {\doibase 10.1073/pnas.1108174108} {\bibfield
  {journal} {\bibinfo  {journal} {Proc. Natl. Acad. Sci. U.S.A.}\ }\textbf
  {\bibinfo {volume} {108}},\ \bibinfo {pages} {12233} (\bibinfo {year}
  {2011})}\BibitemShut {NoStop}%
\bibitem [{\citenamefont {Xie}\ \emph {et~al.}(2024)\citenamefont {Xie},
  \citenamefont {Ghaemi}, \citenamefont {Mitrano},\ and\ \citenamefont
  {Uchoa}}]{TopoExcitonXie}%
  \BibitemOpen
  \bibfield  {author} {\bibinfo {author} {\bibfnamefont {H.-Y.}\ \bibnamefont
  {Xie}}, \bibinfo {author} {\bibfnamefont {P.}~\bibnamefont {Ghaemi}},
  \bibinfo {author} {\bibfnamefont {M.}~\bibnamefont {Mitrano}}, \ and\
  \bibinfo {author} {\bibfnamefont {B.}~\bibnamefont {Uchoa}},\ }\href
  {\doibase 10.1073/pnas.2401644121} {\bibfield  {journal} {\bibinfo  {journal}
  {Proc. Natl. Acad. Sci. USA}\ }\textbf {\bibinfo {volume} {121}},\ \bibinfo
  {pages} {e2401644121} (\bibinfo {year} {2024})}\BibitemShut {NoStop}%
\bibitem [{\citenamefont {Jin}\ \emph {et~al.}(2020)\citenamefont {Jin},
  \citenamefont {Zhang}, \citenamefont {Liu}, \citenamefont {Dai},
  \citenamefont {Shen}, \citenamefont {Wang},\ and\ \citenamefont
  {Liu}}]{NodalRingPRB2020}%
  \BibitemOpen
  \bibfield  {author} {\bibinfo {author} {\bibfnamefont {L.}~\bibnamefont
  {Jin}}, \bibinfo {author} {\bibfnamefont {X.}~\bibnamefont {Zhang}}, \bibinfo
  {author} {\bibfnamefont {Y.}~\bibnamefont {Liu}}, \bibinfo {author}
  {\bibfnamefont {X.}~\bibnamefont {Dai}}, \bibinfo {author} {\bibfnamefont
  {X.}~\bibnamefont {Shen}}, \bibinfo {author} {\bibfnamefont {L.}~\bibnamefont
  {Wang}}, \ and\ \bibinfo {author} {\bibfnamefont {G.}~\bibnamefont {Liu}},\
  }\href {\doibase 10.1103/PhysRevB.102.125118} {\bibfield  {journal} {\bibinfo
   {journal} {Phys. Rev. B}\ }\textbf {\bibinfo {volume} {102}},\ \bibinfo
  {pages} {125118} (\bibinfo {year} {2020})}\BibitemShut {NoStop}%
\bibitem [{\citenamefont {Moon}\ and\ \citenamefont
  {Koshino}(2013)}]{TBGKoshinoPRB2013}%
  \BibitemOpen
  \bibfield  {author} {\bibinfo {author} {\bibfnamefont {P.}~\bibnamefont
  {Moon}}\ and\ \bibinfo {author} {\bibfnamefont {M.}~\bibnamefont {Koshino}},\
  }\href {\doibase 10.1103/PhysRevB.87.205404} {\bibfield  {journal} {\bibinfo
  {journal} {Phys. Rev. B}\ }\textbf {\bibinfo {volume} {87}},\ \bibinfo
  {pages} {205404} (\bibinfo {year} {2013})}\BibitemShut {NoStop}%
\end{thebibliography}%

\end{document}